\documentclass{jfm}
\usepackage{tikz}
\usetikzlibrary{patterns}
\usetikzlibrary{circuits}
\usetikzlibrary{circuits.ee.IEC}
\usetikzlibrary{arrows}
\usetikzlibrary{backgrounds}
\usetikzlibrary{quotes,angles}
\usetikzlibrary{shapes,chains}
\usepackage{verbatim}
\colorlet{lcnorm}{blue!20!black}
\graphicspath{{Figures/}}

\usepackage{xcolor}
\colorlet{lcnorm}{blue}
\usepackage{graphicx}
\usepackage{newtxtext}
\usepackage{newtxmath}
\usepackage{natbib}
\usepackage{hyperref}
\hypersetup{
    colorlinks = true,
    urlcolor   = blue,
    citecolor  = black,
}
\newtheorem{lemma}{Lemma}
\newtheorem{corollary}{Corollary}
\newcommand{\RomanNumeralCaps}[1]
\linenumbers

\title{The emission properties, structure and stability of ionic liquid menisci undergoing electrically-assisted ion evaporation}

\author{Ximo Gallud \aff{1}
  \corresp{\email{ximogc@mit.edu}},
  Paulo C. Lozano\aff{1}}

\affiliation{\aff{1}Massachusetts Institute of Technology, Department of Aeronautics and Astronautics, 77 Massachusetts Avenue, 02139 Cambridge, MA}

\begin{document}
\maketitle

\begin{abstract}
The properties and structure of electrically-stressed ionic liquid menisci experiencing ion evaporation are simulated using an electrohydrodynamic model with field-enhanced thermionic emission in steady state for an axially-symmetric geometry. Solutions are explored as a function of the external background field, meniscus dimension, hydraulic impedance and liquid temperature. Statically stable solutions for emitting menisci are found to be constrained to a set of conditions: a minimum hydraulic impedance, a maximum current output, and a narrow range of background fields that maximizes at menisci sizes of 0.5-3 microns in radius. Static stability is lost when the electric field adjacent to the electrode that holds the meniscus corresponds to an electric pressure that exceeds twice the surface tension stress of a sphere of the same size as the meniscus. Preliminary investigations suggest this limit to be universal, therefore independent of most ionic liquid properties, reservoir pressure, hydraulic impedance or temperature and could explain the experimentally observed bifurcation of a steady ion source into two or more emission sites.  Ohmic heating near the emission region increases the liquid temperature, which is found to be important to accurately describe stability boundaries. Temperature increase does not affect the current output when the hydraulic impedance is constant. This phenomenon is thought to be due to an improved interface charge relaxation enhanced by the higher electrical conductivity. Dissipated Ohmic energy is mostly conducted to the electrode wall. The higher thermal diffusivity of the wall versus the liquid, allows the ion source to run in steady state without heating.
\end{abstract}

\begin{keywords}
NA
\end{keywords}


\section{Introduction}
Electrospraying is a technique to extract charged particles from electrically-conductive liquid surfaces using strong electric fields. This technique can be implemented in various configurations, but most commonly consists of an electrode in the form of a capillary tube, through which fluid flows from a reservoir. A potential difference is then applied between the liquid and a downstream electrode, thus polarizing the liquid exposed at the end of the tube.

A fluid meniscus is formed in the cavity between the electrodes. The surface of the meniscus adopts a geometrical shape that results from the balance of electric, surface tension and hydrodynamic stresses. These forces depend on the applied potential, fluid flow rate, electrode configuration and liquid properties.

Electrospray sources can operate in various emission regimes. The most widely known is the cone-jet mode \citep{Cloupeau1989ElectrostaticMode}, where the meniscus has a conical shape near the contact line with the tube or Taylor cone \citep{Taylor1964DisintegrationField}, and  transitions into a fast-moving liquid jet close to the cone apex  \citep{Zeleny1935ThePressure}. The jet surface is inherently unstable and eventually breaks into droplets due to field-enhanced capillary instabilities \citep{Rayleigh1892XVI.Force}. The cone-jet mode has been widely studied in terms of its governing physics and the resulting spray structure \citep{FernandezdelaMora2007TheCones,Ganan-Calvo2009RevisionFocusing}, from which scaling laws have been derived for metrics such as the jet width, electric current output, and the size and mass per unit charge of resulting droplets \citep{Ganan-Calvo1997CurrentLaws,FernandezDeLaMora1994TheCones}. 

When the fluid flow rate is reduced, the characteristic dimension that controls the size of the jet and resulting droplets decreases, making the electric field, particularly in the cone-jet transition region and the jet termination \citep{Gamero-Castano2000DirectSurfaces,Gamero-Castano2002Electric-Field-InducedLiquid}, to become sufficiently large to trigger direct ion evaporation from the charged interface \citep{Iribarne1976}. The simultaneous ion evaporation from a cone-jet electrospray defines a second operational mode, characterized by the production of a mixed ion-droplet beam \citep{Perel1969ResearchThruster,Gamero-Castano2001ElectrosprayThrusters,Gamero-Castano2000DirectSurfaces}.

Under certain empirical conditions, namely a sufficiently high electric conductivity and surface tension, a further reduction of the fluid flow rate results in the pure emission of ions, characterized by the absence of any droplet current. While no direct visual observation of a stable meniscus in this mode is available, it is likely that the jet is quenched and ion emission occurs from a closed surface at the meniscus apex. According to cone-jet scaling laws \citep{FernandezDeLaMora1994TheCones}, the fluid flow rate corresponding to this regime is too low to support the formation of a stable jet.

The electrospray pure-ion evaporation mode is observed to exist only for a limited set of liquids, namely liquid metals \citep{Swanson1983LiquidApplications}, concentrated sulfuric acid solutions \citep{Perel1969ResearchThruster} and ionic liquids \citep{Romero-Sanz2003SourceRegime,Lozano2005IonicEmitters}. In addition to its interesting phenomenology, the pure ionic regime has recently gained significant attention for its potential applications in high-performance electric space propulsion \citep{Legge2011ElectrosprayMetals,Romero-Sanz2005IonicLiquids}, Focused Ion Beams (FIB) for etching and deposition \citep{Zorzos2008TheApplications,Perez-Martinez2011IonicApplications,Takeuchi2013DevelopmentModification} or ion microscopy \citep{Levi-Setti1985ProgressMicroanalysis,Sugiyama2004AMicroscopy}.

Ionic liquids are a type of molten salts that remain liquid at relatively low temperatures, including room temperature and sometimes much lower. Unlike conventional simple salts, ionic liquids are formed by complex molecular ions, which are poorly coordinated in part due to their asymmetric nature, and therefore require significantly lower temperatures to organize into a solid structure. However, also as in conventional salts, strong ionic interactions between their molecules result in extraordinarily low vapor pressures, allowing them to be exposed to a vacuum in their liquid state, practically without evaporation.

Ionic liquid ion sources (ILIS) are of special interest because they can be made of numerous combinations of organic molecules tailored to the specific requirements of each application \citep{Plechkova2008ApplicationsIndustry}.

Unlike Liquid Metal Ion Sources (LMIS), where space charge plays a primordial role to enhance the stability of the meniscus
by shielding the effects of external electric perturbations \citep{Gomer1979OnSources}, ILIS space charge effects
are less relevant, which makes the stability of the source more susceptible to the specific
properties of the working ionic liquid \citep{Garoz2007TaylorConductivity}, emitter geometry \citep{Castro2009EffectTips} and other perturbations.

Experimental challenges have
hindered a clear understanding of ILIS, specially the role of key operating parameters
such as the external electric field \citep{Krpoun2008AEmitters,Perez-Martinez2015IonMicrotips}, liquid temperature \citep{Lozano2005IonicEmitters}, and other physical and geometrical tip characteristics relevant to passive-type sources, such as the size of the inlet pores \citep{Courtney2015InfluencesSources}, electrode shape or hydraulic impedance of the
feeding material \citep{Castro2009EffectTips} and material dielectric properties \citep{Coffman2013OnSource}. Among these challenges are the current lack of non-destructive techniques to resolve the small scales of ILIS menisci ($\sim$ 1-5 $\mu$m) to interrogate the system in-situ, e.g., to capture the shape of the interface
profile, the nature of fluid interactions with the tip and the characteristics of internal creeping flow while confirming that the source is operating in the pure ionic mode, for example through simultaneous mass spectrometry of the beam. Electron microscopy \citep{Terhune2016Radiation-inducedField} has been attempted to observe the small menisci, however the electron beam interacts strongly with the charged surface making these observations uncertain at best. The lack of empirical evidence, emphasizes the relevance of studying these liquid structures through numerical simulations 

There is a large set of parameters that establish the operational characteristics of electrospray sources. In many ways, empirical determination of these characteristics becomes intractable given the vast number of parameter combinations that are possible. This fact has motivated the development of computational models that aim to improve the understanding of the fundamental physics of the electrospray emission process. In the cone-jet literature, many simulation frameworks have been developed based on the Taylor-Melcher leaky dielectric model \citep{Saville1997ELECTROHYDRODYNAMICS:TheModel}, which have been successful in validating how emission properties and characteristic length scales are accurately represented by universal scaling laws \citep{Pantano1994Zeroth-orderMode,Higuera2003FlowCone,Gamero-Castano2019NumericalMode,Herrada2012NumericalMode,Collins2008ElectrohydrodynamicLiquidcones}.

The Taylor-Melcher leaky dielectric model is valid in the limit when the electric charge relaxation time is very short compared to the scale of the fluid hydrodynamic time, and the charge is relaxed at the meniscus interface, therefore assuming quasi-neutrality in the bulk fluid and fully conductive charge transport. This fact has shown to be not valid for transient ultra-fast flows such as the onset of the electrospray first droplet ejection \citep{Ganan-Calvo2016TheEjection,Pillai2016ElectrokineticsDrops}, where the hydrodynamic timescales become on the order of the charge relaxation time and bulk charge convection becomes relevant.

Furthermore, the Taylor-Melcher leaky dielectric model has not been fully developed to capture the onset of pure ion evaporation from a closed interface. Ion evaporation is a highly non-linear activated process, which is usually modelled in a similar way to classical field-enhanced thermionic emission where a critical electric field is required to reach a state of substantial ion evaporation \citep{Iribarne1976}.

Interfacial charge transport is governed by this activated process and therefore the need for special numerical techniques added to the standard Taylor-Melcher leaky dielectric model to capture its behavior. First efforts introducing surface charge transport for pure ionic emission include the work of \cite{Higuera}, who simulated an ionic liquid drop attached to a flat conducting plate. Equilibrium meniscus shapes were obtained by sequentially solving the Laplace field equation outside and inside the droplet (no space charge was considered) with the activated emission condition derived by \cite{Iribarne1976}. Electric and surface tension stresses were placed as a boundary conditions for a Stokes flow solver. By using the interfacial velocity distributions coming from Stokes flow and a second order Runge-Kutta temporal integration method, Higuera propagated the interface along time-steps towards the equilibrium solution. 

Higuera considered two cases. In the first case of constant meniscus volume, the author was able to sketch out the concept of starting voltage seen in the I-V (current vs voltage) traces, which is experimentally observed \citep{Krpoun2008AEmitters}. The current increase with the electric field yielded a linear behaviour before it got unstable at a particular electric field. The same scaling
relationship is reported by a number of empirical studies and it is believed to be due to the limits in conductive charge transport within ionic liquids \citep{Legge2011ElectrosprayMetals,Lozano2005IonicEmitters,Courtney2012EmissionSources}.

In the second case, Higuera considered an external reservoir capable of pumping fluid with pressure $p_0$ towards the meniscus, and the pressure drop that occurs because of friction of the fluid with the channel walls that connect the reservoir to the external electrodes (hydraulic impedance). The non-dimensional total current emitted versus non-dimensional field was shown to be very dependent on $p_0$ and the hydraulic impedance coefficient, yielding currents with abnormal dissimilar behaviour (up to 3 orders of magnitude difference for relatively similar values of $p_0$ and hydraulic impedance coefficient).

Regardless of the limitations of Higuera's model, the author was able to depict the notion of a maximum external field, which suggests that purely ionic emission might only be permissible within a narrow band of stability. The numerical variability for the current in the second case as a function of $p_0$ and the hydraulic impedance coefficient points out the importance of upstream conditions in determining emission behavior, which is in agreement with experimental work.

\cite{Coffman2016} updated Higuera's model by removing volumetric
constraints, by including a substantial fraction of the liquid feeding system in the computational domain and by introducing Ohmic heating effects, which were predicted to play an important role in the current output. 

Coffman's free volume generalization of the problem initialized by Higuera took three main input parameters, namely the electric field downstream $E_0$, a characteristic meniscus size $r_0$ and an hydraulic impedance coefficient $C_R$. The author's model unveiled a set of sharper family of emitting equilibrium shapes that sustained pure ion evaporation for high values of $E_0$. These solutions exist under a specific set of conditions, namely limited ranges of external $E_0$ and meniscus dimension $r_0$ ($1\sim 5 $ $\mu$m). These ranges would expand if sufficient hydraulic impedance is provided.

Coffman was able to reproduce the constant volume solutions of Higuera (no feeding channel) and categorize them in a set of solutions of particularly small size ($r_0 \sim 250$ nm), a low capillary number and high dielectric constant. This combination of parameters yielded equilibrium solutions that were practically hydrostatic, and with a depleted distribution of surface charge in such a way that the evaporation process was generally decoupled from the balance between the surface tension and the electric stresses.


This extended Higuera's solutions to a higher range of electric fields with stable solutions for relatively large meniscus sizes at sufficient hydraulic impedance, which were reported to exist experimentally by \cite{Castro2009EffectTips} and \cite{Romero-Sanz2003SourceRegime}. Coffman reported an increase of the electric field stability range for higher hydraulic impedance and an inverse proportionality relationship between the hydraulic impedance and total emitted current. The trade-off between the stability increase and the reduction in current throughput was found to be in agreement with the experimental findings in \citep{Lozano2005IonicEmitters}. 

Owing to the size of the problem (more than 10 independent non-dimensional numbers and 5 variables), lack of computational power and the constraints imposed by commercial solvers  (mesh resolution limitations, no parallelization), \cite{Coffman2016StructureIons} only report a moderate exploration of the region of stability as a function of the aforementioned input parameters, does not investigate Ohmic heating effects on stability and current emission, neglect volumetric charge effects due to temperature gradients and couple the hydraulic impedance coefficient to the meniscus size. 


The work presented here leverages the electrohydrodynamic model (EHD) with charge evaporation by \cite{Coffman2016StructureIons} and extends it to include bulk free charges originated by variable conductivity coefficients, presenting the results for a hydraulic impedance coefficient independent of the meniscus size. More importantly, this work provides a detailed exploration of the stability regions and their interdependence on relevant metrics, such as menisci contact angles with the flat electrode and total current emitted. Based on these extensions, it appears that upper stability limits are a result of two competing phenomena. The first one is given by the maximum current output that a static evaporating meniscus can provide, while the second responds to a maximum electric pressure a meniscus can withstand before no static solutions can be found. The bifurcation of a static meniscus could be a possible outcome of this situation, which is reminiscent of what is experimentally observed in this type of ion sources. Numerical results suggest that this presumed bifurcation may represent a universal limit for all working liquids experiencing pure-ion emission with negligible space charge.

Results indicate that an accurate resolution of the aforementioned limits of stability cannot be provided without considering energy effects. In this regard, simulations show how heated menisci can typically access to a higher range of stable electric fields though the increase of electrical conductivity near the emission region. 

A detailed description of the numerical procedure is also provided to find the equilibrium solutions and information regarding the influence of Ohmic heating in relation to the emission properties and stability boundaries. Section \ref{sec:leaky_d_model}, presents the electrohydrodynamic model adapted to tackle charge evaporation and the domain of simulation. Section \ref{sec:numerical_procedure} summarizes the numerical details used to solve the equations of the model. Section \ref{sec:results}, presents and discusses the static stability of the equilibrium solutions found in the model. Finally, the conclusions, future efforts and limitations are presented in section \ref{sec:conclusions}.

\section{Description of the EHD model with electrically assisted charge evaporation}
\label{sec:leaky_d_model}
\subsection{Geometrical domain}

\begin{figure*}
\centering
\begin{tikzpicture}[scale=1.2]
\def\cwidth{3}
\def\cheight{3}
\def\lthick{0.5}
\def\ewidth{12}
\def\eheight{5}
\def\x0{-0.5}
\def\y0{0}
\def\h{0.75}
	\coordinate (origin) at (\x0,\y0);
	\coordinate (symorigin) at (\x0+\cwidth,\y0);
	\coordinate (axissymdown) at (\x0/2+\cwidth/2,\y0);
	\coordinate (axissymup) at (\x0+\cwidth/2,\y0+\eheight+\cheight);
    \draw[thick] (origin) --++(90:3cm);
    \fill[pattern=north west lines] (origin) rectangle ++(-\lthick,\cheight);
    \fill[pattern=north west lines] (\x0-\lthick,\y0+\cheight-\lthick) rectangle ++(-\ewidth/2 + \cwidth/2 + \lthick,\lthick);
    \draw[thick] (symorigin) --++(90:3cm);
    \fill[pattern=north west lines] (symorigin) rectangle ++(\lthick,\cheight); 
	\fill[pattern=north west lines] (\x0+\cwidth+\lthick,\y0+\cheight-\lthick) rectangle ++(\ewidth/2 - \cwidth/2 - \lthick,\lthick); 
    \draw[thick] (symorigin)+(0,\cheight) --++ (\ewidth/2-\cwidth/2,\cheight);
	\draw[thick] (origin)+(0,\cheight) --++ (-\ewidth/2+\cwidth/2,\cheight);
	
    \draw[thick] (axissymup) --++ (-\ewidth/2,0);
    \draw[thick] (axissymup) --++ (\ewidth/2,0);
    \fill[pattern=north west lines] (axissymup) rectangle ++(\ewidth/2,\lthick);
    \fill[pattern=north west lines] (axissymup) rectangle ++(-\ewidth/2,\lthick);

	\draw[thick] plot[smooth,tension=0.5] coordinates{(\x0,\cheight) (\x0+0.4*\cwidth,\cheight+1) (\x0+0.5*\cwidth,\cheight+1.2) (\x0+0.6*\cwidth,\cheight+1) (\x0+\cwidth,\cheight)};
	\draw[] (\x0+0.4*\cwidth,\cheight+1) rectangle (\x0+0.6*\cwidth,\cheight+1.4);

	\draw (origin) parabola bend (\x0+\cwidth/2,0.75) (symorigin);
	
	\foreach \xp in {-0.25,0,...,2.25} {
      \draw[-latex] (\xp,0) -- ++(0,-\xp*\xp/3+2*\xp/3+5/12);
    }
    \coordinate (A) at (\x0-2*\lthick,\cheight);
    \coordinate (B) at (\x0,\cheight);
    \coordinate (C) at (\x0+0.4*\cwidth,\cheight+1);
  	\draw pic["$\theta$",draw=black,latex'-latex',angle eccentricity=1.2,angle radius=0.7cm] {angle=C--B--A};
    \draw[thick] (\ewidth/2-3*\lthick+\cwidth/2,\cheight) -- (\ewidth/2-3*\lthick+\cwidth/2,\cheight+\eheight);
    \draw (\ewidth/2-3*\lthick+\cwidth/2,\cheight+\eheight/2) node [circle, draw, fill=white!30] {V};
   
    \draw [thick,dash dot] (\x0+\cwidth/2,-0.25) -- (\x0+\cwidth/2,\cheight+\eheight+\lthick+0.25);
	\node[] at (\x0+4*\cwidth/6,\lthick+3*\cheight/4) {$\mathbf{\Omega}_l$};
	\node[] at (\x0+\cwidth+\ewidth/8,\cheight+\eheight/2) {$\mathbf{\Omega}_v$};
	\node[] at (\x0+\cwidth+\ewidth/8,\cheight+\lthick/2) {$\Gamma^v_D$};
	\node[] at (\x0+7*\cwidth/8,\cheight/2) {$\Gamma^l_D$};
	\node[] at (\x0+\cwidth/2+\lthick/1.5,\cheight/2) {$\Gamma^l_L$};
	\node[] at (\x0+\cwidth/2,-0.6) {$\Gamma_I$};
	\node[] at (\x0+\cwidth/2+2*\lthick,1.8*\lthick+\cheight) {$\Gamma_M$};
	\node[] at (\x0+\cwidth/2+\lthick/1.5,\cheight+\eheight/2) {$\Gamma^v_L$};
	\node[] at (\x0+\cwidth+\ewidth/8,\cheight+\eheight-\lthick/2) {$\Gamma_U$};
	\node[] at (\x0+\cwidth/2+\ewidth/2-\lthick/1.5,\cheight+\eheight/2+3*\lthick) {$\Gamma_R$};
	
	\draw[-latex,thick] (\x0+1.75*\cwidth+\lthick,\y0 +\cheight+\eheight/2+\lthick) -- (\x0+1.75*\cwidth+\lthick,\y0 +\cheight+\eheight/2+3*\lthick);
	\node[] at (\x0+1.65*\cwidth+\lthick,\y0 +\cheight+\eheight/2+2*\lthick) {$E_0$};
	
	\draw[latex'-latex'] (\x0-2*\lthick,\y0) -- (\x0-2*\lthick,\cheight);
	\draw[latex'-latex'] (\x0+\cwidth/2-\ewidth/2+\lthick,\cheight) -- (\x0+\cwidth/2-\ewidth/2+\lthick,\cheight+\eheight);
	\draw[latex'-latex'] (\x0+\cwidth/2-\ewidth/2,\cheight+\eheight+1.65*\lthick) -- (\x0+\cwidth/2,\cheight+\eheight+1.65*\lthick);
	\draw[-latex] (\x0+\cwidth/2,3*\cheight/4) -- (\x0,3*\cheight/4);
	\draw (\x0+\cwidth/2-\ewidth/2,\cheight+\eheight) --++ (0,2*\lthick);
	\draw (\x0+\cwidth/2,\cheight+\eheight+\lthick+0.25) --++ (0,\lthick/2);
	\draw (\x0+-2*\lthick-0.25,\y0) -- (\x0,\y0);
	\node[] at (\x0+\cwidth/2-\ewidth/2+0.5*\lthick,\cheight+\eheight/2) {$z_0$};
	\node[] at (\x0+\cwidth/2-\ewidth/4,\cheight+\eheight+2.2*\lthick) {$r_p$};
	\node[] at (\x0+\cwidth/4,3*\cheight/4+0.25) {$r_0$};
	\node[] at (\x0+-2*\lthick-0.25,\cheight/2) {$z_p$};
	
	\node[] (emission) at (\x0+1.65*\cwidth-\ewidth/2-0.5*\lthick+0.15,\cheight+\eheight/2+1*\lthick) {\includegraphics[width=.47\textwidth]{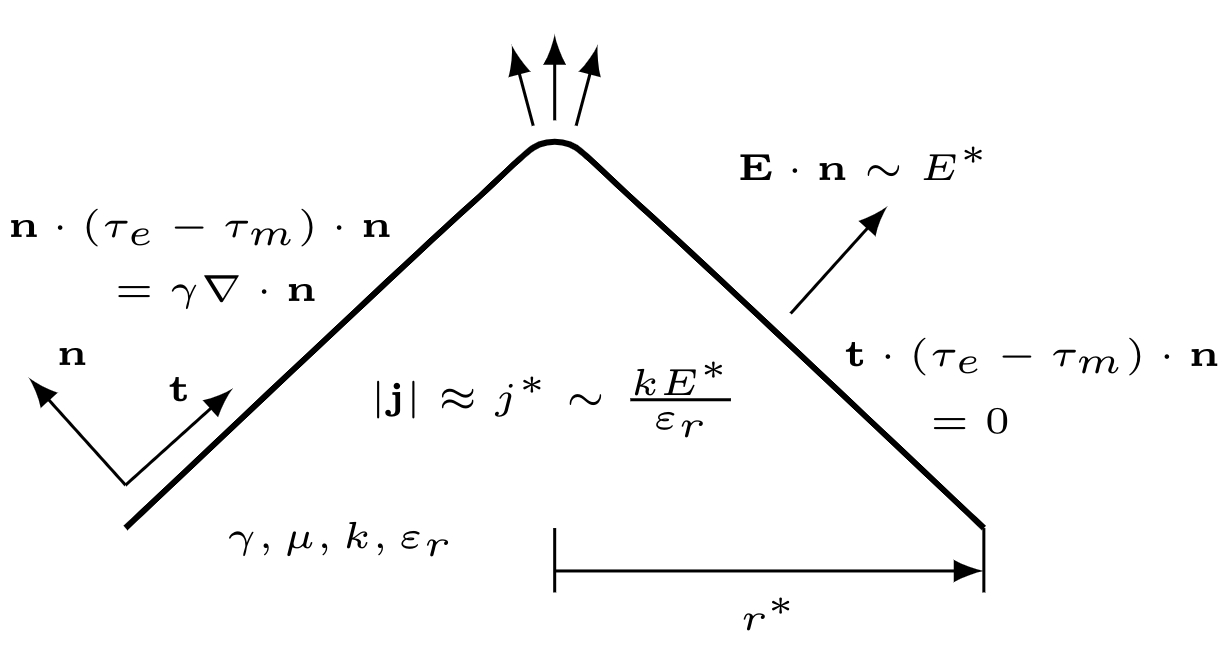}};
	\node[] (tip) at (\x0+0.5*\cwidth,\cheight+1.2){};
	\draw[dashed] (\x0+0.3*\cwidth,\cheight+2.4) -- (\x0+0.6*\cwidth,\cheight+1.4);
	\draw[dashed] (\x0+0.4*\cwidth,\cheight+1) -- (\x0-2*\lthick,\cheight+1.4);
	
\end{tikzpicture}

    \caption{Computational domain diagram, boundary nomenclatures and characteristic dimensions of the problem.}
    \label{fig:computational_domain}
\end{figure*}

The geometry of the computational domain is similar to that considered in \cite{Coffman2016StructureIons} and is shown in figure \ref{fig:computational_domain}. 
The geometry consists of an axially symmetric fluid channel of radius $r_0$ that terminates on a conducting flat electrode ($\Gamma_D$). 
This electrode is biased to a potential difference $\Delta V = -E_0 z_0$ with respect to another downstream flat electrode ($\Gamma_U$) located at a distance  $z_0$  from the fluid channel, where $E_0$ is the downstream electric field. The channel is filled with ionic liquid ($\mathbf{\Omega}_l$). There is a vacuum in the volume between the bottom flat electrode and  liquid surface and the downstream electrode ($\mathbf{\Omega}_v$). A fluid reservoir at pressure $p_r$ feeds liquid into the channel. This reservoir is not treated computationally. The fluid enters the computational domain at $\Gamma_I$, which is at a distance $z_p$ from the downstream electrode, as if it were the outlet of a fully developed pipe flow (Hagen-Poiseuille paraboloidal flow). The fluid meniscus ($\Gamma_M$) separating the vacuum and wetted regions is fixed (pinned) to the rim of the fluid channel and free to adopt any value of $\theta$.
The vacuum region width is large enough ($\frac{r_p}{r_0} = \frac{z_p}{r_0}=  20$) to ensure the downstream electric field remains undisturbed by the meniscus.
\subsection{Physics of pure-ion evaporation}

It is assumed in this work that pure-ion evaporation in high conductivity fluids like ionic liquids can be described as an activated process of the form:
\begin{equation}
j^e_n = \frac{\sigma k_B T}{h} \exp{\left(\frac{-E_a}{k_B T}\right)}
\label{eq:arrhenius}
\end{equation}

Where $j^e_n = \mathbf{j} \cdot \mathbf{n}$ is the local current density emitted at the surface of the meniscus, $E_a$ is the activation energy, $T$ is the liquid
temperature, $\sigma$ is the surface charge on $\Gamma_M$ and $k_B$ and $h$ are the Boltzmann and Planck constants, respectively \citep{Iribarne1976}.

The activation energy can be considered to be a function of the free energy of solvation for the extraction of a specific type of ion $\Delta G$ (of the order
of 1-2 eV for many solvated ions). In the presence of an electric field, it is also a function of the electric field perpendicular to the meniscus interface in the vacuum $E^v_n = \mathbf{E}\cdot \mathbf{n}$. This function $G\left(E^v_n\right)$,  encompasses the effect of the electric field required to bring this ion from an undisturbed region at infinity to the surface. Overall, the activation energy becomes $E_a = \Delta G - G(E^v_n)$.
An image charge argument can be brought into consideration when analyzing the dependence of $G(E^v_n)$ with respect to the normal component of the external electric field. In the limit of a planar interface geometry, this function can be approximated by: 
\begin{equation}
    G(E^v_n) = \sqrt{\frac{q^3 E^v_n}{4\pi\varepsilon_0}}
\end{equation}

Where $q$ is the charge of the ion ejected, and $\varepsilon_0$ is the electric permittivity of vacuum. When $G(E^v_n) \sim \Delta G$, the ion evaporation kinetics (equation \ref{eq:arrhenius}) increases to the level that charges are emitted from the meniscus tip region.
An estimation of the value of the critical electric field at which this occurs is:
\begin{equation}
    E^* = \frac{4 \pi \varepsilon_0 \left(\Delta G\right)^2}{q^3}
    \label{eq:E_critical}
\end{equation}
For typical values of ionic liquids, this critical electric field is on the order of $10^9 \frac{V}{m}$.

This value of electric field can be used to determine the characteristic size of the emission region when neglecting hydrodynamic pressure. The electric pressure in the vicinity of the emission region must balance the surface tension stress of the liquid surface, which is given by a curvature $\left(\frac{2}{r^*}\right)$ when the emission region is approximated as a spherical cap of radius $r^*$. 

Explicitly, the balance of stresses in the normal direction should be:
\begin{equation}
    \frac{1}{2}\varepsilon_0 {E^v_n}^2 - \frac{1}{2}\varepsilon_0 \varepsilon_r {E^l_n}^2 = \frac{2\gamma}{r^*}
    \label{eq:ord_mag_normal_stress}
\end{equation}

Where $E^l_n = \mathbf{E} \cdot \mathbf{n}$, is the local electric field perpendicular to the meniscus surface in the liquid. To a first approximation, the ionic liquid meniscus behavior approaches that of a perfect dielectric fluid where $E^l_n \approx \frac{E^v_n}{\varepsilon_r}$. If the meniscus is emitting, it will adapt its surface shape so that $E^v_n \sim E^*$. Using these two assumptions, the balance of stresses in  (\ref{eq:ord_mag_normal_stress}) yields:
\begin{equation}
    \frac{1}{2}\varepsilon_0 {E^*}^2 \frac{\varepsilon_r -1}{\varepsilon_r} = \frac{2\gamma}{r^*}
\end{equation}
For ionic liquids where $\varepsilon \gg 1$, the characteristic emission radius yields:
\begin{equation}
r^* = \frac{4\gamma}{\varepsilon_0 {E^*}^2}
\label{eq:r_star}
\end{equation}
Where $r^*$ is on the order of $50$ nm.

The total current emitted in the surroundings of $r*$ can be stated as:
\begin{equation}
    I^* \approx j^*A \approx \kappa E^l_n A \approx \frac{\kappa E^*}{\varepsilon_r}\pi r^{*^2} = \frac{16\pi \kappa \gamma^2}{\varepsilon_0^2 \varepsilon_r {E^*}^3} 
    \label{eq:istar}
\end{equation}
Where $j^* \approx \kappa E^l_n \approx \frac{\kappa E^*}{\varepsilon_r}$ is the characteristic current density in the emission region, $\kappa$ is the electrical conductivity and $A = \pi r^{*^2}$ is a characteristic cross section area of the emission region.
For typical ionic liquid ion sources, $I^*$ is on the order of 50 to 500 nA. Mass conservation allows to give an approximate order of magnitude of the velocity in the bulk liquid and near the emission region:
\begin{equation}
    u^* = \frac{j^*}{\rho \frac{q}{m}}
    \label{eq:order_ustar}
\end{equation}
Where $\rho$ is the density of the ionic liquid, and $m$ the mass of the ions ejected. For ionic liquids, $u^*$ is very small, on the order of 0.1 $\frac{\text{m}}{\text{s}}$.

Once they have been emitted, energy conservation can be used to approximate its velocity in the vacuum $\nu^*_e$ right after traveling a distance $r^*$, therefore still very close to the meniscus emission region:
\begin{equation}
    \frac{1}{2} m v^{*^2}_e \approx q \Delta \Phi^* \label{eq:cons_energy_PIR}
\end{equation}

In this case, $\Delta \Phi^* \approx E^* r^*$ is an approximation to the potential drop after this distance. The Poisson equation in this region yields:
\begin{equation}
    \nabla \cdot \left(\varepsilon_0 \mathbf{E} \right) = \rho_{sc}
\end{equation}

Which can be approximated to a first order to give an order magnitude of the field increase due to space charge:

\begin{equation}
\varepsilon_0 \frac{\Delta E}{r^*} \sim \rho_{sc} \sim \frac{j^*}{v_e^*}
\label{eq:poisson_first_order}
\end{equation}

Eq. \ref{eq:poisson_first_order} can be rearranged in relative terms to the critical electric field by using eqs. \ref{eq:r_star}, \ref{eq:istar} and \ref{eq:cons_energy_PIR} as:
\begin{equation}
\frac{\Delta E}{E^*} \sim \frac{\kappa}{\varepsilon_0 \varepsilon_r} \frac{r^*}{\sqrt{\frac{2q}{m}E^* r^*}} \sim \frac{\tau_{p}}{\tau_{e}} \sim \sqrt{\frac{I^*}{8 \pi \frac{q}{m}\gamma} \frac{\kappa}{\varepsilon_0 \varepsilon_r}}
\label{eq:spch_approx}
\end{equation}

Where $\tau_p = \frac{r^*}{\sqrt{\frac{2q}{m}E^* r^*}}$ is the characteristic passing time (time that an ion takes to move past the emission region $r^*$), and $\tau_e = \frac{\varepsilon_0 \varepsilon_r}{\kappa}$ is the characteristic charge relaxation time (time that an ion takes to move from the bulk liquid to the interface where it is ejected due to thermoionic emission).  

For materials such as ionic liquids ($\kappa \sim 1 \; \frac{\text{S}}{\text{m}}$), relatively long charge relaxation times  compared to the ion passing time in the emission region originate negligible modifications of the electric field due to space charge, that is $\frac{\Delta E}{E^*} \sim \frac{\tau_p}{\tau_e}$ is on the order of $10^{-2}$ to $10^{-1}$. High conductivity liquids such as liquid metals have very short charge relaxation times compared to ion passing times and space charge dominates the magnitude of the electric fields near the emission region, thus yielding $\frac{\Delta E}{E^*}$ on the order of $10^0$.

This work uses the surface charge approximation and does not resolve the Debye layer along the meniscus interface. While the structure of the Debye layer is still not totally established in ion evaporation conditions in ionic liquids (electrode-free), the characteristic size of the electrical double layer ($\delta$) in ionic liquids in contact with adjacent electrodes is certainly better known. The Debye layer thickness is molecular in scale, at most $\delta \sim 10^{-9}$ m \citep{Bazant2011DoubleCrowding,Smith2016TheConcentration,Gebbie2015Long-rangeLiquids}. This value is two orders of magnitude larger than the Debye length for ionic liquids when computed with conventional formulations ($\delta_{DL} \sim 10^{-11}$ m), although such sizes do not make much physical sense given the relatively large size of ionic liquid molecules. In any event, these values are at least an order of magnitude smaller than the $r^* \sim 50$ nm that characterize the smallest liquid domain in this problem. Modifications to include Debye layer effects would likely yield more accurate results, yet the surface charge approach performed in this article predicts quite well the magnitude of emitted current, matching what is typically observed in experiments ($I \sim I^*)$, as seen in the following sections.



\subsection{Model equations}
The conditions to generate an emitting free-volume ionic liquid ion source emitting under the aforementioned phyisical characteristic magnitudes in steady state ($E^*$, $r^*$, $I^*$)  are highly dependent on the geometrical characteristics of the electrodes, external field, upstream fluid conditions and physical properties of the source working liquid.

 The fluid comes from a propellant reservoir at pressure $p_r$ and enters the computational domain at a pressure $p = p_r - \Delta p$ at the inlet $\Gamma_I$, where the pressure drop $\Delta p = Q Z$ is modeled using the standard Darcy law in which $Q$ is the total fluid volumetric flow rate and $Z$ is the hydraulic impedance of the channel.  The volumetric flow rate can be written as a function of the emitted current $I$ using the linear transformation $Q = \frac{I}{\rho \frac{q}{m}}$, where $\frac{q}{m}$ is an average charge-to-mass ratio of the evaporated ions.

The current emitted is an indirect result coming from the equilibrium solution shape of the free-volume meniscus for given electrode geometry, physical properties of the liquid, $E_0$, $p_r$ and $Z$.

The incompressible liquid flows along the liquid column ($\mathbf{\Omega}_l$) towards the vicinity of the emission region ($r^*$) when forced by the electric stresses acting on the surface of the meniscus.  Mass is emitted perpendicular to the surface of the meniscus $\Gamma_M$ in the form of a continuous current density of ions $j^e_n = \mathbf{j}\cdot \mathbf{n}$. The conductivity is assumed to depend linearly with temperature:
\begin{equation}
    \kappa(T) = \kappa_0 + \kappa'(T-T_0)
    \label{eq:conductivity}
\end{equation}
Where $\kappa_0$ is the conductivity of the ionic liquid at a reference temperature $T_0$ and $\kappa'$ is a constant sensitivity coefficient of the conductivity to temperature. As the space charge $\rho_{sc}$ for ionic liquid ion sources can be neglected to a first order approximation, the electric stresses are calculated by solving the Laplace equation in the vacuum domain and the Poisson equation and charge conservation equations in the liquid domain. 
The Maxwell-Faraday equation yields for both liquid and vacuum domains:
\begin{equation}
    \nabla \times  \mathbf{E} = 0 \qquad \text{in} \qquad \mathbf{\Omega}_l \cup \mathbf{\Omega}_v
    \label{eq:max_faraday}
\end{equation}
Equation \ref{eq:max_faraday} is equivalent to writing the electric field as the derivative of an electric potential $\mathbf{E} = -\nabla \phi$.
The Laplace and Poisson equations in the vacuum and liquid domains can be expressed as:
\begin{equation}
    \nabla \cdot \left(\varepsilon_0 \mathbf{E} \right) =      -\varepsilon_0\nabla^2 \phi = \rho_{sc} \approx 0 \qquad \text{in} \qquad \mathbf{\Omega}_v
    \label{eq:poisson_vac}
\end{equation}
\begin{equation}
    \nabla \cdot \left(\varepsilon_0 \varepsilon_r \mathbf{E} \right) =  - \varepsilon_0 \varepsilon_r \nabla^2 \phi  = \rho_{m} \qquad \text{in} \qquad \mathbf{\Omega}_l
    \label{eq:poisson_liq}
\end{equation}
Where $\rho_{m}$ is the charge density in the bulk fluid.
\\
The Poisson equation on the interface domain can be expressed as:
\begin{equation}
\varepsilon_0 E^n_v - \varepsilon_0 \varepsilon_r E^n_l = \sigma \qquad \text{on} \qquad \Gamma_M
    \label{eq:interface_charge_condition}
\end{equation}
Where $\sigma$ is the surface charge density along the meniscus interface $\Gamma_M$.  The charge conservation equation is defined for the bulk liquid and the meniscus interface as:
\begin{equation}
    \nabla \cdot \left(\kappa(T) \mathbf{E} + \rho_m \mathbf{u} \right) = 0 \qquad \text{in} \qquad \mathbf{\Omega}_l
    \label{eq:charge_conserv_bulk}
\end{equation}

Eq. (\ref{eq:charge_conserv_bulk})  contains two terms associated to the conductive ($\mathbf{j}_{cond} = \kappa(T)\mathbf{E}$) and convective ($\mathbf{j}_{conv} = \rho_m \mathbf{u}$) bulk charge transport. The bulk convective charge transport term can be neglected due to the fact that $j^* >> u^*$ (eq. \ref{eq:order_ustar}) for typical physical parameters of ionic liquids, namely $\rho \sim O(10^3)$ $\frac{\text{kg}}{\text{m}^3}$, $\frac{q}{m} \sim O(10^6)$ $\frac{\text{C}}{\text{kg}}$. 

If that is the case, an expression can be obtained for $\rho_{m}$ as a function of the electric field in $\mathbf{\Omega}_l$ by substituting $\mathbf{j} = \kappa(T)\mathbf{E}$ into the charge conservation equation (\ref{eq:charge_conserv_bulk}) and subtracting  (\ref{eq:poisson_liq}). This yields:
\begin{equation}
    \rho_m = \frac{-\varepsilon_0 \varepsilon_r \nabla \kappa(T) \cdot \mathbf{E}}{\kappa(T)}
    \label{eq:rho_m}
\end{equation}

Notice from eq. \ref{eq:rho_m} that the breakup of quasi-neutrality is originated by spatial gradients in conductivity. The dependency of the conductivity with temperature (eq. \ref{eq:conductivity}) combined with temperature gradients in the bulk fluid originate this space charge.

Analogously, eq. (\ref{eq:charge_conserv_meniscus}) is the charge conservation equation defined for the meniscus interface, where the interfacial charge convection (left hand side) balances the conductive current density entering the interface, and the evaporated current density (first and second terms of the right hand side, respectively). The operator $\nabla_S$ appearing in the convective charge transport expression is the tangential surface gradient or the gradient of $\sigma$ in the direction tangent to $\Gamma_M$ (see \cite{Saville1997ELECTROHYDRODYNAMICS:TheModel}). 

\begin{equation}
\begin{split}
  \mathbf{u} \cdot \nabla_S \sigma - \sigma \mathbf{n}\cdot \left(\mathbf{n} \cdot \nabla \right)  \mathbf{u} = \kappa \left(T\right)E^n_l - j_n^e \qquad
   \text{on} \qquad \Gamma_M
\end{split}
    \label{eq:charge_conserv_meniscus}
\end{equation}

The rest of the boundary conditions for the electric problem are:
\begin{align}\begin{split}
    \phi = 0 &\qquad \text { on } \qquad \Gamma_I \cup \Gamma^l_D\cup\Gamma^v_D\\
    \phi =-E_0 z_0 & \qquad \text{ on } \qquad \Gamma_U \\ -\nabla \phi \cdot \mathbf{n}=0 & \qquad \text{ on }\qquad \Gamma^v_L \cup \Gamma^l_L \cup \Gamma_R.
\end{split}
\label{eq:Electric_BC}
\end{align}

The dynamics of the fluid are described by the incompressible steady state Navier-Stokes equations. 
\begin{equation}
    \nabla \cdot \mathbf{u} = 0 \qquad \text{in} \qquad \mathbf{\Omega}_l
    \label{eq:mass_cons}
\end{equation}
\begin{equation}
\begin{split}
    \rho \left(\mathbf{u} \cdot \nabla \right) \mathbf{u} &= \nabla \cdot \tau_f + \rho_m \mathbf{E} \quad \text{in} \quad \mathbf{\Omega}_l
    \end{split}
    \label{eq:navier-stokes}
\end{equation}

Where $\rho$ is the ionic liquid density, $\mathbf{u}$ is the fluid velocity and $\tau_f$ is the viscous fluid stress tensor. The fluid stress tensor yields:

\begin{equation}
    \tau_f = -p \mathbf{I} + 2 \mu \mathbf{e} = -p \mathbf{I} + \mu \left( \nabla \mathbf{u} + \nabla \mathbf{u}^T\right)
\end{equation}

Where $p$ is the bulk pressure, $\mu$ is the visosity of the fluid and $\mathbf{e} = \frac{1}{2} \left( \nabla \mathbf{u} + \nabla \mathbf{u}^T\right)$ is the strain rate tensor. It is observed that the product of fluid viscosity $\mu$ and electrical conductivity is weakly dependent of temperature in ionic liquids \citep{Zhang2006PhysicalEvaluation}. That is, $\kappa(T)\mu(T) = \kappa_0 \mu_0$. Viscosity is modeled as follows:

\begin{equation}
    \mu(T) = \frac{\kappa_0 \mu_0}{\kappa_0 + \kappa'(T-T_0)}
    \label{eq:viscosity}
\end{equation} 

to keep the extent of this relationship valid in these simulations, as in \cite{Coffman2016StructureIons}.

The balance of stresses in the normal and tangential direction to the interface $\Gamma_M$ are respectively:
\begin{equation}
\label{eq:equilibrium_stresses_normal}
\begin{split}
    \mathbf{n}\cdot\left(\tau^v_e - \tau^l_e - \tau_f\right)\cdot \mathbf{n} = \gamma \nabla \cdot \mathbf{n}
     \quad \text{on} \quad \Gamma_M
    \end{split}
\end{equation}
\begin{equation}
    \mathbf{t}\cdot\left(\tau^v_e - \tau^l_e - \tau_f\right)\cdot \mathbf{n} = 0
     \quad \text{on} \quad \Gamma_M
    \label{eq:equilibrium_stresses_tangential}
\end{equation}

Where $\gamma$ is the surface tension coefficient and $\tau^l_e$, $\tau^v_e$ are the electric stress tensors in the liquid and vacuum respectively.

The fluid enters the computational domain as fully developed pipe flow at the inlet ($\Gamma_I$), namely constant pressure and negligible shear stress at all the channel cross section:
\begin{align}
\label{eq:Hydraulic_BC}
    \begin{split}
      \mathbf{n}\cdot \tau_f\cdot \mathbf{n}= -p = - \left(p_r - \Delta p\right) \qquad \text{ on } \qquad \Gamma_I \\ \mathbf{t}\cdot \tau_f\cdot \mathbf{n}=0 \qquad \text{ on } \qquad \Gamma_I
    \end{split}
\end{align} 
Where $p_r$ is the pressure at the reservoir and $\Delta p = \frac{I}{\rho \frac{q}{m}} Z$ is the pressure drop caused by the friction of the fluid with the walls.  

The fluid does not slip on the walls, thus:
\begin{equation}
    \mathbf{u}=0 \qquad \text{ on } \qquad  \Gamma^l_D
    \label{eq:non_slip}
\end{equation}

The mass conservation at the interface yields:

\begin{equation}
j^e_n = \rho \frac{q}{m} \mathbf{u} \cdot \mathbf{n} \qquad \text{on} \qquad \Gamma_M
\label{eq:charge_to_mass}
\end{equation}

The temperature in the meniscus is governed by the energy transport equation balancing Ohmic dissipation with conductive and convective transport of heat:
\begin{equation}
    \rho c_p \nabla T \cdot \mathbf{u} = \kappa_T \nabla^2 T + \frac{\mathbf{j}\cdot\mathbf{j}}{\kappa\left(T\right)} + \Phi \qquad \text{in} \qquad \mathbf{\Omega}_l
    \label{eq:energy_transport}
\end{equation}

Where $c_p$ is the heat capacity, $\kappa_T$ is the thermal conductivity and $\Phi$ is the viscous dissipation power per unit volume for the incompressible ionic liquid. The viscous dissipation term takes the following form:
\begin{equation}
    \Phi = 2\mu e^2_{ij}
\end{equation}
Where $e^2_{ij}$ indicates summation over all the elements of the strain rate tensor to the square power.

The rest of the boundary conditions for the energy transport problem are:

\begin{align}\begin{split}
    \nabla T \cdot \mathbf{n} = 0 & \qquad \text{ on } \qquad \Gamma^l_L \\
    T=T_w & \qquad \text{ on } \qquad \Gamma^l_D.
\end{split}\end{align}
Where $T_w$ is the temperature on the wall of the fluid channel.

As a summary, tables \ref{tab:bulk_eq} and \ref{tab:interface_eq} show the set of non-dimensional equations fulfilled in the bulk and interface domains respectively. Non-dimensional numbers are shown in table \ref{tab:nond_numbers}. The reference parameters for the non-dimensionalization are the contact line radius ($r_0$) for the length scale; for the pressure and the stresses, the capillary pressure of a sphere of such radius
$\tau_0 = \frac{2\gamma}{r_0}$; for the electric fields, the corresponding $E_c = \sqrt{\frac{4\gamma}{r_0 \varepsilon_0}}$  whose electric pressure balances $\tau_0$; the current density by $j_c = \kappa_0 E_c$; velocities by $u_c = \frac{j_c}{\rho \frac{q}{m}}$; temperatures by the reference
value $T_0$ at which the conductivity $\kappa$ equals the reference conductivity $\kappa_0$; viscosity is scaled by $\mu_0$ and surface and bulk volumetric charges are scaled by $\sigma_c = \varepsilon_0 E_c$ and $\rho_{m_c} = \frac{\varepsilon_0 E_c}{r_0}$, respectively.

These non-dimensional variable definitions are compiled for the reader in table \ref{tab:nond_variables}.
In order to keep a better equation readability, it is useful to define the non-dimensional conductivity $\hat{K} = \frac{\kappa}{\kappa_0}$ and non-dimensional viscosity $\mu = \frac{\mu}{\mu_0}$ from eqs. \ref{eq:conductivity} and \ref{eq:viscosity} as:
\begin{equation}
\hat{K} = 1 + \Lambda\left(\hat{T}-1\right) 
\label{eq:nond_conductivity}
\end{equation}
\begin{equation}
    \hat{\mu} = \frac{1}{1 + \Lambda\left(\hat{T}-1\right)}
\label{eq:nond_viscosity}
\end{equation}
Where $\Lambda = \frac{k' T_0}{\kappa_0}$ is the non dimensional sensitivity of the electric conductivity to changes in temperature.

While this non-dimensionalization has been mostly used in the numerical procedure to keep consistency with existing literature \citep{Coffman2019ElectrohydrodynamicsField}, it has been noticed that dimensionless magnitudes referencing the emission region ($\frac{E_0}{E^*}$, $\frac{I}{I^*}$, $\frac{r_0}{r^*}$) provide very useful physical interpretations. Non-dimensionalization referencing the emission region can be easily obtained by postprocessing solutions without modifying any physical result. 

A relevant non-dimensional number in this paper comes from the non-dimensional form of the boundary conditions in (\ref{eq:Hydraulic_BC}). This yields:

\begin{align}
\label{eq:Hydraulic_BC_nond}
    \begin{split}
      \mathbf{n}\cdot \hat{\tau}_f\cdot \mathbf{n} = -\left(\hat{p}_r - \hat{I} \hat{R}^\frac{5}{2} \hat{Z} \right)  \qquad \text{ on } \qquad \Gamma_I \\ \mathbf{t}\cdot \hat{\tau}_f\cdot \mathbf{n}=0 \qquad \text{ on } \qquad \Gamma_I
    \end{split}
\end{align} 

Where $\hat{I} = \int_{d\Gamma_M} \hat{\mathbf{j}}\cdot \mathbf{n} \; d\Gamma_M$ is the non-dimensional current, $\hat{R} = \frac{r_0}{r^*}$ is the non-dimensional contact line radius and $\hat{Z} = \frac{Z}{Z^*}$, $Z^* = \frac{2\gamma \rho \frac{q}{m}}{\kappa_0E^* r^{*^3} }$ is the non-dimensional value of the tip hydraulic impedance $Z$.


\begin{table}
	\centering

	\caption{Non-dimensionalized bulk equations}
	\label{tab:bulk_eq}
    \def~{\hphantom{0}}
    \renewcommand{\arraystretch}{2}
	\begin{tabular}{p{0.25\linewidth} p{0.55\linewidth} p{0.1\linewidth}}
		\textbf{Equation Name} & \textbf{Equation} & \textbf{Domain}\\
		Vacuum Maxwell-Poisson & $\hat{\nabla} \cdot \hat{\mathbf{E}} = 0$ &  $\Omega_v$\\
		Liquid Maxwell-Poisson & $\hat{\nabla} \cdot \left(\varepsilon_r \hat{\mathbf{E}}\right) = \hat{\rho}_m$ &  $\Omega_l$\\
		Maxwell-Faraday & $\hat{\nabla} \times \hat{\mathbf{E}} = 0 \rightarrow \hat{\mathbf{E}} = -\hat{\nabla} \hat{\phi}$ & $\Omega_l \cup \Omega_v$\\
		Charge conservation & $\hat{\nabla} \cdot \hat{\mathbf{j}} = \hat{\nabla} \cdot \left(\hat{K} \hat{\mathbf{E}}\right) = 0$ & $\Omega_l$\\
				Mass conservation & $\hat{\nabla} \cdot \hat{\mathbf{u}} = 0$ & $\Omega_l$\\
		Momentum conservation & $\varepsilon^2_r We \left(\hat{\mathbf{u}} \cdot \hat{\nabla} \right) \hat{\mathbf{u}} = \hat{\nabla} \cdot \left(-\hat{p}\mathbf{I} + \frac{\varepsilon_r Ca \hat{\mu}}{\hat{R}^{\frac{1}{2}}}\left(\hat{\nabla}\hat{\mathbf{u}} + \hat{\nabla}\hat{\mathbf{u}}^T\right) \right) + 2 \hat{\rho}_m \hat{\mathbf{E}}$ & $\Omega_l$\\
				Energy conservation & $ \frac{ Gz}{\varepsilon_r H \sqrt{\hat{R}}}\; \hat{\mathbf{u}} \cdot \hat{\nabla} \hat{T} = \frac{\hat{\nabla}^2\hat{T}}{\varepsilon^2_r H\hat{R}} + \frac{ \left(\hat{\mathbf{j}}\cdot\hat{\mathbf{j}}\right)}{\hat{K}} + \frac{Ca K_C \varepsilon_r\hat{\mu}}{\hat{R}^2}\hat{e}^2_{ij} $ & $\Omega_l$\\
		\end{tabular}
	\end{table}
	
\begin{table}
	\centering
	\caption{Non-dimensionalized equations fulfilled on the meniscus interface $\Gamma_M$}
	\label{tab:interface_eq}

	\def~{\hphantom{0}}
    \renewcommand{\arraystretch}{2}
	\begin{tabular}{p{0.32\linewidth} p{0.65\linewidth}}
		\textbf{Equation Name} & \textbf{Equation}\\
			Charge conservation & $ K_C \hat{R}^{-\frac{3}{2}} \left(\hat{\mathbf{u}} \cdot \hat{\nabla}_S \hat{\sigma} - \hat{\sigma} \mathbf{n}\cdot \left( \mathbf{n} \cdot \hat{\nabla} \right) \hat{\mathbf{u}}\right) = \hat{K}\hat{E}^l_n  - \hat{j^e_n} $ \\
		Surface charge jump condition & $\hat{\sigma} =  \hat{E}^v_n - \epsilon_r \hat{E}^l_n$ \\
			Equality of tangential components of the electric field & $ \hat{E}^v_t =\hat{E}^l_t$ \\
		Kinetic law for charge evaporation & $\hat{j^e_n} = \frac{\hat{\sigma} \hat{T}}{\varepsilon_r \chi} \exp{\left(-\frac{\psi}{\hat{T}}\left(1 - \hat{R}^{-\frac{1}{4}}\sqrt{\hat{E}^v_n}\right)\right)}$\\
		Equilibrium of stresses in the tangential direction & $\frac{\varepsilon_r Ca \hat{\mu}}{\hat{R}^{\frac{1}{2}}}\mathbf{t}\cdot \left(\hat{\nabla}\hat{\mathbf{u}} + \hat{\nabla}\hat{\mathbf{u}}^T\right)\cdot \mathbf{n} = \hat{\sigma}\hat{E}_t$\\
			Equilibrium of stresses in the normal direction &  $-\hat{p}+ \frac{\varepsilon_r Ca \hat{\mu}}{\hat{R}^{\frac{1}{2}}}\mathbf{n}\cdot \left(\hat{\nabla}\hat{\mathbf{u}} + \hat{\nabla}\hat{\mathbf{u}}^T\right)\cdot \mathbf{n} = \hat{E}^{v^2}_n - \varepsilon_r \hat{E}^{l^2}_n + \left(\varepsilon_r -1\right)\hat{E}^2_t-\frac{1}{2}\hat{\nabla}\cdot\mathbf{n}$\\
		Mass conservation of ions evaporated & $\hat{\mathbf{u}}\cdot \mathbf{n} =  \hat{j}^e_n$ \\
		Thermal insulation & $\mathbf{n}\cdot\hat{\nabla} \hat{T} = 0$ \\
		\end{tabular}
	\end{table}
\begin{table}
\centering
    \def~{\hphantom{0}}
    \renewcommand{\arraystretch}{2}
    \caption{Set of non-dimensional numbers}
    	\label{tab:nond_numbers}

\begin{tabular}{p{0.33\linewidth} p{0.33\linewidth} p{0.33\linewidth}}
$We = \frac{\rho {u^*}^2r^*}{2\gamma}$. Weber number. Ratio of characteristic inertial fluid stresses to surface tension stresses in the emission region. & $Ca = \frac{\mu_0 u^*}{2 \gamma}$. Capillary number. Ratio of viscous drag stresses to surface tension stresses in the emission region.  
& $\Lambda = \frac{k' T_0}{\kappa_0}$. Non dimensional sensitivity of the electric conductivity  to changes in temperature.\\
$K_c = \frac{\varepsilon_0 \varepsilon_r u^*}{\kappa_0 r^*}$. Ratio of the charge relaxation time $\left(\frac{\varepsilon_0 \varepsilon_r}{\kappa_0}\right)$ to the characteristic residence time of liquid $\left(\frac{r^*}{u^*}\right)$ in the meniscus tip.
& $\hat{R} = \frac{r_0}{r^*}$. Ratio between the radius of the fluid channel $r_0$ and the characteristic emission size $r^*$.
& $\chi = \frac{h\kappa_0}{k_B T_0 \varepsilon_0 \varepsilon_r}$. Ratio of the kinetic emission time $\left(\frac{h}{k_BT_0}\right)$ to the characteristic charge relaxation time in the liquid $\left(\frac{\varepsilon_0 \varepsilon_r}{\kappa_0}\right)$.\\
$\psi = \frac{\Delta G}{k_B T_0}$. Ratio of solvation energy $\Delta G$ and characteristic thermal molecular energy $k_B T_0$. 
& $Gz = \frac{\rho c_p u^*r^*}{k_T}$. Graetz number. The ratio of characteristic convective  $\left(\frac{\rho c_p u^* T_0}{r^*}\right)$ and conductive $\left(\frac{k_T T_0}{r^{*^2}}\right)$ heat transfer magnitudes.
& $H = \frac{\left(j^* r^*\right)^2}{\kappa_0 k_T T_0}$. Ratio of the order of magnitude of Ohmic heat dissipation $\left(\frac{j^{*^2}}{\kappa_0}\right)$ and that of the conductive heat transfer.\\
\end{tabular}
\end{table}
\begin{table}
	\centering

	\caption{Non-dimensional variables}
	\label{tab:nond_variables}
    \def~{\hphantom{0}}
    \renewcommand{\arraystretch}{2}
	\begin{tabular}{p{0.25\linewidth} p{0.55\linewidth} }

		\textbf{Variable Name} & \textbf{Dimensionless form}\\

	Length & $\hat{r} = \frac{r}{r_0}$, $\hat{z} = \frac{z}{r_0}$\\

		Pressures and stresses & $\hat{p}=\frac{p}{p_c}$, $\hat{\tau} = \frac{\tau}{p_c}$, $p_c = \frac{2\gamma}{r_0}$\\

		Electric fields & $\hat{\mathbf{E}} =\frac{\mathbf{E}}{E_c}$, $E_c = \sqrt{\frac{4\gamma}{\varepsilon_0 r_0}}$\\

				Surface charge & $\hat{\sigma} = \frac{\sigma}{\sigma_c}$, $\sigma_c = \varepsilon_0 E_c$ \\
				
				Bulk charge & $\hat{\rho}_m = \frac{\rho_m}{\rho_{m_c}}$, $\rho_{m_c} = \frac{\varepsilon_0 E_c}{r_0}$ \\
		Current density &  $\hat{\mathbf{j}} = \frac{\mathbf{j}}{j_c}$, $j_c = \kappa_0 E_c$\\

		Total emitted current &  $\hat{I} = \frac{I}{I_c}$, $I_c = j_c r^2_0$\\

				Velocity & $\hat{\mathbf{u}}=\frac{\mathbf{u}}{u_c}$, $u_c = \frac{j_c}{\rho \frac{q}{m}}$\\

		Temperature & $ \hat{T} = \frac{T}{T_0}$\\

		\end{tabular}
	\end{table}
\section{Numerical procedure
 }
 \label{sec:numerical_procedure}
 \subsection{Iterative Solver Description}
 The solver is initialized with a reasonable guess of the axisymmetric contour $\left(\Gamma^0_M\right)$, which is generally not in equilibrium. 
 
The initial guess is perturbed across several
$k$ iterations with information obtained by solving equations in tables \ref{tab:bulk_eq} and \ref{tab:interface_eq} sequentially. These perturbations will approach the meniscus interface at each iteration ($\Gamma_M^k$) towards its equilibrium position. A detailed description of this iterative procedure is exposed in this section.
\\
In a single iteration, the EHD model is solved in three different steps, each of which comprising the equations of a relevant physics, namely the electric, fluid, and energy transport problems. 

The electric part of the solver yields the non-dimensional potential $\left(\hat{\phi}^k\right)$ in $\mathbf{\Omega}_v \cup \mathbf{\Omega}_l$ and the surface charge $\left(\hat{\sigma}^k\right)$ on $\Gamma_M$ at iteration $k$ by solving equations \ref{eq:arrhenius}, \ref{eq:poisson_vac}, \ref{eq:poisson_liq}, \ref{eq:charge_conserv_bulk} (or equivalently \ref{eq:rho_m}, if neglecting bulk charge convection),  \ref{eq:interface_charge_condition} and \ref{eq:charge_conserv_meniscus} by assuming a known distribution of non-dimensional temperature $\hat{T}^{k-1}$ and convective current density $\hat{j}^{k-1}_{conv}$ from the previous iteration (left hand side of eq. \ref{eq:charge_conserv_meniscus}).

These distributions are interpolated from the previous iteration domain $\mathbf{\Omega}^{k-1}_l$ and $\Gamma^{k-1}_M$ to $\mathbf{\Omega}^k_l$ and $\Gamma^{k}_M$ using standard linear mapping. An expression can be obtained for the surface charge $\hat{\sigma}^k$ as a function of the potential derivatives by substituting (\ref{eq:arrhenius}) in (\ref{eq:charge_conserv_meniscus}). This yields for iteration $k$:
\begin{align}
\begin{split}
    \hat{\sigma}^k = \frac{\varepsilon_r \chi}{\hat{T}^{k-1}}\exp{\frac{-\psi}{\hat{T}^{k-1}}\left(1-\hat{R}^{-\frac{1}{4}}\sqrt{-\hat{\nabla} \hat{\phi}^{v^k} \cdot \mathbf{n}}\right)}\cdot \left(\hat{K^{k-1}}\left(-\hat{\nabla} \hat{\phi}^{l^k} \cdot \mathbf{n}\right)+\hat{j}^{k-1}_{conv}\right)
\end{split}
\label{eq:surfCharge}
\end{align}
Where $\hat{\nabla} \hat{\phi}^{l^k}$, $\hat{\nabla} \hat{\phi}^{v^k}$ are the potential gradients evaluated in $\mathbf{\Omega}_l$ and  and $\mathbf{\Omega}_v$ at iteration $k$, respectively. 
Expression \ref{eq:surfCharge} can be used together with equations \ref{eq:poisson_vac} and \ref{eq:poisson_liq} to derive a variational form solvable by standard Finite Element methods (see annex \ref{sec:Annex_Variational_Forms}).

Alternatively, the non-dimensional surface charge jump condition (\ref{eq:interface_charge_condition}) can be used  to write  (\ref{eq:surfCharge}) as a function of the external electric field $-\hat{\nabla}\hat{\phi}^{v^k}$ only:
\begin{equation}
    \hat{\sigma}^k=\frac{\hat{K^{k-1}}\left(-\hat{\nabla}\hat{\phi}^{v^k} \cdot \mathbf{n} \right) + \varepsilon_r \hat{j}^{k-1}_{conv}}{\hat{K}^{k-1} + \frac{\hat{T}^{k-1}}{\chi}\exp{\left(\frac{-\psi}{\hat{T}^{k-1}}\left(1-\hat{R}^{-\frac{1}{4}}\sqrt{\left(-\hat{\nabla}\hat{\phi}^{v^k} \cdot \mathbf{n} \right)}\right)\right)}}
    \label{eq:sigma_evac}
\end{equation}
Where $\hat{K}^{k-1}$ is non-dimensional electric conductivity at the iteration $k-1$, $\hat{K} = 1 + \Lambda \left(\hat{T}^{k-1}-1\right)$.
It is found in this work that form \ref{eq:sigma_evac} is more stable, numerically.

This EHD model goes beyond the standard Taylor-Melcher leaky dielectric formulation in the inclusion of bulk volumetric charges $\rho_m$ in the electric problem. These also become part of the solution process, since they depend on conductivity gradients with temperature. The interfacial charge $\sigma$ and $\rho_m$ are part of the same charge distribution, but $\sigma$ appears as an integrated value of this distribution across a differential disk-like volume of control of the width of the Debye layer \citep{Mori2018FromLimit,Schnitzer2015TheDescription}. In the Taylor-Melcher model, and in this model, the Poisson equation in the Debye layer region is reduced to eq. \ref{eq:interface_charge_condition}, and the charge conservation equation to eq. \ref{eq:charge_conserv_meniscus}. The surface charge approximation is a very useful tool to avoid the calculation of the charge distribution in the Debye layer, since at that region the charge density varies largely. Formally, the joint calculation of $\rho_m$ and $\sigma$ could be interpreted as described in annex \ref{sec:Annex_Interpretation_rhom}. 

From the solution of \ref{eq:weak_form_electric_def}, we obtain the non-dimensional electric stress tensors on $\Gamma_M$ $\left(\hat{\mathbf{\tau}}^{v^k}_{\mathbf{e}}, \hat{\mathbf{\tau}}^{l^k}_{\mathbf{e}}\right)$, the distribution of current density evaporated at the surface $j^{e^k}_n = \hat{\mathbf{j}}^k\cdot\mathbf{n}$, and the total current evaporated $\left(\hat{I}^k = \int_{\Gamma^k_M} \hat{\mathbf{j}}^k\cdot\mathbf{n} \; d\Gamma^k_M\right)$.

The fluid solver yields the non-dimensional velocity field $\left(\hat{\mathbf{u}}^k\right)$, non-dimensional pressure distribution $\left(\hat{p}^k\right)$ along the surface of the meniscus and normal component of the viscous stress tensor $\mathbf{n}\cdot \hat{\mathbf{\tau}}_{f}\cdot \mathbf{n}$. It takes as inputs the difference of the tangential component of the electric stress tensors in both $\mathbf{\Omega}_v$ and $\mathbf{\Omega}_l$ at iteration $k$: $\mathbf{t}\cdot \left(\hat{\mathbf{\tau}}^{v^k}_e-\hat{\mathbf{\tau}}^{l^k}_e\right) \cdot \mathbf{n}$, the distribution of current density $j^{e^k}_n$ on $\Gamma^k_M$, and $\hat{T}^{k-1}$. The fluid problem solves the Navier-Stokes equations subject to the inlet and wall boundary conditions in (\ref{eq:Hydraulic_BC}) and (\ref{eq:non_slip}). The boundary conditions for the Navier-Stokes flow along $\Gamma_M$ are Neumann for the tangential direction (eq. \ref{eq:equilibrium_stresses_tangential}) and Dirichlet for the normal direction (eq. \ref{eq:charge_to_mass}). This mixed boundary condition on irregular domains is enforced weakly using Lagrange multipliers as in \cite{Verfurth1986FiniteI}. Details of the weak form used are shown in section  \ref{eq:weak_form_fluid_def}.  

The energy transport solver yields the temperature distribution along the computational domain $\left(\hat{T}^k\right)$. The temperature plays a substantial role in both the fluid and electric problems, as the electrical conductivity $\left(\kappa\right)$ and fluid viscosity $\left(\mu\right)$ are strong functions of the temperature. It takes the current density in $\mathbf{\Omega}_l$, $\mathbf{j}^{k}$ as input. The variational form used can be seen in section \ref{eq:weak_energy}.

Lastly, the solver uses the previously calculated tensor distributions and current to guess another $\Gamma^k_M$ that is closer to the equilibrium condition.

At this stage of the solving process, a guess of the meniscus surface profile $\Gamma^k_M$ has been considered. It is assumed that the surface is in equilibrium in the tangential direction (\ref{eq:equilibrium_stresses_tangential}), and the total evaporated current density (\ref{eq:arrhenius}) is directly proportional to the normal velocity distribution along $\Gamma^k_M$ through a mass-to-charge scaling constant (see table \ref{tab:interface_eq}). The equilibrium of stresses in the normal direction (\ref{eq:equilibrium_stresses_normal}) has yet to be enforced. Therefore, for a given surface $\Gamma^k_M$, the distribution of stresses in the normal direction along $\Gamma_M$ will not be 0, but a distribution of residuals $\mathbf{R}^k = \left[r^k_1,r^k_2, ...,r^k_i,...,r^k_{N_R}\right]$, where $N_R$ is the total number of points in the discretization of $\Gamma^k_M$. Eq. (\ref{eq:equilibrium_stresses_normal}) at iteration $k$ yields:
\begin{equation}
    \mathbf{R}^{k} = \mathbf{n}\cdot \left(\hat{\tau}^{v^k}_e - \hat{\tau}^{l^k}_e - \hat{\tau}^k_f \right)  \cdot \mathbf{n} - \frac{1}{2} \hat{\nabla} \cdot \mathbf{n}^k 
    \label{eq:normal_equilibrium_residue}
\end{equation}
The objective of the problem is to drive a representative scalar metric of the residue to 0, $\|\mathbf{R}^k\| \rightarrow 0$ for increasing values of $k$. This process is described next.

\subsection{Stopping Criterion}
In a problem of this nature, it is essential to define the numerical criterion to terminate the simulations when no statically stable solutions can be found. 
\subsubsection{Stopping condition.}
The stability condition used in this work is the same as that introduced by \cite{Coffman2016}. Let's define the relative residual $\mathcal{R}^k =  \left[\alpha^k_1,\alpha^k_2, ...,\alpha^k_i,...,\alpha^k_{N_R}\right]$ where:

\begin{equation}
    \alpha_i = \max \left( \frac{|r_i|}{|\left(\mathbf{n}\cdot \left( \hat{\tau}^{v^k}_e - \hat{\tau}^{l^k}_e  \right) \cdot \mathbf{n}\right)_i|}, \frac{|r_i|}{|\left(\mathbf{n}\cdot \hat{\tau}^k_f   \cdot \mathbf{n}\right)_i|}, \frac{|r_i|}{|\left(\frac{1}{2}\hat{\nabla} \cdot \mathbf{n}\right)_i|}\right)
\end{equation}

That is, $\alpha_i$ is the maximum absolute relative magnitude of the residue at point $i$ with respect to the three relevant stresses (electric, fluid and surface tension).

Static stability is assumed if:
\begin{equation}
    \max \mathcal{R}^k \leq \epsilon
    \label{eq:residue_tol}
\end{equation}

The solver stops at the first $k$ when condition \ref{eq:residue_tol} is met. Similar to \cite{Coffman2016}, a value of $\epsilon = 0.01$ is used here. A very slight deviation of the external conditions (e.g, $\Delta \hat{E}_0 = 0.01$, $\Delta \hat{R}^k = 0.001$) will originate $\mathcal{R} \sim O(1)$ for initial in-equilibrium surface shapes. For this reason, $\epsilon = 0.01$ leads to a reasonable stopping condition for static equilibrium solutions.
\subsubsection{Stopping criteria for no solutions found.}
\label{sec:stopping_no_stable}
A different stopping criterion is required when a maximum number of iterations is reached without convergence, that is $k>k_{max}$ and $\mathcal{R} > \epsilon$. A value of $k_{max} = 1500$ is used here. 

It is useful to define the signed metric $A\left(\mathcal{R}^k\right)$:

\begin{equation}
   A(\mathcal{R}^k) = \text{sign}(\max{\mathcal{R}^k})\mathcal{R}^k
\end{equation}
Where $\text{sign}(\max{\mathcal{R}^k})$ is 1 if the electric stress is higher than the sum of surface tension and fluid stress, and -1 if otherwise.
Once the maximum iterations are reached, the metric $A\left(\mathcal{R}^k\right)$ along $k$, behaves in two ways:
\begin{itemize}
    \item $A\left(\mathcal{R}^k\right)$ oscillates along $k$ between a positive and negative number. The amplitude of the oscillations is static or grows with $k$. Each $k$ that leads to a maximum or minimum of $A\left(\mathcal{R}^k\right)$ shares a very similar associated $\hat{y}^k$. This behaviour often happens on the limits of stability for small $\hat{Z}$ and electric fields smaller than $\hat{E}_{max}$. 
    \item $A\left(\mathcal{R}^k\right)$ is static and does not change when $k$ increases. This may suggest the existence of a solution that is marginally stable, thus very close to the boundaries of instability. This situation happens often for electric fields closer to $\hat{E}_{max}$ at sufficient $\hat{Z}$ prior to the disappearance of the conical shape and at the lower end field limit $\hat{E} = 0.513$ when the electrified droplet becomes unstable preceding the onset of emission. Near these regions, the equilibrium solutions present turning points, or limit points at which a family of solutions turns back on itself. This fact is a physical symptom of instability, as discussed in the literature of instability for electrified droplets \citep{Basaran1992EffectField,Basaran1989AxisymmetricField,Basaran1989AxisymmetricDrops,Basaran1990AxisymmetricField}. 
\end{itemize}
\subsection{Surface Update}
\label{sec:surface_update}
The methodology used to update the surface each iteration is similar to that in \cite{Coffman2016}. Let $\hat{y}^k\left(\hat{r}\right)$ be a parametrization of the meniscus interface $\Gamma^k_M$ as a function of $\hat{r}$. 

Let $\hat{y}^{k'},\hat{y}^{k''}, ...$ be the successive derivatives with respect to $\hat{r}$, (e.g, $\hat{y}^{k'} = \frac{d\hat{y}}{d\hat{r}}, ...$). The normal vector can be put as:
\begin{equation}
    \mathbf{n}^k = \frac{1}{\sqrt{1+ \hat{y}^{{k'}^2}}}\left(-\hat{y}^{k'},1 \right)
    \label{eq:normal_vector}
\end{equation}
For a given $\hat{y}^k$, equation (\ref{eq:surf_tension}) can be used to write an expression of the non-dimensional surface tension stress $\hat{\tau}^k_{st}$ along the meniscus:
\begin{equation}
\hat{\tau}^k_{st} =
   \frac{1}{2} \hat{\nabla} \cdot \mathbf{n}^k = \frac{1}{2} \frac{\left(1 + \hat{y}^{{k'}^2}\right)\hat{y}^{k'} + r \hat{y}^{k''}}{\hat{r}\left(1 + \hat{y}^{{k'}^2}\right)^{\frac{3}{2}}}
    \label{eq:surf_tension}
\end{equation}
Conversely, for a given $\hat{\tau}^k_{st}$, the shape $\hat{y}^{k'}$ can be found that satisfies:
\begin{equation}
   \hat{r}\left(1 + \hat{y}^{{k'}^2}\right)^{\frac{3}{2}} \hat{\tau}^k_{st} -  \frac{1}{2} \left(1 + \hat{y}^{{k'}^2}\right)\hat{y}^{k'} - \frac{1}{2} \hat{r} \hat{y}^{k''} = 0
   \label{eq:surface_tension_integration}
\end{equation}
The surface is relaxed towards equilibrium iteratively by taking a fraction of the residue distribution at past iterations to update the surface tension at each iteration, then integrate (\ref{eq:surface_tension_integration}) to find $\hat{y}^k$. Two alternatives for the surface update are:
\begin{equation}
    \hat{\tau}^{k+1}_{st} = \hat{\tau}^{k}_{st} + \beta \; \mathbf{R}^k
    \label{eq:update_surf}
\end{equation}
\begin{equation}
     \hat{\tau}^{k+1}_{st} = \hat{\tau}^{k}_{st} + \beta \; \mathbf{R}^k - \frac{(\mathbf{R}^k - \mathbf{R}^{k-1}) \cdot \mathbf{R}^k}{\|\mathbf{R}^k - \mathbf{R}^{k-1} \|^2} (\hat{\tau}^{k}_{st} + \beta  \mathbf{R}^k - \hat{\tau}^{k-1}_{st} - \beta \mathbf{R}^{k-1})
     \label{eq:update_surf_2nd}
\end{equation}
Equation \ref{eq:update_surf} is a standard numerical relaxation scheme, with the $\beta$ coefficient being a numerical relaxation parameter ($\beta \in (0,1]$).  Eq. \ref{eq:update_surf_2nd} includes information from the residual of past iterations (up to $k-1$) and can originate a higher order convergence. This method is known as the Anderson extrapolation method \citep{Anderson1965IterativeEquations}. 
 Intuitively, the closer $\beta$ is to unity, the more \textit{information} will be added to the surface update from the current iteration and the faster convergence will be. However, because of the characteristic non-linearity of the problem, $\beta$ cannot be chosen arbitrarily close to unity. This non-linearity is accentuated at large $\hat{R}$, for which the numerical solver is very prone to fail for $\beta\sim 1$ due to current runaway \citep{Gallud2019AThesis}. For this reason, conservative values of $\beta$ are selected in the range $\beta = 0.01 \sim 0.1$, depending on $\hat{R}$.
With the value $\hat{\tau}^{k+1}_{st}$, the integration of \ref{eq:surface_tension_integration} can be performed considering the axisymmetric boundary condition and the pinning of the meniscus to the rim of the fluid channel:
\begin{align}
\begin{split}
    \hat{y}^{{k+1}'} = 0 \quad \text{on} \quad \hat{r} = 0 \\
    \hat{y}^{k+1} = 0 \quad \text{on} \quad \hat{r} = 1
    \end{split}
\end{align}

After obtaining the new interface profile $\hat{y}^{k+1}$, the stresses are recomputed by iterating on the three problems described in this section.

\begin{figure}
\centering
\begin{tikzpicture}[%
    >=triangle 60,              
    start chain=going below,    
    node distance=14mm and 80mm, 
    every join/.style={norm},   
    ]
\tikzset{
  base/.style={draw, on chain, on grid, align=center, minimum height=5ex},
  circ/.style={base, circle, text width=4em},
  proc/.style={base, rectangle, text width=8em},
  bigproc/.style={base, rectangle, text width=17em},
  test/.style={base, diamond, aspect=1.5, text width=3em},
  term/.style={proc, rounded corners},
  bigterm/.style={bigproc, rounded corners},
  coord/.style={coordinate, on chain, on grid, node distance=6mm and 25mm},
  nmark/.style={draw, cyan, circle, font={\sffamily\bfseries}},
  norm/.style={->, draw, lcnorm},
  free/.style={->, draw, lcfree},
  cong/.style={->, draw, lccong},
  it/.style={font={\small\itshape}}
}
\node [term] (r0) {\textbf{Input}\\
					$\Gamma^0_M$,$\hat{E}_0$,$\hat{p}_r$,$\hat{Z}$,$\hat{R}$ \\ Ionic liquid properties};
\node [circ,join] (p0) {\textbf{Start}};
\node [bigterm, join] (d0) {\textbf{Remesh domain} $\Gamma^k_M$};
\node [bigterm, join] (e0) {\textbf{Get Preliminary Information}
							\begin{itemize}
							\item Solve a simplified version of system \ref{eq:weak_form_electric_def} (equipotential meniscus) or interpolate from $\hat{\phi}^{k-1}$ to get an initial guess $\hat{\phi}^{k}_0$ for the electric problem.
			  \item Interpolate $\hat{j}^{j-1}_{conv}, \hat{T}^{k-1}$ to the new domains $\mathbf{\Omega}_l$ and $\mathbf{\Omega}_v$.		\end{itemize}};

\node [bigterm, join] (e1) {\textbf{Solve Electric Problem}
							\begin{itemize}
							\item Solve system \ref{eq:weak_form_electric_def} using a Newton algorithm and $\hat{\phi}^k_0$ as initial guess \citep{Gallud2019AThesis}. Get $\hat{\phi}^k, \hat{\sigma}^k$.
							\item Project electric fields: $\hat{\mathbf{E}}^k = -\hat{\nabla}\hat{\phi}^k$
							\end{itemize}};
							
\node [bigterm, join] (e2) {\textbf{Solve Fluid Problem}
							\begin{itemize}
							\item Solve system \ref{eq:weak_form_fluid_def}
							\end{itemize}};
							
\node [bigterm, right=of e2] (f0) {\textbf{Solve Energy Transport Problem}
							\begin{itemize}
							\item Solve system \ref{eq:weak_energy}
		\end{itemize}};
							
\node [bigterm, right=of e1] (f1) {\textbf{Update Surface}
							\begin{itemize}
							\item Solve  (\ref{eq:surface_tension_integration}) for $\hat{y}^{k+1}$ using $\hat{\tau}^{k+1}_{st}$ from  (\ref{eq:update_surf}) or (\ref{eq:update_surf_2nd}).
							\end{itemize}};

\node [test, right=of e0,yshift = -1em] (f3) {Is $\max{\mathcal{R}^k} < \epsilon$ ?}; 
\node [test, right=of d0] (f2) {Is $k>k_{max}$ ?}; 
\node [circ,right=of p0,yshift = 1.5em] (q0) {\textbf{No solution found}};
\node [circ,node distance=14mm and 65mm, right=of f3, xshift=-10em] (q1) {\textbf{Static solution found}};

\path (e0) to node [near start,xshift=3em,yshift=-0.75em] {$\hat{j}^{k-1}_{conv}, \hat{T}^{k-1}$} (e1);
\path (e1) to node [near start,xshift=5em,yshift=-0.75em] {$\hat{j}^{{n}^k}_e, \hat{T}^{k-1}, \mathbf{t}\cdot\hat{\mathbf{\tau}}_e^{k}\cdot\mathbf{n}$} (e2);
\path (e2) to node [near start,xshift=3em,yshift=1em] {$\hat{\mathbf{u}}^k,\hat{\mathbf{E}}^k$} (f0);
\path (f0) to node [near start,xshift=2.5em,yshift=1em] {$\hat{I},\hat{\mathbf{\tau}}_e^{k},\hat{\mathbf{\tau}}_f^{k}$} (f1);
\path (f1) to node [near start,xshift=2em,yshift=1em] {$\Gamma^{k+1}_M$} (f3);

\draw [->,lcnorm] (e2.east) -- (f0);
\node [coord, left=25mm of f2] (c2) {};
\path (f2.west) to node [near start, yshift=-1em] {$n$} (c2);
 \path (c2) to node [near end, yshift=-2.5em, xshift=3.5em] {mod$\left(k,k_r\right)\neq 0$} (e0);
 \path (c2) to node [near end, yshift=-4em, xshift=2.5em] {$k=k+1$} (e0);
 \draw [o->,lcnorm] (f2.west) -- (c2) |- (e0);
  \path (c2) to node [near start, yshift=0.75em, xshift=-1.75em] {mod$\left(k,k_r\right)=0$} (d0); 
  \path (c2) to node [near start, yshift=-0.75em, xshift=-2.75em] {$k=k+1$} (d0);
  \draw [->,lcnorm] (c2) |- (d0);
  \draw [->,lcnorm] (f0.north) |- (f1.south);  
  \draw [->,lcnorm] (f1.north) |- (f3.south);
  \path (f3.east) to node [near start, yshift=1em,xshift=-0.8em] {$y$} (q1);
  \draw [->,lcnorm] (f3.east) |- (q1.west);
  \path (f2.north) to node [near start, xshift=1em] {$y$} (q0); 
  \draw [o->,lcnorm] (f2.north) |- (q0.south);
  \path (f3.north) to node [near start, xshift=1em] {$n$} (f2); 
  \draw [o->,lcnorm] (f3.north) |- (f2.south);

\end{tikzpicture}

    \caption{Numerical procedure diagram for obtaining an equilibrium surface for given $\hat{E}_0, \hat{p}_r,\hat{Z},\hat{R}$ and an initial guess $\Gamma^0_M$.}
    \label{fig:numerical_procedure}
\end{figure}
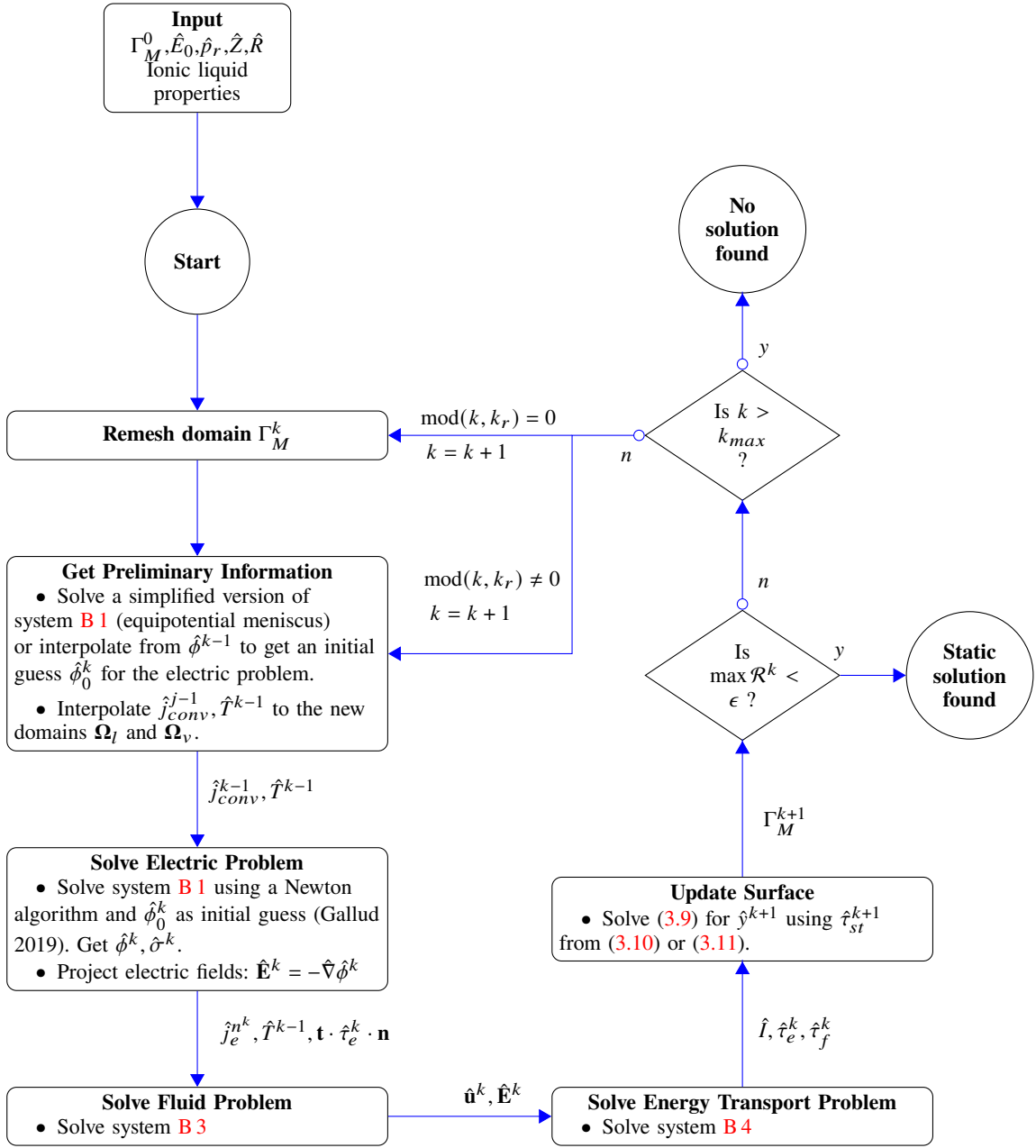
\section{Results and discussion}
\label{sec:results}
\subsection{Ionic liquid physical properties and model inputs}
 
 The results presented in this section follow the same characteristic non-dimensional numbers based on the properties of standard ionic liquids as defined in \cite{Coffman2019ElectrohydrodynamicsField}.
 These properties are similar to those of $\text{EMI}-\text{BF}_4$, which is a widely used ionic liquid in the literature of pure ion evaporation \citep{Legge2011ElectrosprayMetals,Romero-Sanz2003SourceRegime}.

The physical properties are $\kappa_0 = 1$ 
$\frac{\text{S}}{\text{m}}$, $\kappa' = 0.04$ $\frac{\text{S}}{\text{m K}}$, $\frac{q}{m} = 10^6$  $\frac{\text{C}}{\text{kg}}$, $\mu_0 = 0.037$ Pa s,
$\kappa_T = 0.2$ $\frac{\text{W}}{\text{m K}}$, $c_p = 1500$ $\frac{\text{J}}{\text{kg K}}$, $\gamma = 0.05$ $\frac{\text{N}}{\text{m}}$, $\Delta G =
1$ eV, $\rho = 10^3$ $\frac{\text{kg}}{\text{m}^3}$ and $\varepsilon_r = 10$. These properties determine most of the non-dimensional parameters shown in tables \ref{tab:bulk_eq} and \ref{tab:interface_eq}, namely $\Lambda = 12$, $\psi = 38.6$, $\chi = 1.81 \cdot 10^{-3}$, $We = 2.26 \cdot 10^{-6}$, $Ca = 0.026$, $Gz = 0.024$, $K_c = 1.32 \cdot 10^{-4}$, and $H =
0.176$. 

The reported results contain variations of parameters that are mostly external to the physical properties of the working ionic liquid. The space of independent variables that are numerically explored are $\hat{E}_0$, $\hat{R}$, and $\hat{Z}$. The reservoir pressure is taken to be $\hat{p}_r = 0$, since this is the most common case for operation of passively-fed emitters.

\subsection{Diagram of the regions of static stability}
\begin{figure}
    \centering
    \includegraphics[width=\textwidth]{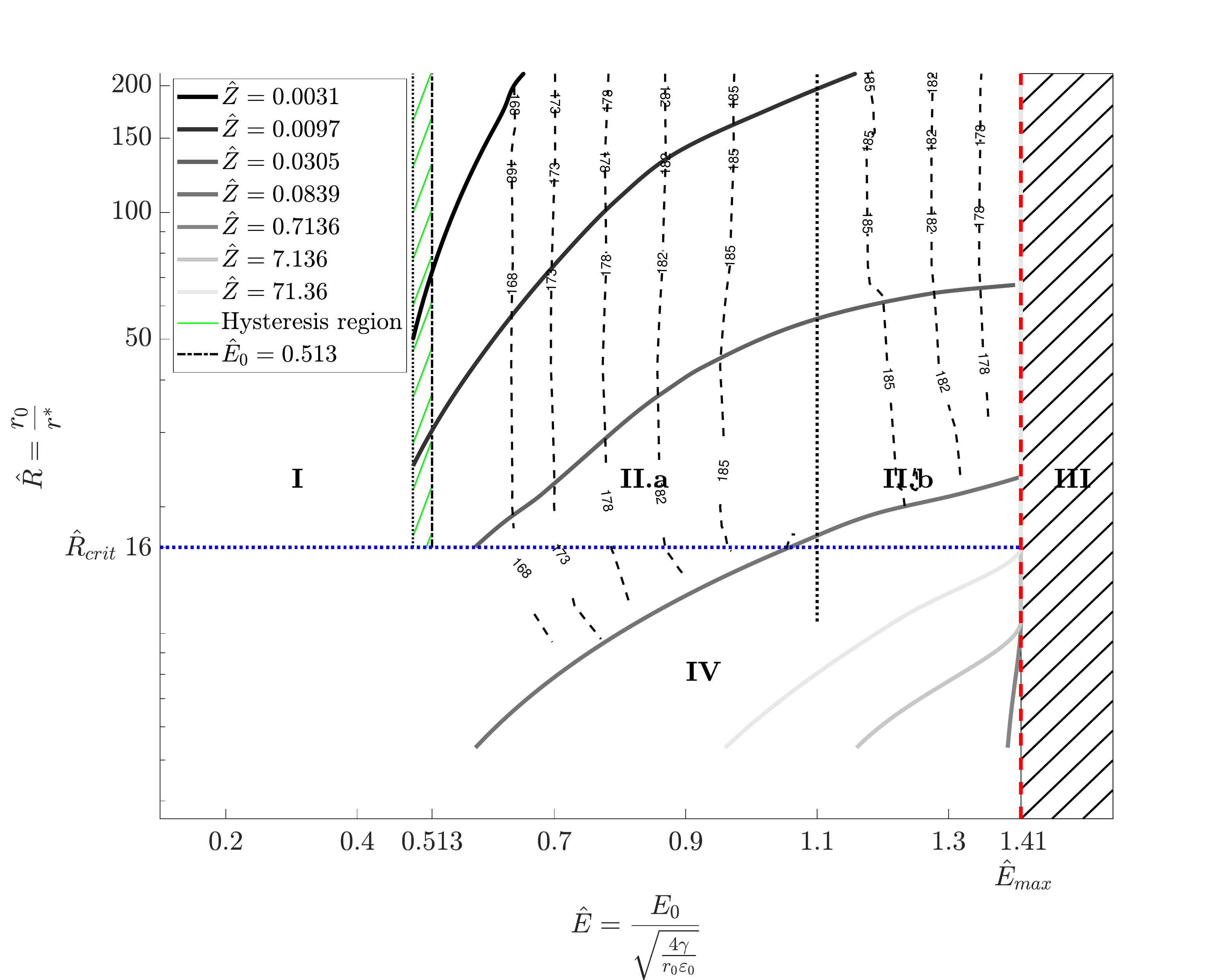}
    \caption{Map of the stability boundaries as a function of the non-dimensional external electric field $\hat{E}$ and non-dimensional contact line radius $\hat{R}$ for seven hydraulic impedance coefficients. Static solutions exist at a given $\hat{R}$ for external electric fields smaller than the limit boundary for the aforementioned impedance. Dashed lines show the regions of the stability diagram that share the same contact line angle $\theta$ with the electrode for $\hat{Z} = 0.0839$. Contact angle values can be extrapolated to the other hydraulic impedance coefficients.}
    \label{fig:stability_contact}
\end{figure}

\label{sec:stabilityDiagram}
A more detailed version of the stability diagram presented in \cite{Coffman2016} is presented in this section. In particular, this analysis extends the range of exploration of solutions from an interval of non-dimensional contact line radius $\hat{R} \in \left[10,110\right]$ in \cite{Coffman2016} to $\hat{R} \in \left[6,210\right]$. 

Figure \ref{fig:stability_contact} shows the combinations of non-dimensional external electric field $\hat{E}_0$ and contact line radius $\hat{R}$ that yield statically stable menisci. Static equilibrium solutions are found at a given $\hat{R}$ for combinations of electric fields outside the black stripped region above $\hat{E}_{max}$ and below the solid grey lines at their correspondent value of non-dimensional hydraulic impedance coefficient.
According to the characteristics of the equilibrium solutions, the stability diagram is divided in four regions.

Region $\mathbf{I}$ spans the set of non-dimensional contact line radii above the critical $\hat{R}_{crit} \approx 16$ and external fields below $\hat{E}_0 \approx 0.513$. The region $\mathbf{I}$ is characterized by a lack of meaningful current output. This family of hyperboloid-like equilibrium solutions is well known in the literature \citep{Basaran1989AxisymmetricField} and out of the scope of discussion in this paper. These non-emitting equilibrium shapes experience turning solutions when going past the field $\hat{E}_0 = 0.513$. As mentioned in section \ref{sec:stopping_no_stable}, solutions turn back on themselves as a symptom of imminent instability at turning points.


The existence of a critical radius below which no turning point exists ($\hat{R}_{crit}$) suggests that the disparity between $r*$ and $r_0$ is important for stability. On the limit where $r_0 >> r^*$ (high $\hat{R}$) the non-dimensional critical electric field scales as $\frac{E^*}{E_c} = \hat{E}^* \sim \hat{R}^{\frac{1}{2}}$ (see the non dimensional kinetic law for charge evaporation in table \ref{tab:interface_eq}).
The invariance of the turning point at $\hat{E}_0 = 0.513$ at high $\hat{R}$ confirms that the associated instability is not driven by the activated emission process, but by standard Rayleigh instability. In other words, if the evaporation process were significant in this loss of stability, the maximum local electric field in the vicinity of the meniscus tip would be on the order of the critical field. Instead, equilibrium surfaces on the verge of the turning point instability ( $\hat{E}_0 < 0.513$) are observed to be mostly independent of $\hat{R}$ and $\hat{Z}$, and the local electric fields at the menisci tip are more than one order of magnitude smaller than $\hat{E}^*$. 

 The lack of ion emission precludes any ion transport and $\frac{\varepsilon_r \hat{j}_{conv}}{\hat{K}}$ can also be neglected. The surface charge expression in (\ref{eq:sigma_evac}) can therefore be reduced to $\hat{\sigma} = -\hat{\nabla}\hat{\phi}^{v} \cdot \mathbf{n}$. The latter expression indicates the surface charge can be considered to be fully relaxed, and the meniscus behaves like a conductor. 

\cite{Beroz2019StabilityDroplets} showed that the static stability of a conducting axisymmetric droplet exposed to an external electric field and pinned or
sliding on a conducting surface or free floating follows a scaling law of the form:
\begin{equation}
    \label{eq:Beroz}
    \frac{r^3_0}{V} > \frac{\pi \varepsilon_0 E_0^2}{\frac{2 \gamma}{r_0}}
\end{equation}

Where $r_0$ is the pinning radius and $V$ is the volume of the droplet.
This scaling law predicts the stability limits obtained numerically by \cite{Basaran1990AxisymmetricField}  for the cases of negligible hydrostatic pressure inside the droplet.

Using the reference magnitudes, the non-dimensional form of (\ref{eq:Beroz}) becomes:
\begin{equation}
    \frac{1}{\hat{V}} > 2 \pi \hat{E}^2_0
    \label{eq:nondBeroz}
\end{equation}

\begin{figure}
    \centering
    \includegraphics[width=\textwidth]{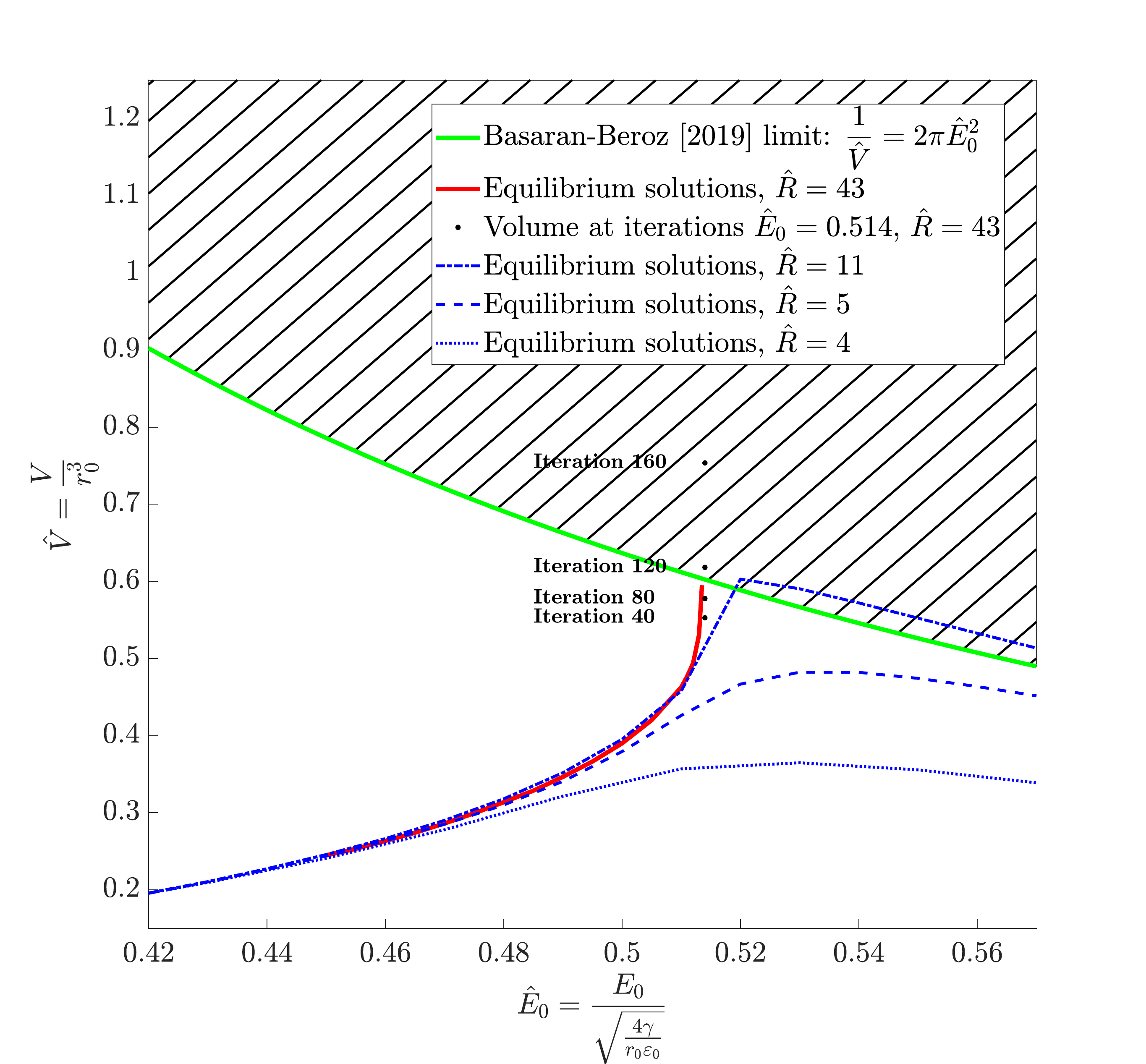}
    \caption{Non-dimensional volume of the equilibrium shapes in region \textbf{I} of the stability diagram. Comparison with the Basaran-Beroz limit \citep{Beroz2019StabilityDroplets} in green. Solutions for $\hat{R}$ greater than $\hat{R}_{crit}$ are shown in red, whereas solutions at smaller $\hat{R}$ are shown in dashed blue. The volume of the shapes at selected iterations for the first unstable $\hat{E}_0$  are shown in the black markers, where the volume can be seen to grow exponentially before breaking the numerical procedure.}
    \label{fig:Beroz_limit}
\end{figure}

The non-dimensional volume in the region of non-dimensional electric fields close to the lower turning point is shown in figure \ref{fig:Beroz_limit}. It is observed that increasing the electric field yields equilibrium shapes of higher volume. The convergence criteria (\ref{eq:residue_tol}) was reached for non-dimensional electric fields up to $\hat{E}_0=0.513$. As seen in figure \ref{fig:Beroz_limit} for electric fields slightly higher than this limit, and contact line radii higher than $\hat{R}_{crit}$, the volume of the shapes along the successive iterations approaches the Basaran-Beroz stability boundary until the volume is large enough to trigger the Rayleigh instability. It is worth mentioning that the derivative of the volume with respect to the external field becomes singular at the instability, as expected by its turning point nature.
\\

Region $\mathbf{II}$ spans non-dimensional contact line radii greater $\hat{R}_{crit} \approx 16$ and fields greater than $\hat{E}_0 \approx 0.485$. These high electric field solutions are characterized by menisci with substantial charge evaporation.

Figure \ref{fig:stability_contact} shows the combination of electric fields and contact radius $\hat{R}$ where statically stable emitting solutions were found in region $\mathbf{II}$ for seven different non-dimensional hydraulic impedance coefficients ($\hat{Z}$). Upper limits for increasing values of $\hat{Z}$ are shown in brighter grey-shaded hard lines. 

As shown in figure \ref{fig:stability_contact} for a given $\hat{R}$, the range of electric fields where static solutions were found increases for higher hydraulic impedance coefficients until a maximum range 
ending at $\hat{E}_{max} \approx 1.414 \sim \sqrt{2}$. The upper limit of stability corresponding to $\hat{Z} >= 0.0305$ collapses at $\hat{E} = \hat{E}_{max}$ for $\hat{R} > \hat{R}_{crit}$.

Figure \ref{fig:stability_contact} also shows the meniscus contact angle isolines with the downside electrode $\Gamma^v_D$, $\theta$, for the different combinations of $\hat{R}$ and $\hat{E}_0$. Simulations show that $\theta$ is very weakly dependent on the $\hat{Z}$ and $\hat{R}$ in this region. Contact angle isolines in figure \ref{fig:stability_contact} correspond to $\hat{Z} = 0.0839$ and they could be extrapolated to other values of $\hat{Z}$ within either region of static stability.

The dependence of the contact angle on the external electric field $\hat{E}_0$ is distinct enough that solutions in region \textbf{II} can be  classified further in two subregions.

Subregion \textbf{II.a} is limited to electric fields below $\hat{E}_0 \approx 1.1$ and characterized by equilibrium shapes that increment their contact line angle $\theta$ and decrease their volume for increasing values of the electric field.

\begin{figure}
    \centering
    \includegraphics[width=\textwidth]{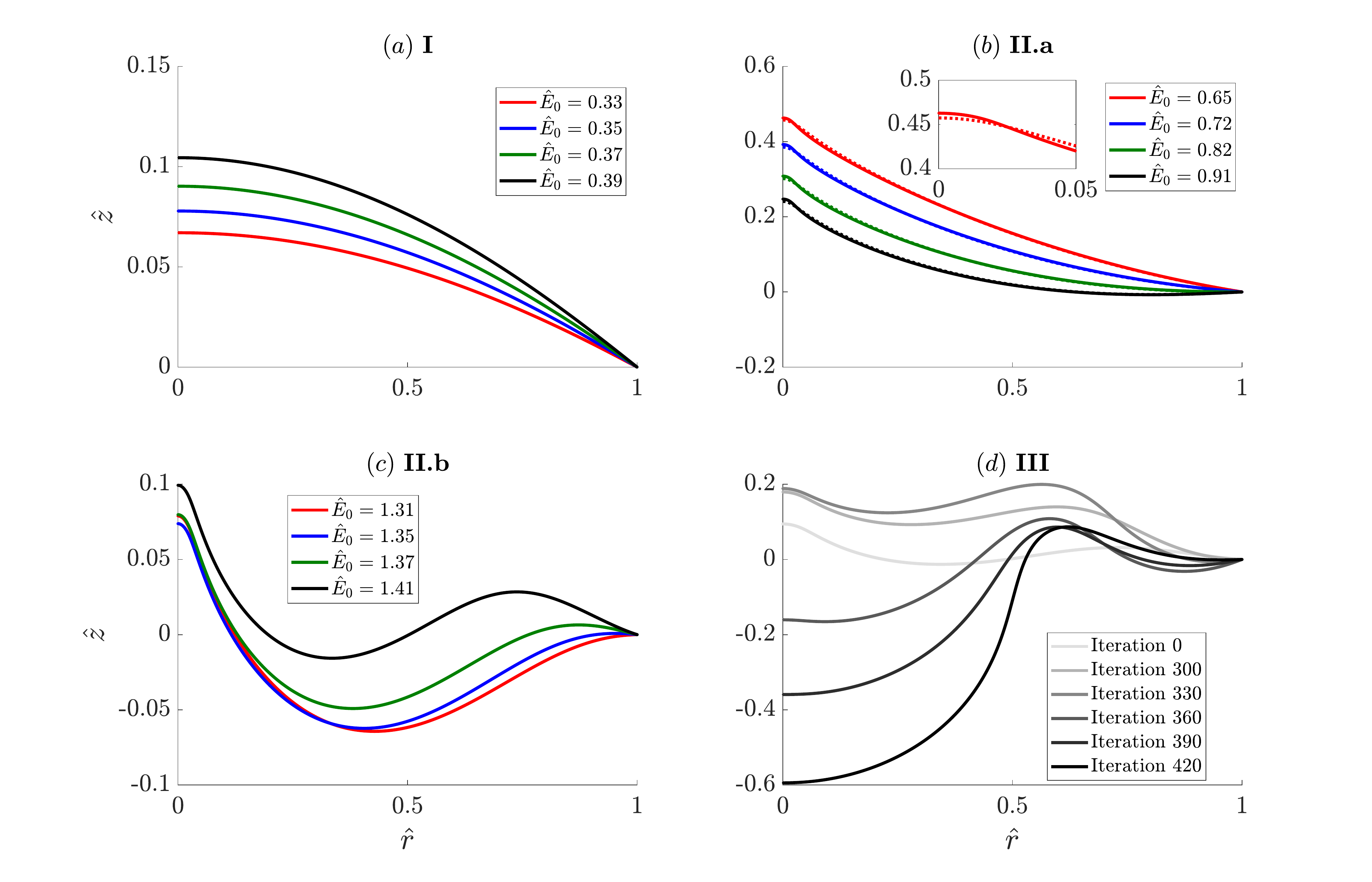}
    \caption{Characteristic equilibrium shapes of representative regions identified in the stability diagram. Equilibrium shapes in region \textbf{I} are depicted in (a) with $\hat{Z} = 0.0839$ and $\hat{R} = 43$. Region \textbf{II.a} characteristic equilibrium shapes are in (b) with $\hat{Z} = 0.0147$ in solid and $\hat{Z} = 0.147$ in dotted lines for $\hat{R} = 54$. Region \textbf{II.b} contains shapes depicted in (c) for $\hat{Z} = 0.1586$, $\hat{R} = 54$. Shapes along iterations for a combination of $\hat{E} = 1.43$, $\hat{R} = 32$ and $\hat{Z} = 0.1586$ in region \textbf{III} are shown in (d). Equilibrium was not reached in the latter simulation.}
    \label{fig:shapes_efield}
\end{figure}

\begin{figure}
    \centering
    \includegraphics[width=0.7\textwidth]{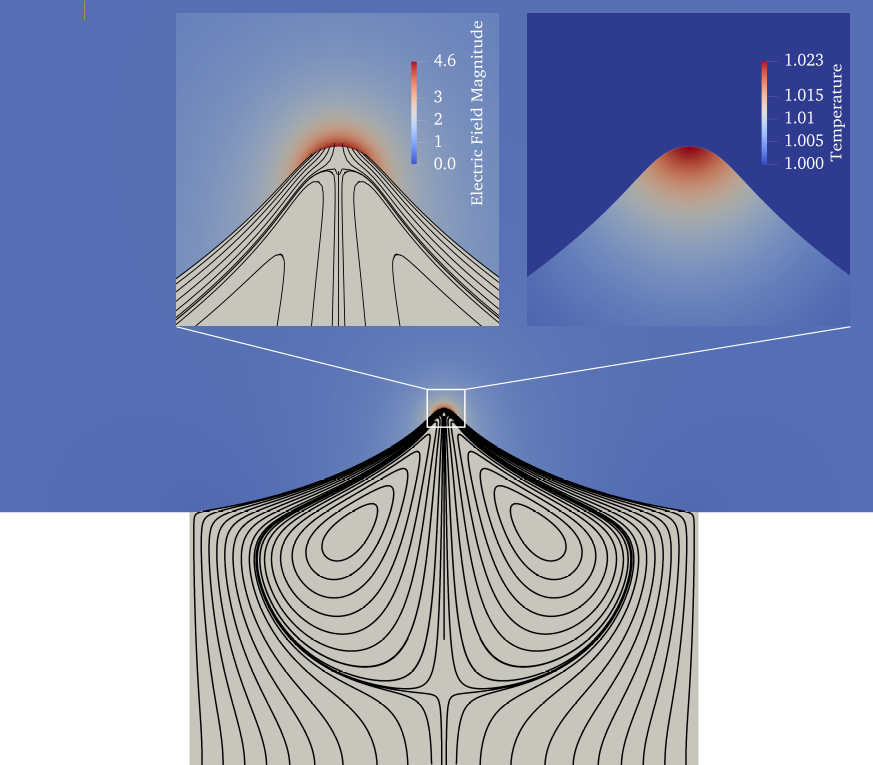}
    \caption{Prototypical pure-ion menisci internal flow structure. Operational space parameters used in this figure correspond to $\hat{R} = 43$, $\hat{E}_0 = 0.7$, $\hat{Z} = 0.0839$, $\hat{p}_r = 0$. The non-dimensional magnitude of the electric field is shown on the left. Field intensity is on the order of $E^*$ near the tip, where the evaporating fluid velocity streamlines end. The effect of Ohmic heating transport near the tip is represented in the temperature plot on the right side subfigure.}
    \label{fig:2D}
\end{figure}

Solutions within this moderate field range were explored by \cite{Coffman2016StructureIons} and showed a sharper interface than the hyperboloidal menisci in \textbf{I}. Prototypical interface geometries can be seen in figure \ref{fig:shapes_efield}b. These static menisci have a characteristic emission region of non-dimensional size $\frac{r^*}{r_0} = \hat{R}^{-1}$, where the non-dimensional electric fields are on the order of the critical field $\hat{E}^* \sim \hat{R}^{\frac{1}{2}}$. The surface charge on these menisci is not relaxed and the temperature is around a $3-5\%$ higher than in the bulk ionic liquid due to heating by Ohmic dissipation \citep{Coffman2019ElectrohydrodynamicsField}. Figure \ref{fig:2D} includes the flow structure of a prototypical equilibrium interface in \textbf{II.a}. Streamlines show the recirculation cells occupying a large volume of the meniscus. This could be related to the low characteristic flow rates of menisci in the pure ion mode \citep{Herrada2012NumericalMode}.  The emission region is amplified on the top of figure \ref{fig:2D}, where electric fields on the order of the $E^*$ are found.
\\
    \begin{figure}
    \centering
    \includegraphics[width=\linewidth]{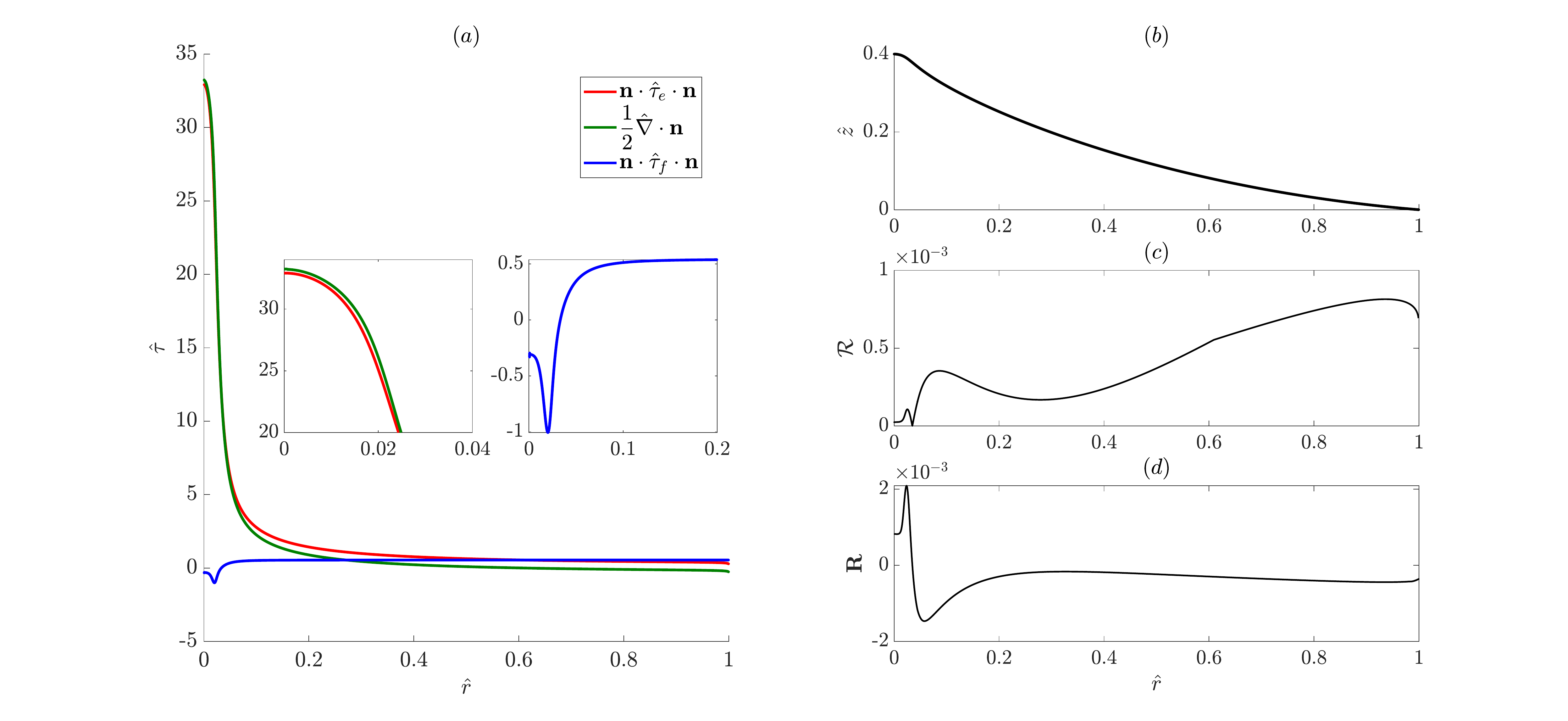}
    \caption{Subfigure a) shows the distribution of dimensionless normal stresses for a prototypical equilibrium shape in region \textbf{II.a} ($\hat{E} = 0.71$, $\hat{R} = 64.2$, $\hat{Z} = 0.0305$). Values at $\hat{r} = 0$ correspond to the stresses onto the meniscus axis of symmetry. Values at $\hat{r} = 1$ correspond to stresses onto the meniscus contact line with the electrode. Electric stresses in red, surface tension in green, hydrodynamic fluid stresses in blue. The corresponding equilibrium shape is shown in subfigure b). The relative residual used as a criterion of convergence is shown in c). The absolute residual is shown in d).}
    \label{fig:stress_IIIa}
\end{figure}
The balance of stresses in the normal direction of a prototypical equilibrium shape in region \textbf{II.a} are shown in figure \ref{fig:stress_IIIa}. Equilibrium shapes in this region look similar to a flattened Taylor cone, with a closed small region at the apex, where the meniscus is emitting. Near the emitting region, the curvature is high enough to sustain the majority of the electric stress needed for pure-ion evaporation. Near the contact line region, the meniscus does not emit. In this regard, the velocity field is negligible and the pressure is mostly that from the boundary conditions in eq. \ref{eq:Hydraulic_BC_nond}, or the one originated due to friction of the fluid with the walls upstream. In this region near the contact line, the meniscus tends to a planar geometry, therefore the electric stress is compensated mostly by the hydrostatic pressure.

\begin{figure}
    \centering
    \includegraphics[width=\linewidth]{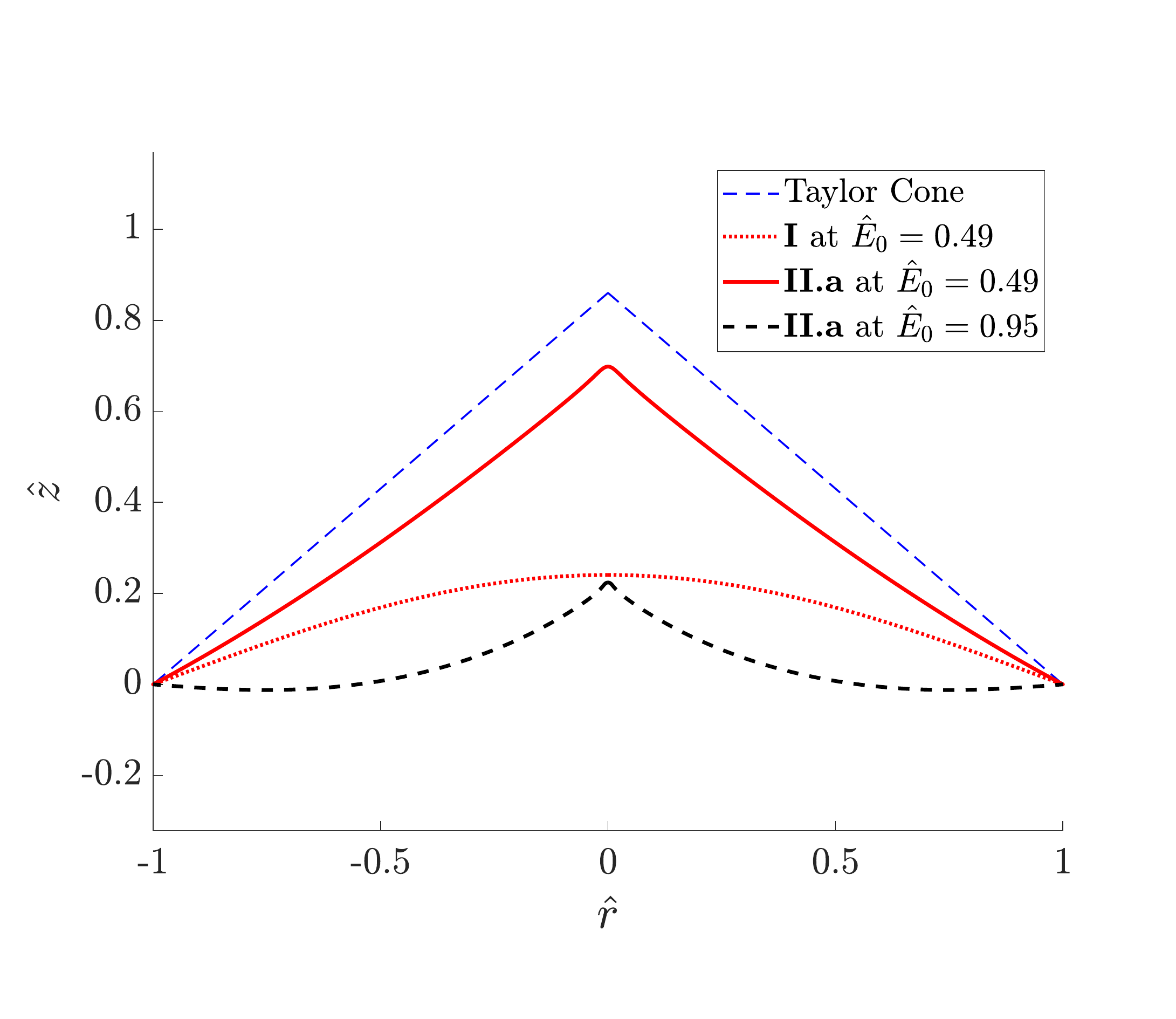}
    \caption{Equilibrium shapes in the hysteresis region for the emitting case (solid red) and non-emitting case (dotted red). Taylor cone geometry and characteristic emitting meniscus at higher stable fields are shown for cross-reference.}
    \label{fig:hysteresis}
\end{figure}

Regions \textbf{I} and \textbf{II.a} overlap in a narrow range of electric fields between $\hat{E}_0 \sim 0.485$ and the turning point in $\hat{E}_0 \sim 0.513$ (green zone in figure \ref{fig:stability_contact}). Whether the solver converges to an emitting equilibrium shape of region \textbf{II.a} or non-emitting equilibrium shape in region \textbf{I} depends on the initial guess provided to the solver. Figure \ref{fig:hysteresis} shows emitting ($\textbf{II.a}$) and non-emitting ($\textbf{I}$) solutions existing for the same external field $\hat{E}_0 = 0.49$. The current diminishes when the electric field is decreased with a starting solution from the emitting region \textbf{II.a}. Current being very small at these field magnitudes undermines the relative importance of the hydrodynamic stress with respect to the surface tension and the electric stress. In this sense, the equilibrium shapes tend to resemble the canonical Taylor solution with negligible static pressure. The exact Taylor conical shape cannot be recovered with this setting due to the planar electrode geometry sustaining the meniscus and the hydrostatic suction pressure originated by the small but non-zero  current flow.

This hysteresis behavior is well documented experimentally for liquid metal ion sources (\cite{Forbes1997UnderstandingWorks}), where the extinction voltage is typically smaller than the one needed for the onset of pure-ion emission.

The turning point nature of the instability when approaching region  \textbf{II.a} from non-emitting interfaces in region \textbf{I},  
suggests the existence of a dynamic mechanism with mass ejection that cannot be described by the time-independent meniscus model with a closed interface presented in this paper. It is difficult to speculate what the emission outcome would be in this transition. It is clear, however, that a significant meniscus volume needs to be shed during it. An option for this could be droplet breakup that might be preceded by both cone-jet formation and ion evaporation. If such a cone-jet were to exist in this region, it would be reasonable to infer a substantial deviation of its interface shape from the Taylor solution due to the high hydraulic impedance of capillaries feeding pure-ion menisci. This shape would change rapidly, resembling a ``suctioned" Taylor cone with a volume that would decrease at higher values of the electric field until the field was high enough to sustain steady ion emission. 

The reduction of meniscus volume in region \textbf{II.a} due to the increase of external field is accompanied with a rise in the contact angle $\theta$ with the downside electrode. It is known that electric fields could exhibit unbounded singular behaviors near sharp corners when these corners are greater than $180^{\circ}$ \citep{Li2000SingularitiesProblems}. The corner sharpens as the values of $\theta$ reach approximately $185^\circ-186^\circ$  and the equilibrium geometric shapes augment their curvature to compensate for the stronger electric stress that appears near the singularity. 

This curvature increase manifests as a small bump appearing near the contact line for external fields higher than $\hat{E}_0 \approx 1.1$. This point marks the beginning of subregion \textbf{II.b}.

Subregion \textbf{II.b} is only accessible when sufficient hydraulic impedance is provided. Equilibrium shapes contain this cylindrical bump near the contact line as seen in figure \ref{fig:shapes_efield}c. The shapes also reduce their contact line $\theta$ and rise their bump amplitude for increasing values of the external electric field $\hat{E}_0$. The cylindrical bump does not emit any charge for the span of electric fields simulated in this region.
    
It should be emphasized that the model presented in this paper is axisymmetric and static. This prevents a determination of the effects of possible three-dimensional disturbances on the surface of this cylindrical bump that resembles a toroid. Disturbances like this originate capillary pinch-off instabilities and the eventual break-up of similar toroidal interfaces into smaller menisci \citep{Fragkopoulos2017Toroidal-dropletCharge,Mehrabian2013CapillaryTorus}. The determination of the dynamic stability of the equilibrium shapes in this subregion is beyond the scope of this study. However, it is certainly relevant to fully understand the structure and behavior of these menisci and should be studied in detail.
    
Region \textbf{II} terminates at external electric fields $\hat{E}_0$ higher than $\hat{E}_{max} \approx \sqrt{2}$, when sufficient hydraulic impedance is provided.
    
In dimensional form, the previous statement can be recast as a function of a reference electric pressure. It is helpful to define such pressure as a function of the electric field downstream from the emission region. In the case of a planar electrode such as the one studied in this paper, this reference field is taken as the external field $E_0$:
    \begin{equation}
          \label{eq:upper_limit}
      \frac{1}{2}\varepsilon_0 E^2_0 > 2 \left(\frac{2 \gamma}{r_0}\right)
    \end{equation}

It is then seen that the pure ion emission cannot be sustained by a meniscus of pinning radius $r_0$ when the reference electric pressure is higher than approximately two times the surface tension stress of a liquid sphere of the same radius.

    
 When $\hat{E}_0 > \hat{E}_{max}$, menisci in region \textbf{III} exhibit a sharp transition towards instability depicted in figure \ref{fig:shapes_efield}d: the cylindrical contact line bump grows to such an extent that the electric field on its crest becomes on the order of the critical field, while the central emission region protuberance shrinks  progressively until it disappears. At this point, the cylindrical bump transforms into an emitting corona with a significantly larger emission area, thus producing a dramatic increase in the current output that in turn, produces a large pressure drop through the feeding channel. This pressure drop induces a sudden suction on the meniscus interface near the axis of symmetry, quickly terminating the simulation as the numerical procedure cannot track these changes. 
 
\begin{figure}
     \centering
     \includegraphics[width=0.8\linewidth]{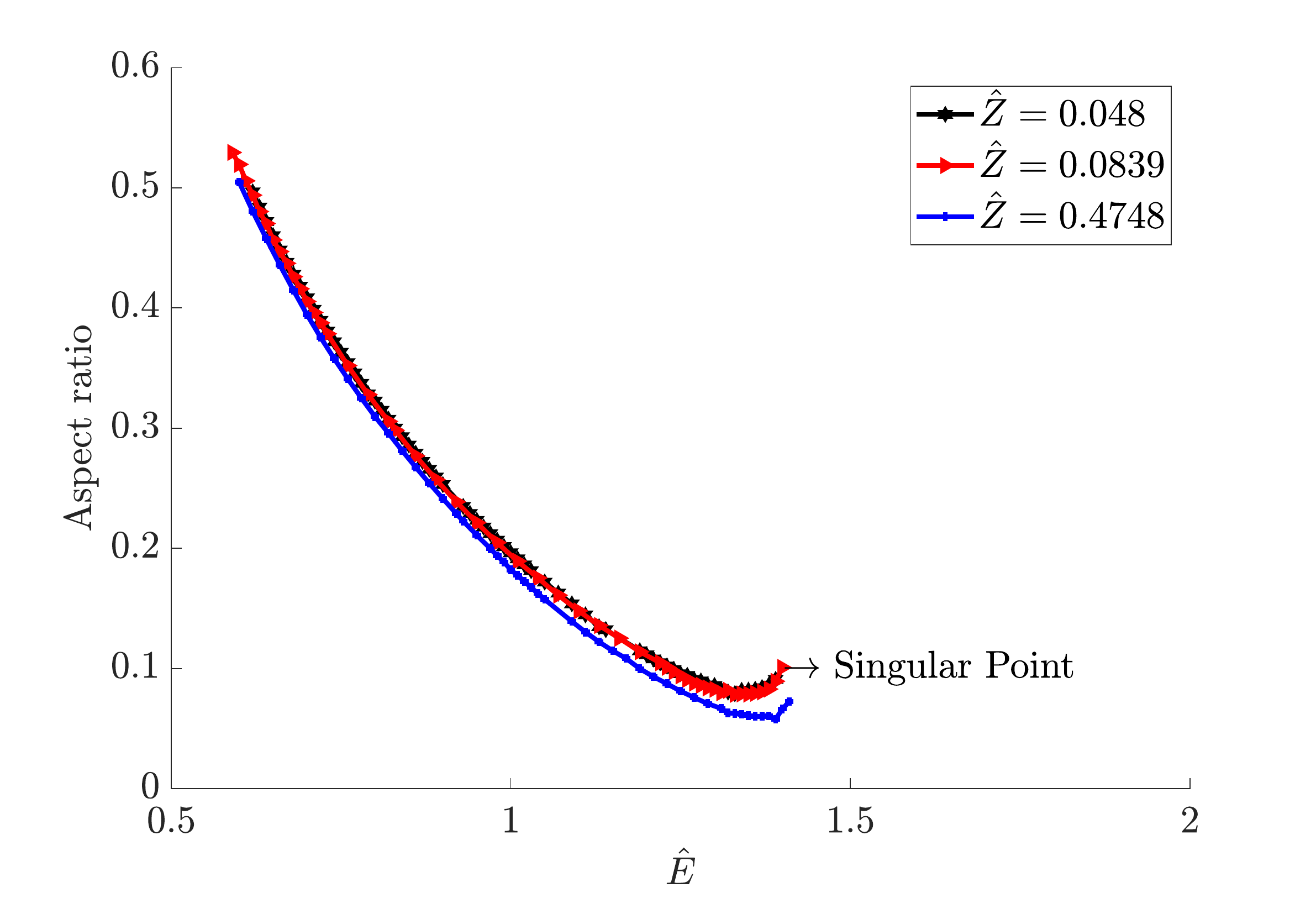}
     \caption{Aspect ratio of equilibrium shapes in region \textbf{II} at different hydraulic impedances.}
     \label{fig:aspect_ratio}
 \end{figure}

 At $\hat{E}_0 = \hat{E}_{max}$ point, the equilibrium interfaces turn on themselves when increasing the values of the electric field in a similar way described in \cite{Basaran1992EffectField} for the electrified menisci in region $\mathbf{I}$. This can be seen in figure 
 \ref{fig:aspect_ratio}, where the aspect ratio of the equilibrium shapes obtained exhibits this singularity.
    
    The scaling in (\ref{eq:upper_limit}) appears to be independent of all parameters of the operational space considered in this study, namely $\hat{p}_r$, $\hat{Z}$ and $\hat{R}$ (when $\hat{R} > \hat{R}_{crit}$) and cannot be described in detail with the axisymmetric and static model implemented for the same dynamic instability reasons mentioned previously.
    
    Regardless, reporting the existence of this sharp transition could be informative for future investigations of menisci bifurcation phenomena that are known to exist in the operation of pure ion emission sources. Bifurcation is observed when the applied voltage increases over a critical value that depends on source geometry and liquid properties \citep{Perez-Martinez2015IonMicrotips}. Such critical voltage would correspond to a non-dimensional field that, according to the results presented here, cannot exceed the upper bound field value of the stability range. This is an important empirical validation point that requires more in-depth work with versions of this model based on source geometries and domains similar to those used in experiments.

\begin{figure}
    \centering
    \includegraphics[width=\textwidth]{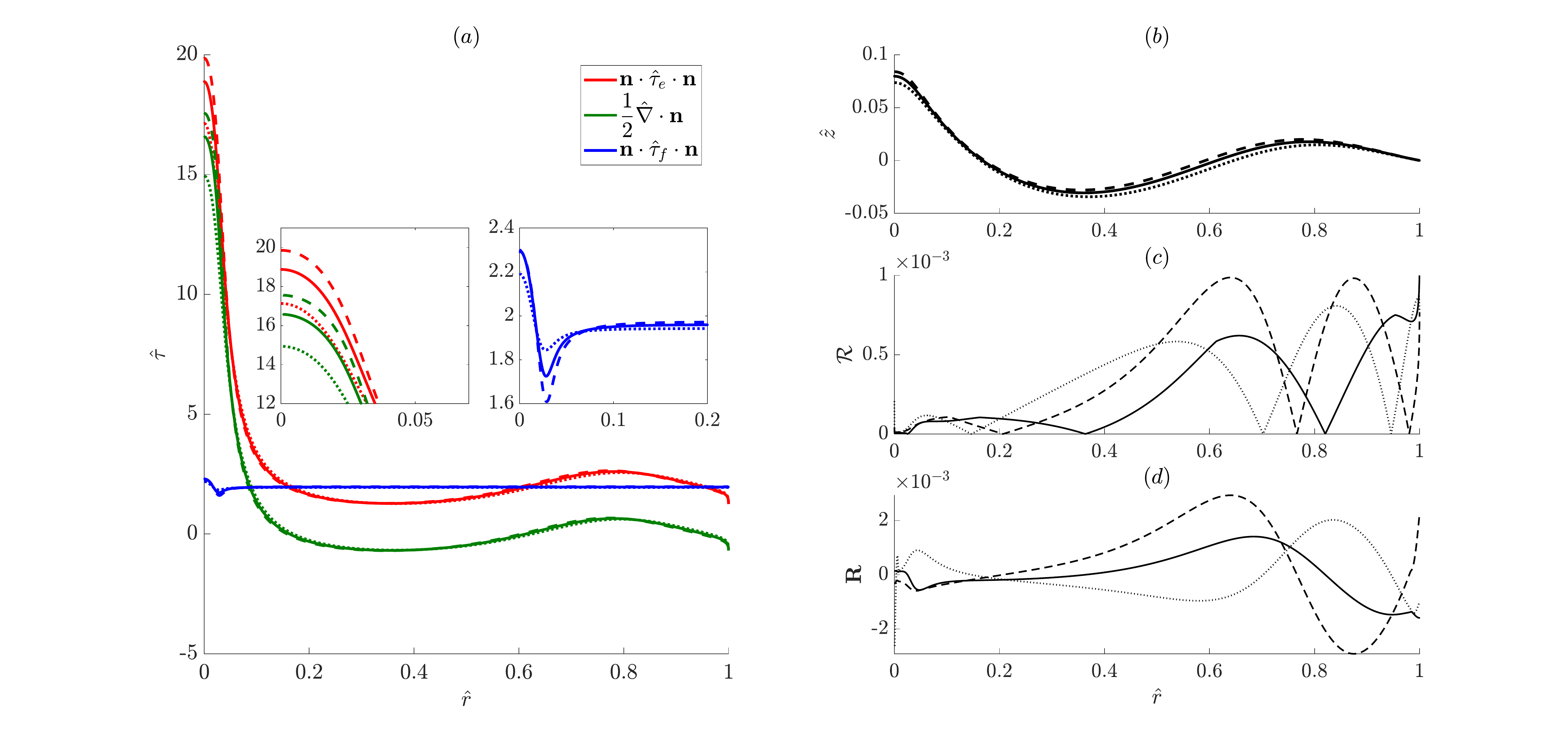}
    \caption{Subfigure a) shows non-dimensional normal stresses and equilibrium shapes for solutions at $\hat{E}_0 = 1.41$, $\hat{R} = 43$, $\hat{Z} = 0.8394$ as a function of the non-dimensional radial coordinate $\hat{r}$.  Stress solutions with three different reservoir pressures are shown in dashed, solid and dotted lines corresponding to $\hat{p}_r = 1,0$ and $-1$ respectively. 
    Electric stress distribution in red, surface tension stress in green and fluid hydrodynamic stress in blue. Subfigure (b) shows corresponding equilibrium shapes. The relative and absolute residuals are shown in subfigures c) and d) respectively.}
    \label{fig:stress_limit_stability}
\end{figure}

    Figure \ref{fig:stress_limit_stability}a shows the stress distributions along the meniscus interface for $\hat{E} = 1.41$, thus very close to the instability boundary (\ref{eq:upper_limit}). Solutions for three different reservoir pressures $\hat{p}_r = -1,0$ and 1 are shown in dotted, solid and dashed lines, respectively. The non dimensional currents emitted are $\hat{I} = 0.920 \cdot 10^{-4}, 1.971 \cdot 10^{-4}$ and $2.978 \cdot 10^{-4}$ respectively (if non-dimensionalized by the characteristic emitted current, $\frac{I}{I^*} =     0.0814, 0.175$ and $0.264$, respectively).  Differences in the stress distributions are concentrated in the vicinity of the emission region, where electric fields need to increase to accommodate higher current outputs at higher reservoir pressures. When emission is irrelevant, such as in the vicinity of the contact line where the bump forms (figure \ref{fig:stress_limit_stability}b) and $\hat{\sigma}$ is relaxed, stress distributions are a function of the external electric field only and directly independent from any parameter resultant from the emission. At this locaction, the only stress that would contain direct information from the emission region is the fluid hydrodynamic stress, where the local pressure equals that from the drop in the channel, thus proportional to the total emitted current. However, simulations show that this pressure near the contact line $\hat{p} = \hat{p}_r-\hat{I} \hat{R}^{\frac{5}{2}} \hat{Z}$ is mostly invariant from $\hat{p_r}$, $\hat{Z}$, $\hat{T}$ on $\Gamma^l_D$, $\psi$ and $\varepsilon_r$ therefore mostly a function of $\hat{E}_0$. This results in a set of equations that locally resemble the equilibrium of a perfect non-emitting conductor subject to an upstream suction stress, but with a sole degree of freedom or $\hat{E}_0$. This fact confers the limit observed in (\ref{eq:upper_limit}) some sense of universality and independence from ionic liquid physical properties, other than $\gamma$.
    
\begin{figure}
    \centering
    \includegraphics[width=\textwidth]{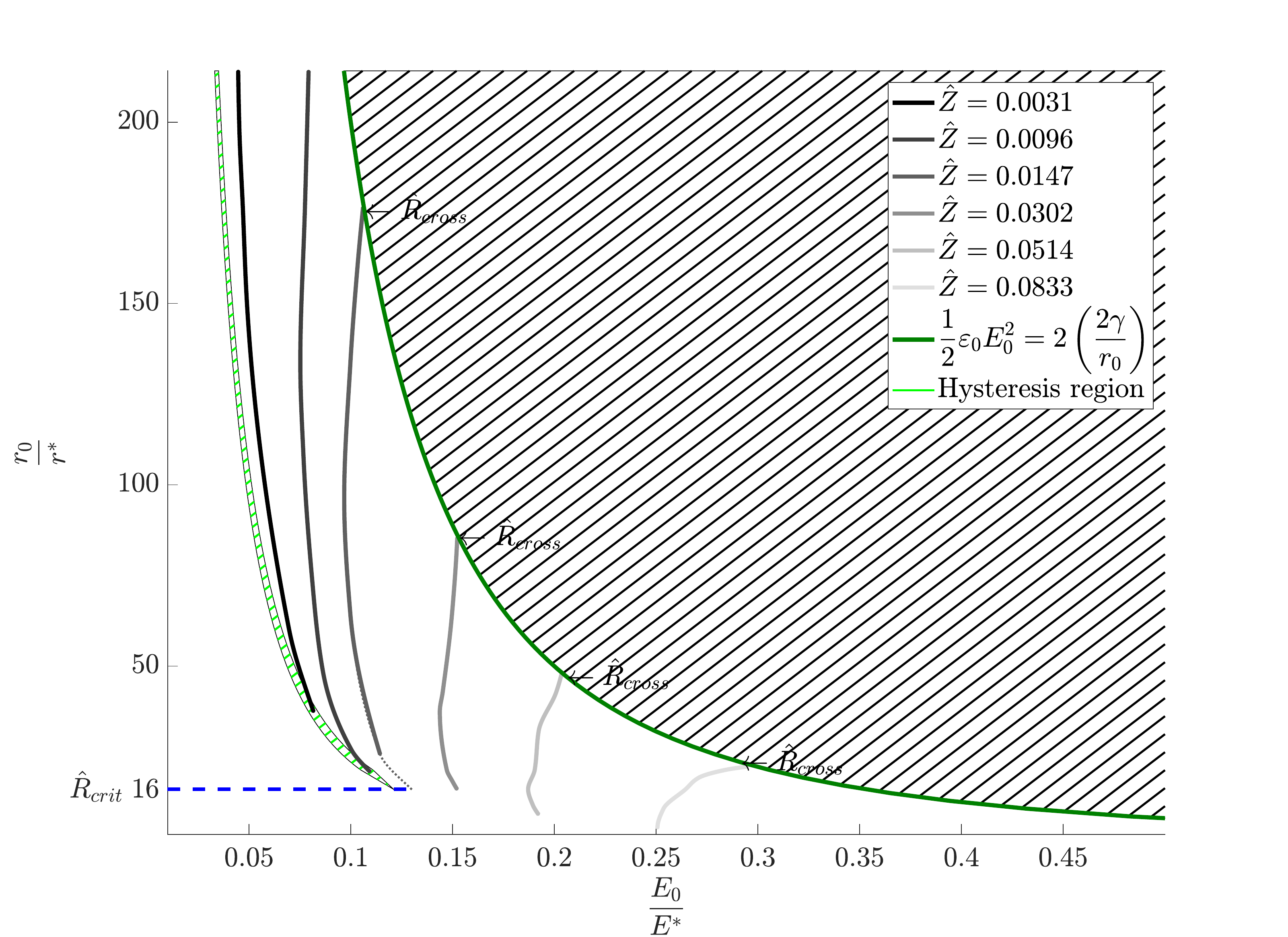}
    \caption{Boundaries of stability as a function of the external field non-dimensionalized by the critical field. Boundaries are shown for different dimensionless hydraulic impedance values $\hat{Z}$. The minimum non-dimensional impedance for the existence of emitting solutions in the range of $\hat{R}$ displayed in the figure is shown in black. Limits for increasing values of $\hat{Z}$ are shown in grey. Extrapolated values are shown in dotted lines. The hypothetical bifurcation point is shown in green. For the analyzed impedance values greater than  $\hat{Z} = 0.0096$, the maximum current limit crosses the presumable bifurcation limit at $\hat{R}_{cross} \approx 180,75,40$ and $25$ for $\hat{Z} = 0.0147,0.0302,0.0514$ and $0.0833$, respectively.}
    \label{fig:stability_Z}
\end{figure}

In cases where the hydraulic impedance is not sufficiently high, statically unstable solutions appear at values below $\hat{E}_{max}$. This can be seen in figure \ref{fig:stability_Z}. The diagram is similar to the one shown in figure \ref{fig:stability_contact}, but instead of using the nominal non-dimensionalization used in this paper, results in this analysis are presented with reference values of the field relating to the emission region ($E^*$).  Recall that the critical electric field depends exclusively on the ionic liquid properties and not on the source geometry, whereas nominal field $E_c$ is a function of the non-dimensional contact line radius $r_0$. For this reason, this alternative non-dimensionalization is more useful for relating simulation results to experimental data. In this non-dimensionalization, the maximum electric pressure limit decays with the field (green line), instead of being a vertical line. 

First, the need of a minimum hydraulic impedance of $\hat{Z} \approx 0.0031$ for static solutions to exist can be noticed for any of the $\hat{R}$ in the simulated range. The corresponding dimensional impedance is approximately $Z = 4.32 \cdot 10^{18}$ $\frac{\text{Pa}}{\text{m}^3/\text{s}}$ for the ionic liquid EMI-BF$_4$. This impedance is very close to that observed by \cite{Romero-Sanz2003SourceRegime} for achieving the pure-ion regime in capillary tubes of similar diameter as those reported here. The value of this impedance was predicted to be $Z \approx 4 \cdot 10^{18}$ $\frac{\text{Pa}}{\text{m}^3/\text{s}}$ by \cite{Perez-Martinez2016EngineeringApplications}.

Second, it can be seen that the stability ranges are widened in figure $\ref{fig:stability_Z}$ for increasing values of $\hat{Z}$.

\begin{figure}
    \centering
    \includegraphics[width=\textwidth]{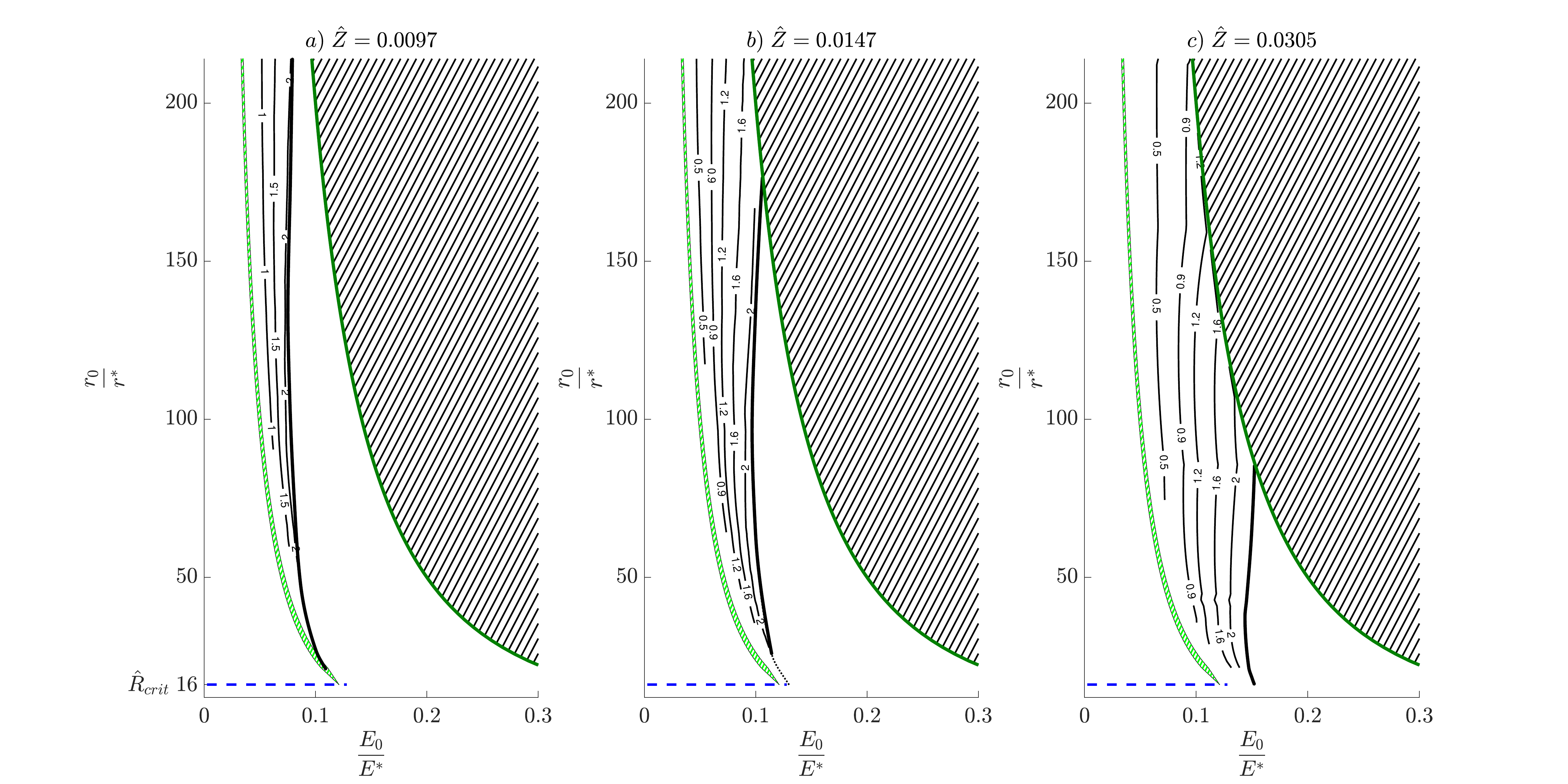}
    \caption{Dimensionless iso-current maps as a function of the contact line radius and external field referenced to $r^*$ and $E^*$ respectively. Results shown for 3 different values of hydraulic impedance. Hypothetical bifurcation point is shown in green. Hypothetical maximum current limit is shown in hard black for each of the $\hat{Z}$ displayed. Extrapolations are shown dotted.}
    \label{fig:stability_current}
\end{figure}

Figure \ref{fig:stability_current} shows the isocurrent lines at three values of $\hat{Z}$. The limits of stability for each $\hat{Z}$ are also shown with bolder lines. 

Notice how the increase of the stability boundaries is at the expense of a lower current output at fixed $\frac{E_0}{E^*}$ and $\hat{R}$. This trade-off between current output and meniscus stability is well known in the experimental pure ion evaporation literature \citep{Castro2009EffectTips,Krpoun2009TailoringElectrodes,Hill2014High-ThroughputStructures}.

Figure \ref{fig:shapes_efield}b shows how equilibrium shapes adapt to this current reduction when changing the hydraulic impedance at fixed $\hat{R}$ and $\hat{E}$. For the higher impedance case (dotted line), equilibrium shapes are smoother in the neighborhood of the emission region. In this case, local electric fields are less intense because of the lower current throughput demand. Therefore, surface tension can balance the electric stress with larger radii of curvature. Equilibrium shapes near the region close to the contact line are practically invariant with the increase of $\hat{Z}$. 

Third, the limit of stability for every hydraulic impedance shown in figure \ref{fig:stability_current} resembles an isocurrent line of about $\frac{I}{I^*} \sim$ 2.2 for all the $\hat{Z}$ shown in figure \ref{fig:stability_current}. This suggests that the static stability of a meniscus in the pure ion mode is linked to a limit in current throughput, when $\hat{E}_0 < \hat{E}_{max}$.

The existence of a maximum current appears to be related to a reduction in the area of emission at the apex of the meniscus. The contraction of the emission area is linked to a decrease in the radius of curvature that is needed to compensate for the higher electric stress. This trade-off between the reduction of the emission area and growth of the current density appears to limit the current that can be extracted from the meniscus for increasing values of $\hat{E}_0$ (see Appendix \ref{sec:Annex_Lumped_Model}). This phenomenon was predicted to exist also for viscousless Liquid Metals \citep{Forbes2004LiquidFollowed}.   

From the data shown in figures \ref{fig:stability_Z} and \ref{fig:stability_current} at a given value of $\hat{Z}$, the two competing instability phenomena will occur at different ranges of $\hat{R}$. Menisci would loss their stability by a presumably bifurcation phenomena if their size $\hat{R} > \hat{R}_{cross}$, and will be limited by a maximum current throughput when $\hat{R} < \hat{R}_{cross}$.

Interestingly, $\hat{R}_{cross}$ provides the largest span of stable electric fields. As seen in figures \ref{fig:stability_Z} and \ref{fig:stability_current}, this $\hat{R}_{cross}$ decreases when more hydraulic impedance is provided, and the range of fields widens.



 For representative values of $r^*$ in ionic liquids ($\sim 50$ nm) and impedances greater than $Z  = 10^{19}$ $\frac{\text{Pa}}{\text{m}^3/\text{s}}$,  $r_{0_{cross}} = \hat{R}_{cross} \cdot r^*$ is found to be below 3 $\mu$m in dimensional form $(\hat{R}_{cross} \sim 100)$.
 
 If the range of stable fields was a measure of the probability of finding the meniscus at any $\hat{R}$, then $\hat{R}_{cross}$ would be good estimation of this value. As mentioned previously, the scale of $\hat{R}_{cross}$ is close to the diffraction limit of standard optical observation systems, thus explaining in part the reason why non-invasive direct observation of pure-ion emitting menisci has not been reported by the scientific community. 
 
 The characteristic small meniscus sizes where the static stability ranges are maximum ($\hat{R}_{cross}$) are not in contradiction with the findings of \cite{Castro2006CapillarySources}, \cite{Garoz2007TaylorConductivity} or \cite{Romero-Sanz2005IonicLiquids}, where the pure ion regime is achieved for substantially larger diameter capillaries between $40$ and $200 \; \mu$m. The results in figures \ref{fig:stability_contact},  \ref{fig:stability_Z} and \ref{fig:stability_current}  only show the predicted static stability ranges for menisci of non-dimensional radius between $\hat{R} = 4$ and $\hat{R} = 210$. For the $r^*$ of EMI-BF$_4$, these ranges correspond to radii in between $0.1$ to $10 \; \mu$m. If the maximum field limit (eq. \ref{eq:upper_limit}) is extrapolated to these radii, stable menisci are still found, yet at lower range of electric fields. It is worth mentioning that having direct observation of these menisci could be very valuable, particularly to discard any emission process governed by smaller ill-anchored menisci at the rim of the capillary channel.

\begin{figure}
    \centering
    \includegraphics[width=\textwidth]{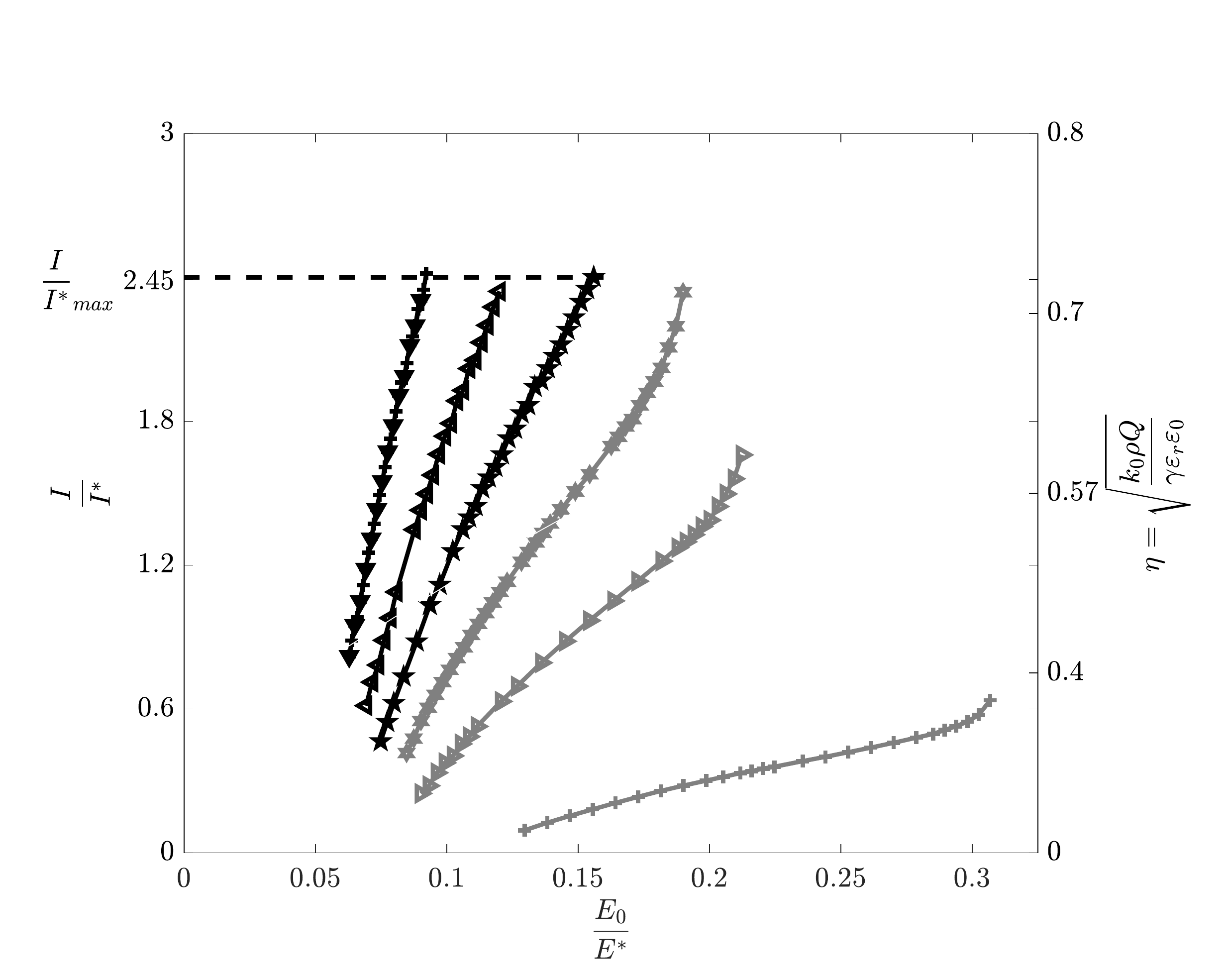}
   \caption{Current emitted as a function of the external field for contact line radius in region \textbf{II}. From left to right, the non-dimensional radius of the curves correspond to $\hat{R} = 100, 75, 64, 54,  43$ and $21$. In that order, the non dimensional hydraulic impedance coefficients correspond to $\hat{Z} = 0.0096, 0.021, 0.030, 0.045, 0.083$ and $0.47$. For the radius depicted in grey, the presumably bifurcation point was reached before $\frac{I}{I^*}_{max}$.}
    \label{fig:IV-CR}
\end{figure}

The effect of the two mechanisms that lead to static instability on the current is shown in figure \ref{fig:IV-CR}. Figure \ref{fig:IV-CR} shows the current-field curves for different pairs of $\hat{R}$ and $\hat{Z}$. The curves with smaller radii and higher hydraulic impedance are shown in grey. The maximum current achieved in these cases corresponds to an external field $\hat{E}_0 = \hat{E}_{max}$, therefore losing stability by the presumed bifurcation of the meniscus. 
These results show how the maximum currents achieved for such bifurcating menisci are typically smaller than the current limit of $\frac{I}{I^*}_{max} \approx 2.4$ obtained for the cases of lower $\hat{Z}$ and higher $\hat{R}$. In these latter cases shown in black, stability is lost when reaching that current. Notice how in curves of such lower impedances, the current emitted per unit field is higher. This effect is well known in the literature \citep{Krpoun2009TailoringElectrodes}. 

The dimensionless flow parameter $\eta=\sqrt{\frac{\rho \kappa_0 Q}{\gamma \varepsilon_r \varepsilon_0}}$ defined by \cite{FernandezDeLaMora1994TheCones} is also shown on a right vertical axis in figure \ref{fig:IV-CR}. Unlike electrosprays in the mixed droplet-ion regime, where decreasing values of $\eta$ are typically needed for achieving higher currents \citep{Lozano2002ExperimentalRegime}, electrosprays in the pure-ion mode exhibit larger current throughput at increasing values of $\eta$. It is also interesting to notice that while conventional cone-jet electrosprays become unstable when approaching $\eta\sim 1$ from higher flow rates, the results in this work suggest that pure-ion electrosprays also become unstable near $\eta\sim 1$, but when approached from lower flow rates.

\begin{figure}
    \centering
    \includegraphics[width=0.5\textwidth]{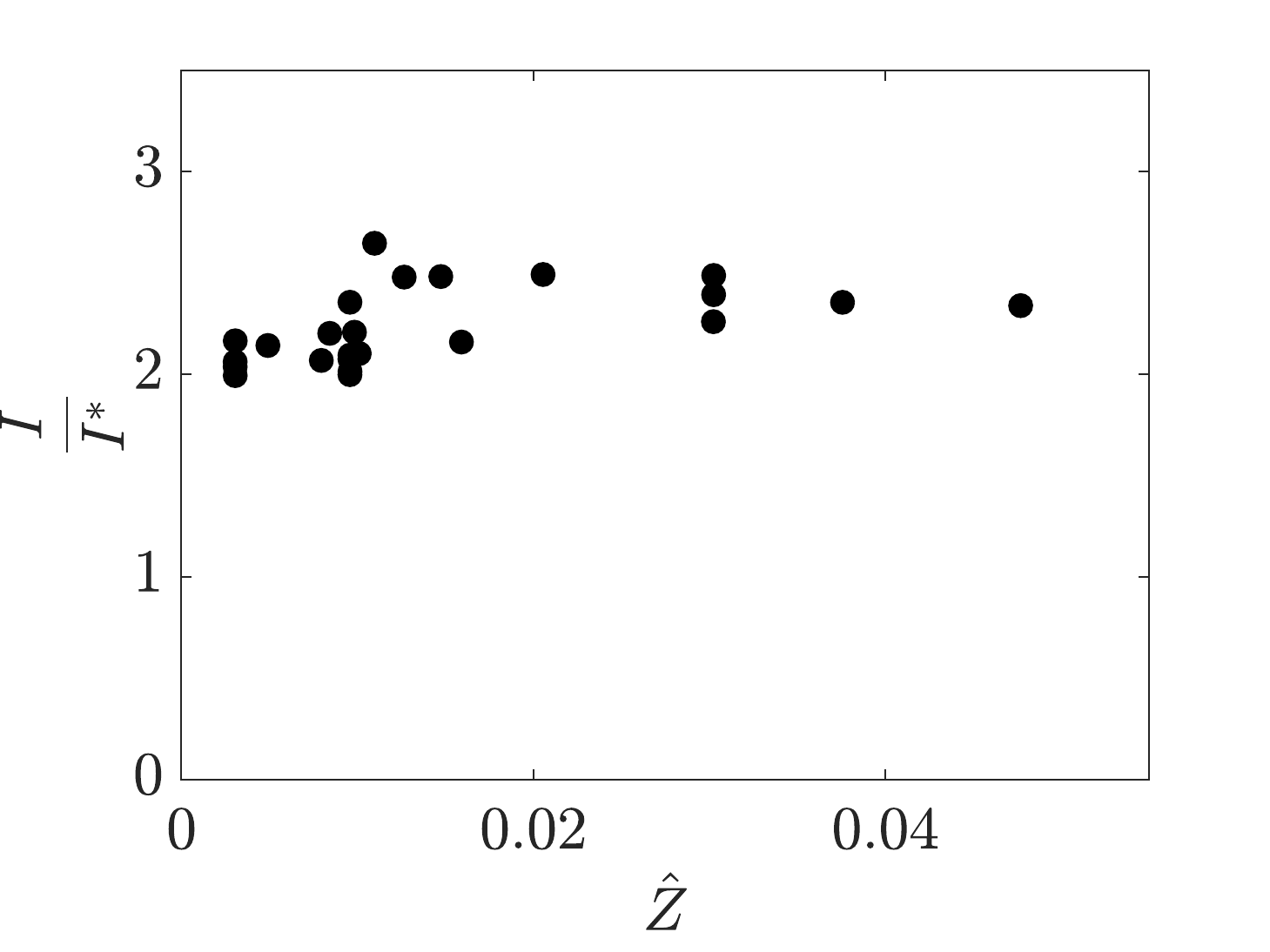}
    \caption{Maximum values of the current reached for 34 different non-dimensional field, radius and hydraulic impedance. Values of radius and external fields are chosen in region \textbf{II}. Values of the  hydraulic impedance are chosen low enough for not triggering the bifurcation point at $\hat{E}_0 = \hat{E}_{max}$.}
    \label{fig:Z_current}
\end{figure}

The current limit of stability appears to hold in a wide range of hydraulic impedances and radii. Figure \ref{fig:Z_current} shows the current emitted in the limit of stability for 35 different pairs of $\hat{Z}$ and $\hat{R}$ when the hydraulic impedance is not sufficient to trigger the bifurcation process. The range of maximum currents is between $2.1-2.4$ times $I^*$ for all the simulated values.

\begin{figure}
    \centering
    \includegraphics[width=\textwidth]{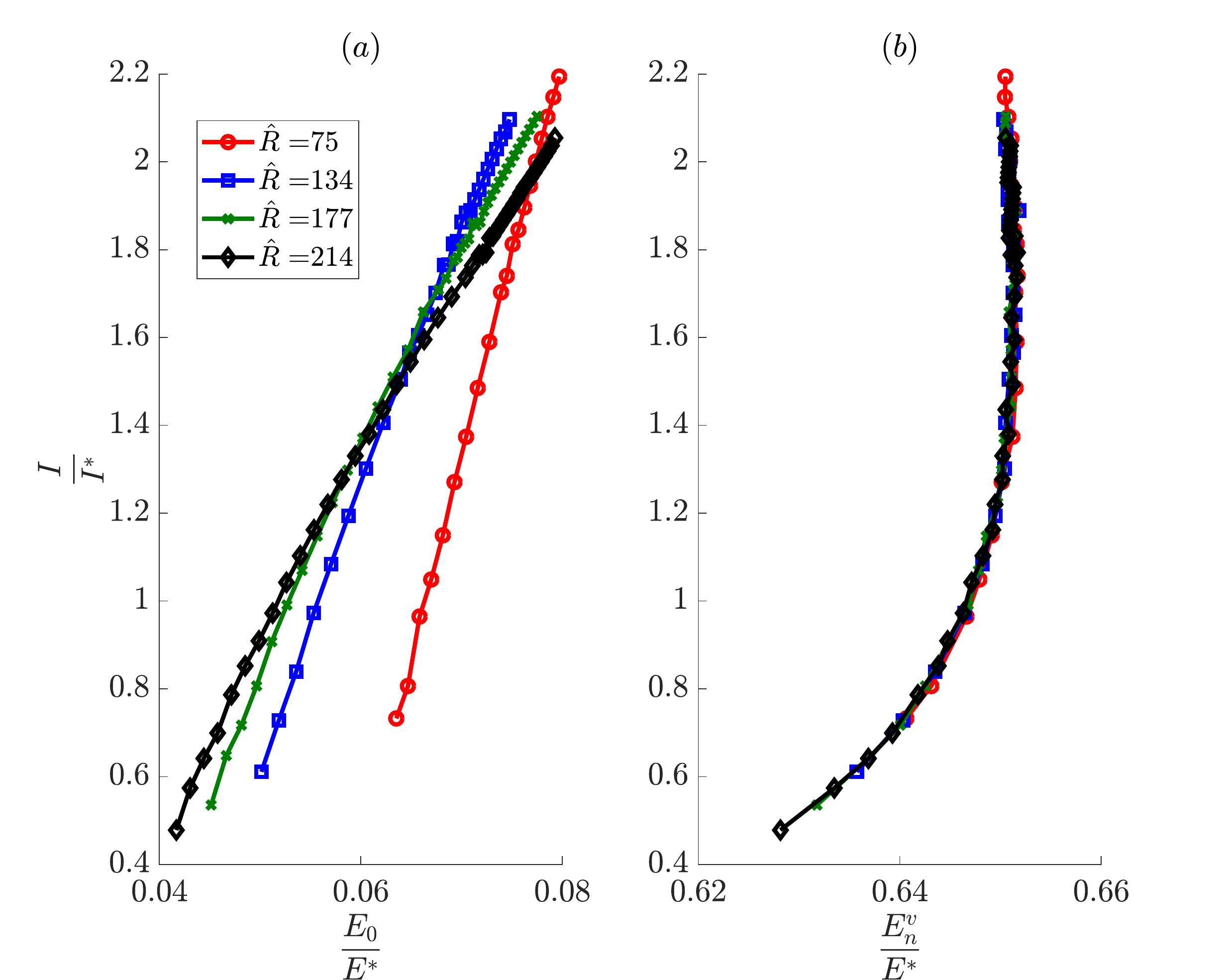}
    \caption{Figure shows the non-dimensional current emitted by differently sized meniscus at $\hat{Z} = 0.0096$ as a function of $\frac{\hat{E}_0}{\hat{E^*}}$ (a) and an average of the $\frac{E^v_n}{\hat{E^*}}$ fields in the neighbourhood of the meniscus tip, or $\hat{r} = 0$ (b). The average is performed as follows $\frac{1}{A_0}\int_{A_0} \frac{E^v_n}{\hat{E^*}} \; d A_0 $ , where $A_0 = \int^{\Delta \hat{r}}_0 2\pi \hat{r} \sqrt{1 + \hat{y}'^2} d\hat{r}$ for the portion of the meniscus $\Delta \hat{r}$ such that $\hat{j}^e_n |_{\hat{r} = \Delta \hat{r} } = 0.99 \hat{j}^e_n |_{\hat{r} = 0}$.}
    \label{fig:IV-EXP}
\end{figure}

The effect of the meniscus geometry ($\hat{R}$) at fixed $\hat{Z} = 0.0096$ are shown in figure \ref{fig:IV-EXP} for different values of $\hat{R}$. For all cases investigated in this figure, the values of $\hat{Z}$ and $\hat{R}$ are not sufficient to trigger the presumed bifurcation and static equilibrium solutions were found yielding a current outputs below $\frac{I}{I^*} \approx 2.1$. Figure \ref{fig:IV-EXP}a shows the current output as a function of the non-dimensional external field $\frac{E_0}{E^*}$. Unlike electrospray cone-jets, where the liquid profile and emitted current is a function of the operational parameters and mostly independent from the electrode geometry \citep{Gamero-Castano2019NumericalMode,FernandezDeLaMora1994TheCones}, menisci in the pure ion mode are typically smaller and more sensitive to changes in the electric field, as their emission region is comparatively closer to the electrodes and the space charge in the ion plume is negligible. The effect of this is seen in the higher steepness of the current-field slope for the smaller menisci. 

It is worth mentioning that, when the current emitted is plotted against an average of the normal fields in the vacuum near the tip of the meniscus ($\frac{E^v_n}{E^*}$), the results nearly collapse into a single curve (figure \ref{fig:IV-EXP}b). This reinforces the notion that current throughput could be regarded as a function of the local values of the electric fields, including the mechanism behind a possible limitation in current, such as the one described in Appendix \ref{sec:Annex_Lumped_Model}. 

    Region $\mathbf{IV}$ is defined for contact line radii below $\hat{R}_{crit}$ and it is characterized by the lack of a transition gap. Equilibrium menisci in this region evolve smoothly from a non-emitting configuration to an emitting configuration for increasing values of $\hat{E}_0$. 
    
    Simulations of equilibrium shapes have also been performed for contact line radii above $r_0 \approx 250$ nm ($\hat{R} \approx 6$). The continuum approach below this length scale is likely no longer valid due to the role that discrete molecular effects start to play.
    
    Menisci in this region resemble those explored by \cite{Higuera}. As discussed by \cite{Coffman2019ElectrohydrodynamicsField}, the non-dimensional critical electric field $\hat{E}^* = \hat{R}^{\frac{1}{2}}$ is on the order of those found near the apex of the hyperboloidal shapes described by \cite{Basaran1990AxisymmetricField}. The pressure drop created by the evaporation process compensates for the electric stress before the Rayleigh instability is triggered. This phenomenon can be seen in figure \ref{fig:Beroz_limit}. For the cases where $\hat{R} < \hat{R}_{crit}$ (blue lines), the pressure drop reduces the volume increase due to the action of the electric field to shapes that lie within the Basaran-Beroz limit \citep{Beroz2019StabilityDroplets}. At this point, the meniscus is no longer hydrostatic, the surface charge is not fully depleted and the channel pressure drop is significant, making the Basaran-Beroz limit no longer valid.
    
If the electric field is increased further for emitting shapes with  $\hat{R}< \hat{R}_{crit}$, then the hydraulic pressure drop becomes more relevant than the surface tension in compensating for the electric stress pull over the meniscus interface. 

It is observed that at at very high hydraulic impedance coefficients ($\hat{Z} > 0.7136$), the instability described in \textbf{III} is triggered at lower electric fields $\hat{E}<\hat{E}_{max}$. Somewhat against intuition, for $\hat{R} < \hat{R}_{crit}$, this instability occurs at increasingly lower external electric fields when $\hat{Z}$ increases. 

\begin{figure}
    \centering
    \includegraphics[width=\textwidth]{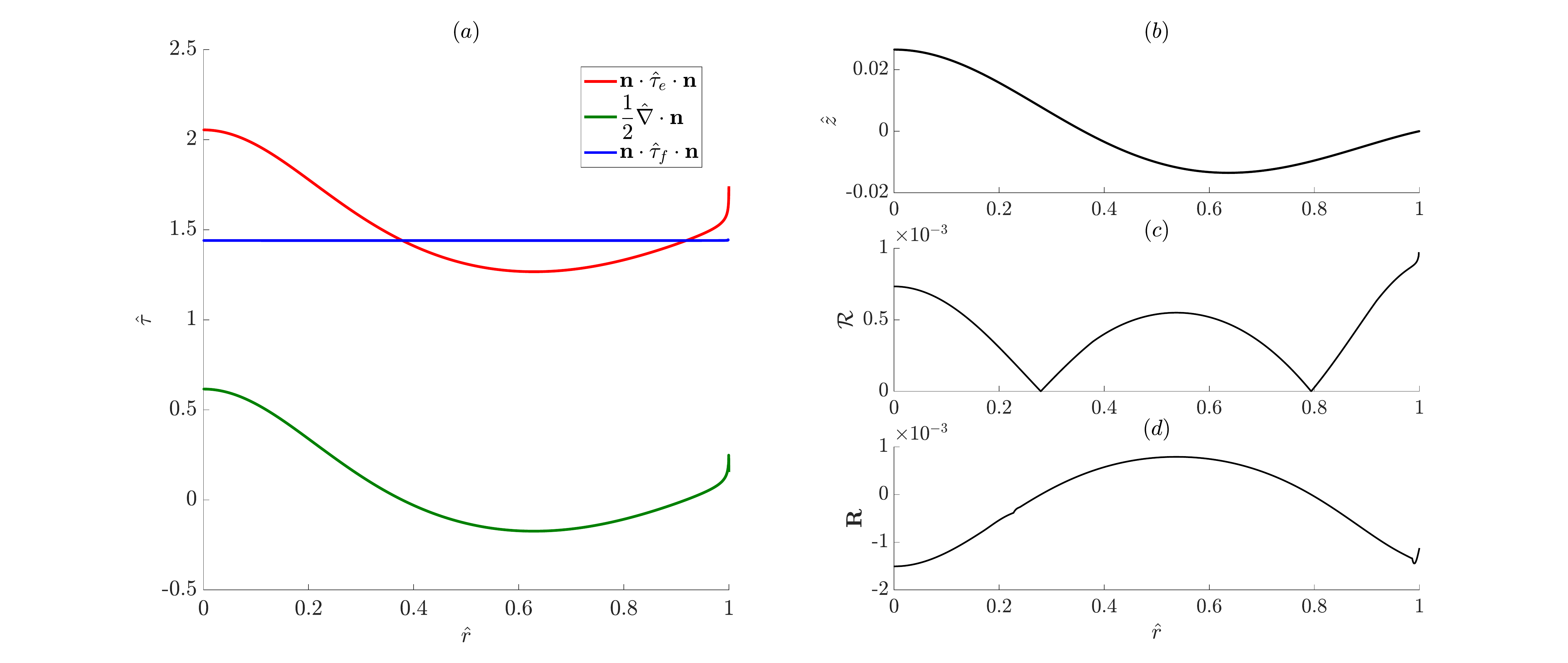}
    \caption{Distribution of the normal component of the stress to the meniscus interface for an equilibrium solution in region ($\mathbf{IV}$) close to the electric field of instability ($\hat{E}_0 = 1.2$, $\hat{R} = 10.7$, $\hat{Z} = 71.4$. The meniscus axisymmetric interface profile $\hat{z}$ is shown in subfigure b). Normal electric and fluid stresses are shown in red and blue respectively, surface tension stress in green. Relative and absolute residuals are shown in subfigures c) and d), respectively.}
    \label{fig:higueraStresses}
\end{figure}

Unlike the distribution of stresses of the equilibrium solutions in region $\mathbf{III}$, where most of the electric stress is balanced by the surface tension, solutions in region $\mathbf{IV}$ are somewhat planar when the electric field downstream approaches the limit of stability. Figure \ref{fig:higueraStresses} shows the normal stress distributions for an equilibrium solution in $\mathbf{IV}$ very close to the instability limit. The meniscus is practically hydrostatic in this region (fluid flow stress is negligible). The electric field stress is practically counteracted by the hydraulic pressure drop due to current evaporation. Without the surface tension playing a relevant role, the hydraulic impedance coefficient controls the sensitivity of the balance to the electric stress. It is observed that when the electric field remains close enough to the stability boundary, the suction pressure due to the hydrostatic drop grows beyond the value of the electric stress and turns the meniscus inside out, thus making it adopt a concave form which was considered to be unstable due to the aforementioned three-dimensional effects not captured in the axially-symmetric formulation.

\subsection{Influence of the liquid bulk temperature on emission and stability properties}

The physical properties of ionic liquids depend on temperature, sometimes in a significant way. It is therefore expected that temperature variations will have an effect on the static stability of menisci investigated in this work.

Ohmic  dissipation, as described by the energy transport equation (\ref{eq:energy_transport}) is the driving mechanism behind the increase of temperature in the liquid, specifically in the vicinity of the emission region where the current density is the highest.

\begin{figure}
    \centering
    \includegraphics[width=\textwidth]{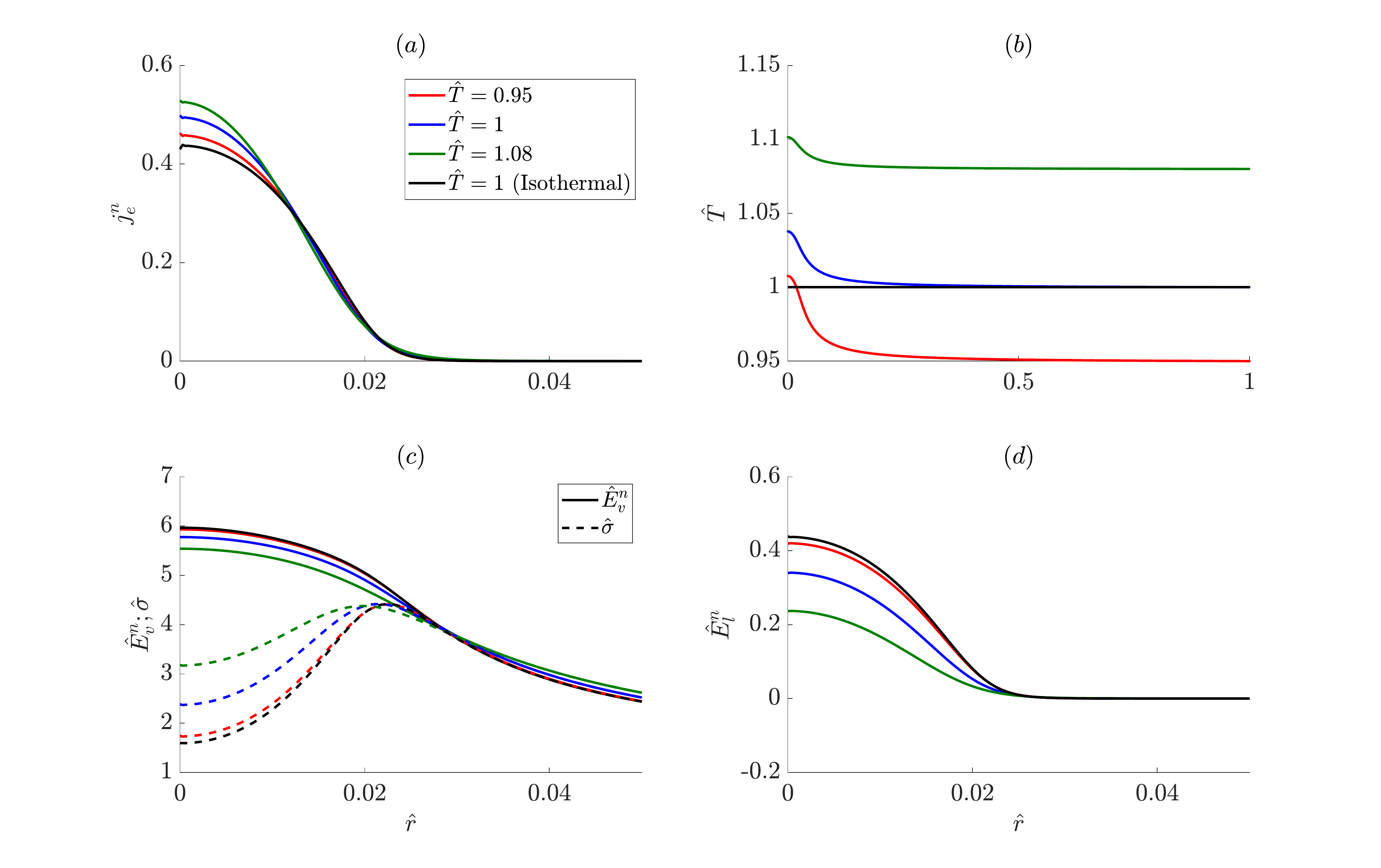}
    \caption{(a) shows the current density distribution along the meniscus interface in the vicinity of the emission region ($\hat{r}$ near $0$). Temperature distribution along the interface from the emission region to the contact line is shown in $(b)$. The non-dimensional electric fields normal to the meniscus interface in the vicinity of the emission region are shown in (c) and (d) for the vacuum and liquid, respectively. (c) also shows the non-dimensional surface charge distribution (dashed line). Results are shown for three different ionic liquid bulk temperatures and the isothermal case for comparison reasons. Simulation data corresponds to $\hat{R} = 64$, $\hat{E}_0 = 0.78$ and $\hat{Z} =0.0144$.}.
    \label{fig:TempData}
\end{figure}

\begin{figure}
    \centering
    \includegraphics[width=\textwidth]{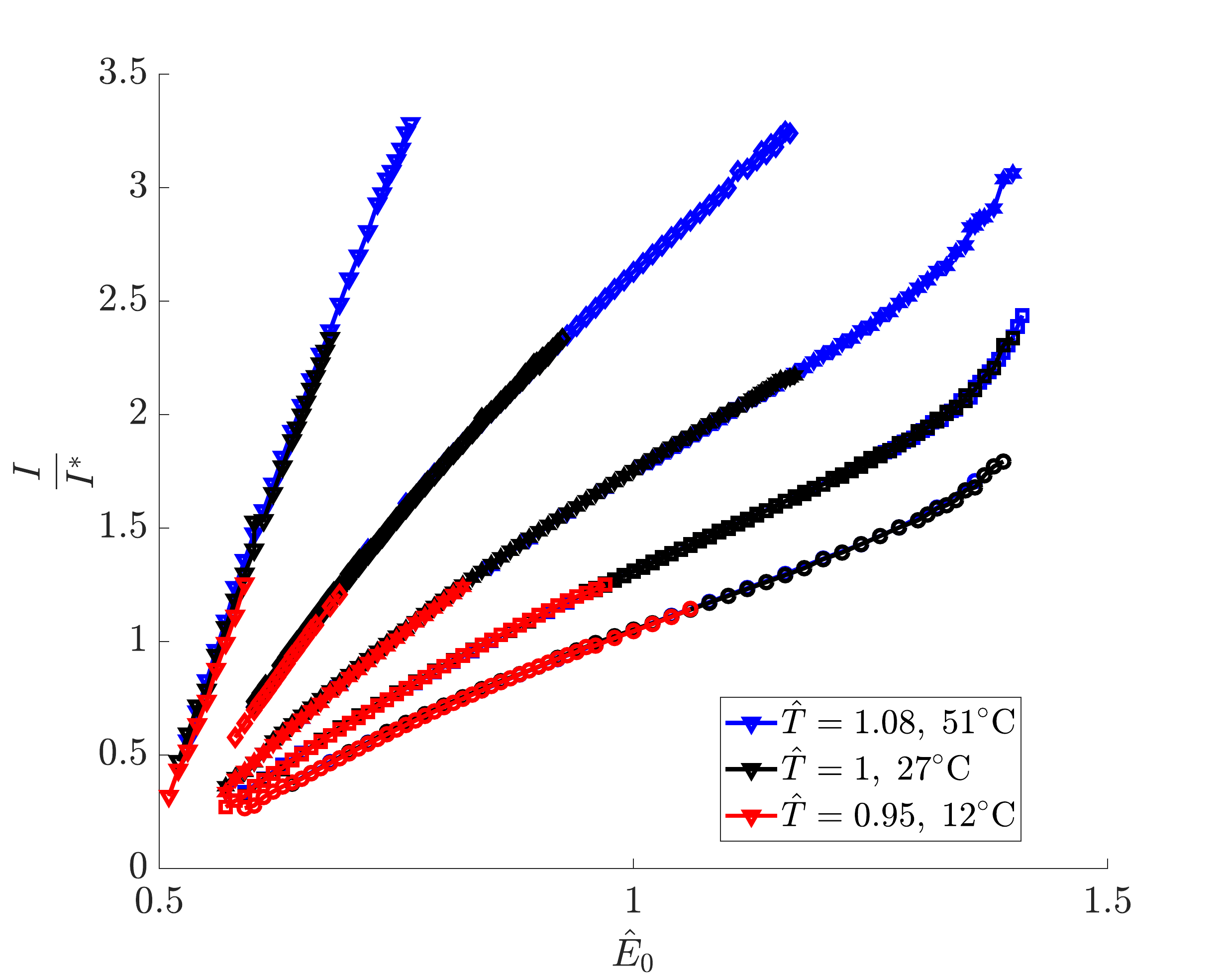}
    \caption{Current emitted scaled to $I^*$  as a function of the non-dimensional electric field. Results are shown for three different temperatures at $\hat{Z} = 0.0302$. From left to right, meniscus sizes are $\hat{R} = 107, 86, 64, 43$ and $21$}
    \label{fig:IV-TEMP}
\end{figure}

The mechanical balance of stresses on the meniscus is affected through changes in electric conductivity $\kappa\left(T\right)$ and  fluid viscosity $\mu\left(T\right)$, and through a modification of the activation law for ion evaporation (\ref{eq:arrhenius}). The global effect of a temperature increase on the emission characteristics can be seen in figures \ref{fig:TempData} and \ref{fig:IV-TEMP}. 

Intuitively, the rise of the electric conductivity due to a temperature increase may incur in more current throughput (for a meniscus with negligible convective charge transport, $\mathbf{j}  = \kappa(T)\mathbf{E}$). However, figure \ref{fig:TempData}a shows that the current distribution along the meniscus interface in the neighborhood of the emission region unexpectedly remains constant despite the temperature increase (and conductivity) along $\Gamma_M$.  

Notice that, for the linear conductivity model with temperature used in this paper and the values of the parameters simulated ($\chi = 1.81 \cdot 10^{-3}$ and $\lambda = 12$),  a higher conductivity also increases the ratio between the characteristic emission time ($\tau_e \sim \frac{h}{k_B T}$) and the charge relaxation time ($\tau_r \sim \frac{\varepsilon_r\varepsilon_0}{\kappa\left(T\right)}$). For moderate increases of temperature, namely $\hat{T} \approx 1.04$, the increase of the ratio $\frac{\tau_e}{\tau_r} $ is about $40\%$, where:
\begin{equation}
    \frac{\tau_e}{\tau_r} = \frac{\chi \left(1+\Lambda \left(\hat{T}-1\right)\right)}{\hat{T}}
    \label{eq:characteristicTimes}
\end{equation}
And $\chi = \frac{h\kappa_0}{k_BT_0\varepsilon_0\varepsilon_r}$ (see table \ref{tab:nond_variables}).

In this case, the meniscus is able to relax surface charge faster than  the rise of emission timescale at higher bulk temperatures. This phenomenon can be seen in figure \ref{fig:TempData}c, where a more relaxed surface charge distribution ($\sigma \sim \varepsilon_0 E^v_n$) is observed. 

This over-relaxation of $\hat{\sigma}$ will tend to reduce the internal electric field, given the assumption that the external electric field $E^v_n$ has a weak dependence on the temperature (Figure \ref{fig:TempData}d). This can be observed by using the interface field condition (\ref{eq:interface_charge_condition}) to  write the internal field as a function of $\sigma$:

\begin{equation*}
    E^l_n = \frac{\varepsilon_0 E^v_n - \sigma}{\varepsilon_0 \varepsilon_r}
\end{equation*}  

The validity of this assumption (see figure \ref{fig:TempData}) is supported by the fact that larger variations in the external electric field would affect exponentially the current output through (\ref{eq:arrhenius}).  

The dependence of the emitted current density on the two phenomena can be better appreciated when writing it as an explicit function of the normal electric field acting on $\Gamma_M$, $E^n_v$:
\begin{equation}
    \begin{split}
        j^e_n = \kappa(T)E^n_l = \kappa_0\left(1+\Lambda\left(\hat{T}-1\right)\right)\frac{\varepsilon_0 E^n_v-\sigma }{\varepsilon_0 \varepsilon_r} \\
        = \frac{\kappa_0\left(1+\Lambda\left(\hat{T}-1\right)\right)\frac{E^n_v}{\varepsilon_r}}{1+\frac{\tau_e}{\tau_r}\exp{\frac{\psi}{\hat{T}}\left(1-\sqrt{\frac{E^n_v}{E^*}}\right)}}
        \label{eq:current_Envac}
    \end{split}
\end{equation}
Where Eqs. (\ref{eq:arrhenius}) and (\ref{eq:interface_charge_condition}) have been used to relate $\sigma$ to $j^e_n$ and $E^n_v$.

Given these results, an anticipated increase of current due to a higher conductivity coefficient is canceled out by a reduction of the electric field inside the meniscus due to charge relaxation. This effect can be seen in figure \ref{fig:IV-TEMP}, which shows the negligible effect of the liquid temperature on the extracted total current for a given $\hat{Z}$ and $\hat{R}$ as a function of $\hat{E}_0$, even if an isothermal meniscus was considered ($\frac{I}{I^*} = 0.317$, for the four cases in figure \ref{fig:TempData}). 

These results support the hypothesis of \cite{Lozano2005IonicEmitters}, where the experimental increase of current at higher temperatures is associated to a decrease of the hydraulic impedance due to the lower viscosity of the ionic liquid.


Another effect linked to an increase of the bulk temperature of the liquid is shown in figure \ref{fig:IV-TEMP}. It can be observed that the the maximum current limit occurs at higher values for lower $\hat{T}$, as predicted by the lumped parameter model in Appendix \ref{sec:Annex_Lumped_Model}. It is worth mentioning that when the hydraulic impedance is sufficiently high, the meniscus reaches the presumed bifurcation point at the same $\hat{E}_{max} \approx \sqrt{2}$ as predicted by the simulations with $\hat{T} = 1$, and before $\hat{I}_{max}$. The case with $\hat{T} = 0.95$ is particularly interesting, since the reduced $\hat{I}_{max}$ allows lower impedance menisci ($\hat{R} = 42.8
$) to reach instability before the bifurcation point. 

\begin{figure}
    \centering
    \includegraphics[width=\textwidth]{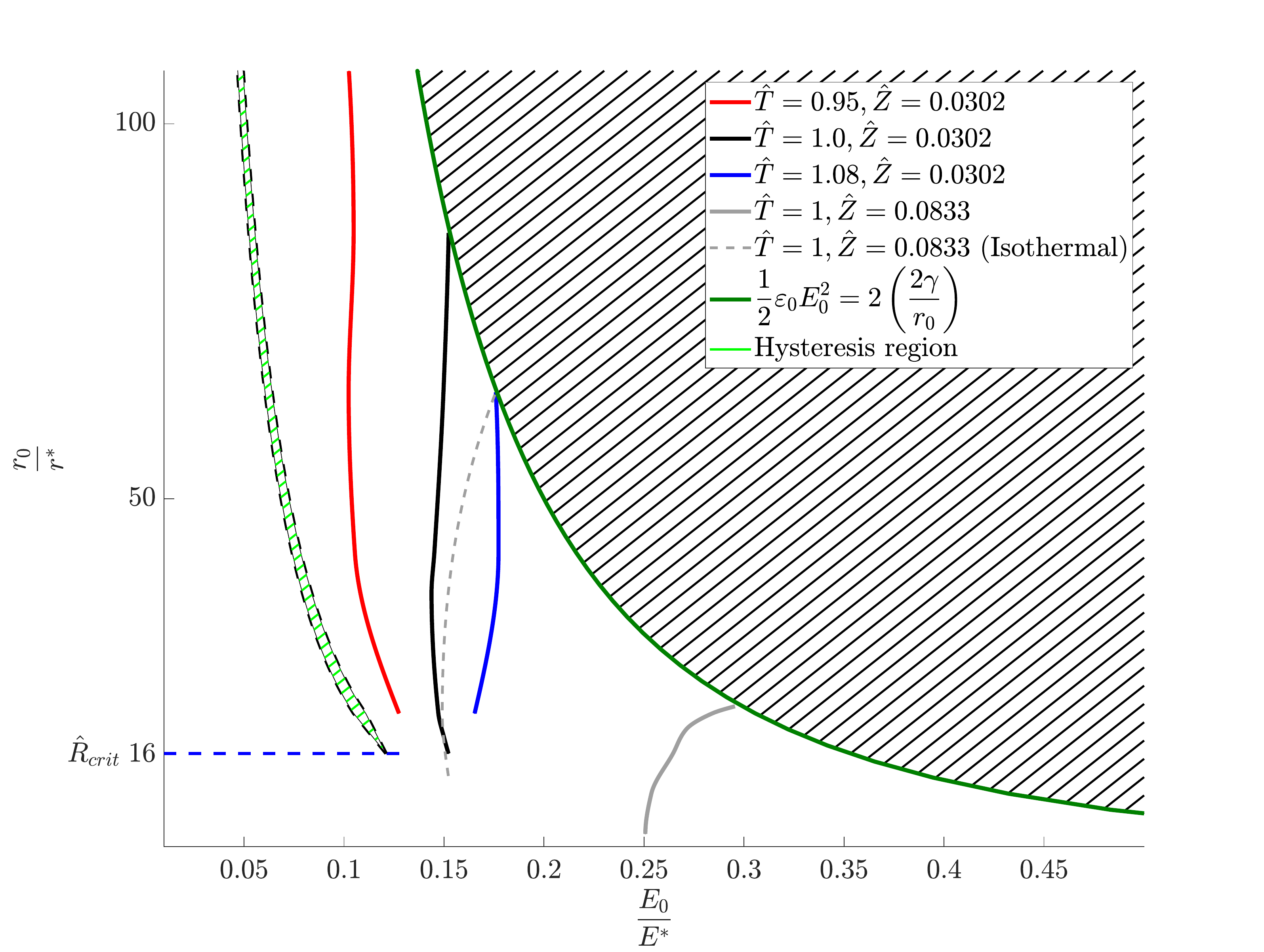}
    \caption{Stability diagram obtained for different ionic liquid bulk temperatures at a constant hydraulic impedance coefficient of $\hat{Z} = 0.0302$. Stability boundary is also shown for a higher impedance $\hat{Z} = 0.0833$ in grey. Stability computed considering an isothermal meniscus is shown with a dashed line for reference. For the latter case, no energy equation was solved, and bulk temperature was set to $\hat{T} = 1$.}
    \label{fig:stability_temp}
\end{figure}

The effect of increasing the temperature is widening the range of electric fields where pure-ion emission is statically stable, irrespective of the meniscus radii $r_0$. The expansion is reflected in the increase of $\hat{I}_{max}$ at higher bulk temperatures (figure \ref{fig:stability_temp}a).  However, it is true that this range cannot increase without limit. According to the findings in this paper, the maximum range is determined by the upper limit electric field above which pure-ion emission cannot be sustained with a single axisymmetric meniscus (eq. \ref{eq:upper_limit}).

Regardless, menisci operating at electric fields below eq. \ref{eq:upper_limit} that are not stable at a given impedance could stabilize if heated, while keeping the same impedance. This could give insight into explaining the temperature thresholds needed for achieving the pure ionic regime in capillary tubes of smaller impedance than porous tips \citep{Garoz2007TaylorConductivity,Romero-Sanz2005IonicLiquids}. 

Figure \ref{fig:stability_temp}a also shows that taking energy conservation into consideration is very relevant in describing the stability boundaries. Dashed lines show how much narrower the stability field range would look like for $\hat{Z} = 0.0833$, when considering an isothermal meniscus (i.e, without taking into account any heating effects). In fact, no statically stable solutions were found at $\hat{Z} = 0.0302$ for the isothermal case. This effect is consistently related to the fact that heated menisci are more accessible to higher maximum currents at similar values of $\hat{Z}$. 

\begin{figure}
    \centering
    \includegraphics[width=\linewidth]{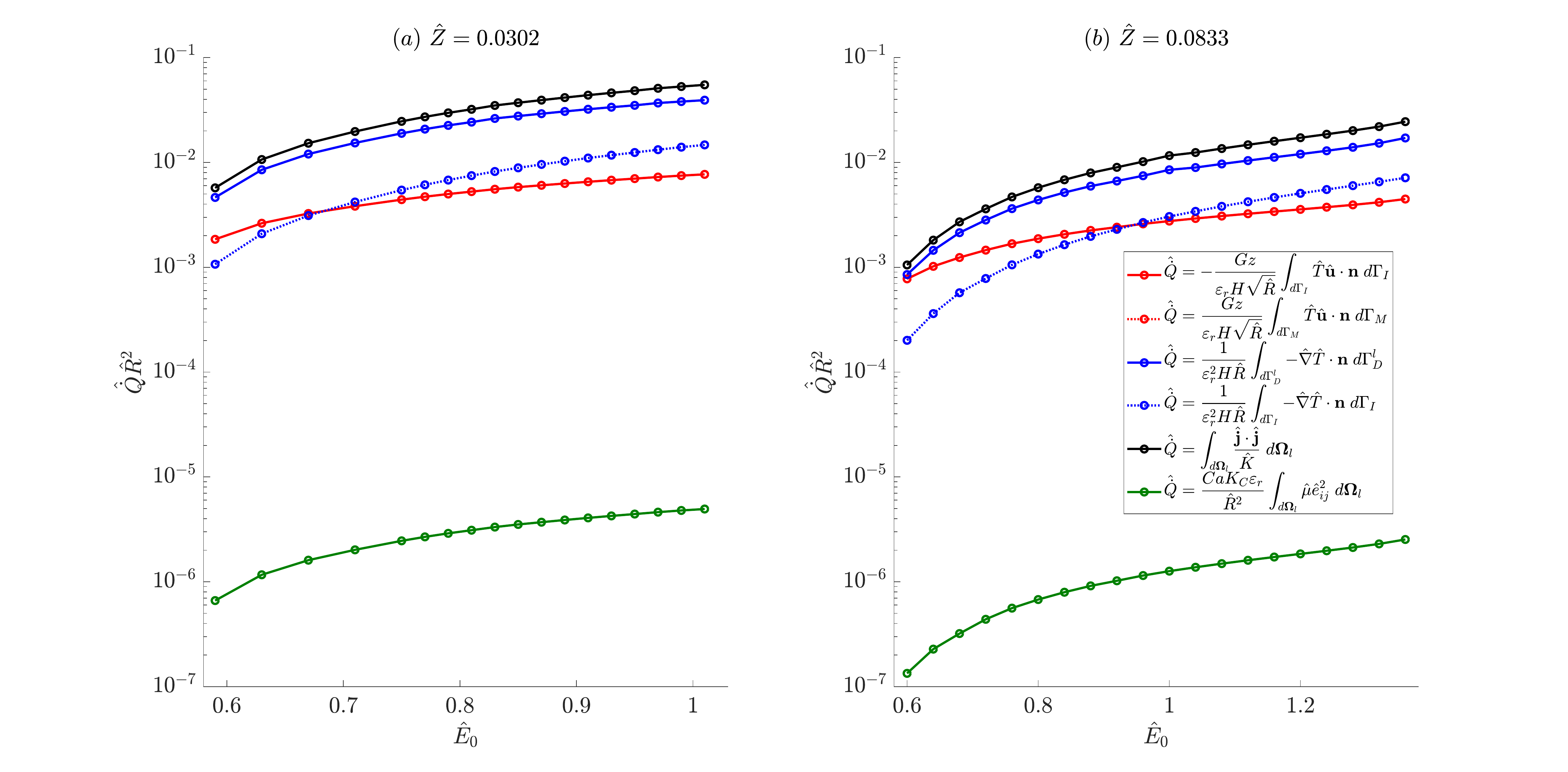}
    \caption{  Non-dimensional power transported by conduction through the channel walls (blue, solid) and the channel inlet (blue, dashed). Power transported by convection into the meniscus through the inlet (red, solid) and out from the meniscus through the meniscus interface (red, dashed).}
    \label{fig:HeatTransport}
\end{figure}

The energy transport results are shown in figures \ref{fig:HeatTransport}a and \ref{fig:HeatTransport}b as a function of the external electric field $\hat{E}_0$. Two different hydraulic impedances are considered corresponding to $\hat{Z} = 0.0302$ and $\hat{Z} = 0.0833$. A contact line radii of $\hat{R} = 64.23$  (3 $\mu$m for EMI-BF$_4$, respectively). Figure \ref{fig:HeatTransport} shows $\hat{\dot{Q}}\hat{R}^2$, which is the non-dimensional power transported in and out of $\mathbf{\Omega}_l$, normalized by the ionic liquid physical properties ($E^*, r^*, k_0$). The first fact to notice is that the enthalpy convected into $\mathbf{\Omega}_l$ through $\Gamma_I$ (red solid line) is practically balanced by the enthalpy convected out of $\mathbf{\Omega}_l$ through ion evaporation on the meniscus interface (red dashed line).
The scale of the Ohmic dissipation and conduction through the walls tends to dominate over the convected power at larger fields. It is shown also that viscous dissipation (in green) is negligible over Ohmic heating (4 orders of magnitude less).

Most of the steady state Ohmic heating is transported via conduction through the channel walls (blue solid line) and the channel inlet (blue dashed line). A rough first order of magnitude estimation of the impact of heat dissipation by conduction to a perfect thermally conducting emitter structure could be stated as follows:
\begin{equation}
    \label{eq:heat_balance}
    E^{*^2} r^{*^3} \kappa_0   \; \hat{\dot{Q}} \hat{R}^2 \approx \rho^e V_D c_p \frac{\Delta T}{\Delta t}
\end{equation}
Where $\rho^e$ is the density of the emitter material, $V^e_D$ is the dry volume of the emitter and $c^e_p$ is its specific heat. Using the values of $E^* \approx 6.95 \cdot 10^{8}$ $\frac{\text{V}}{\text{m}}$, $r^* \approx 46.7$ nm and $\kappa_0 \approx $ 
1 $\frac{\text{S}}{\text{m}}$ and a dry volume of $V^e_D = 0.5$ mm$^3$ per emitter, yields $\frac{\Delta T}{\Delta t} \approx 221 \; \hat{\dot{Q}} \hat{R}^2  \; \frac{\text{K}}{\text{hour}}$ for a carbon emitter ($c^e_p \approx 710 \; \frac{\text{J}}{\text{Kg K}}$, $\rho^e \approx 2260$  
$\frac{\text{Kg}}{\text{m}^3}$) 
and  $\frac{\Delta T}{\Delta t} \approx 137 \; \hat{\dot{Q}} \hat{R}^2 \; \frac{\text{K}}{\text{hour}}$ 
for a tungsten emitter ($c^e_p \approx 134 \; \frac{\text{J}}{\text{Kg K}}$, $\rho^e \approx 19300 \; \frac{\text{Kg}}{\text{m}^3}$). 
For a moderately sized meniscus and electric field value in between the two shown in figure \ref{fig:HeatTransport}, $\hat{\dot{Q}} \hat{R}^2 \approx 5\cdot 10^{-3}$ and $\frac{\Delta T}{\Delta t} \approx 1.11$ $\frac{\text{K}}{\text{hour}}$ and   $0.69$ $\frac{\text{K}}{\text{hour}}$ for a carbon and tungsten emitter, respectively. 

The latter is a worst case estimation of the heating in a floating emitter. Generally speaking, the part of the emitter that captures the heat has substantially higher thermal diffusivity ($\alpha \sim 2.165 \cdot 10^{-4}$ $\frac{\text{m}^2}{\text{s}}$ for carbon and $6.69 \cdot 10^{-5}$  $\frac{\text{m}^2}{\text{s}}$ for tungsten) than the ionic liquid ($1.33 \cdot 10^{-7}$ $\frac{\text{m}^2}{s}$), therefore able to dissipate heat with ease if connected to a thermal reservoir through a similar interface size. These scalings reinforce the notion that the emitter runs fundamentally \textit{cold} in steady state operation, and that stability of the source could be described with accuracy with the constant room temperature boundary condition at the channel walls $\Gamma^l_
D$.

\subsection{Other ionic liquids}
The model presented in this paper is non-dimensional. Due to the similarities in scale for many non-dimensional numbers of ionic liquids numbers, these results are generalizable to other ionic liquids.  

In particular, from the results presented in this paper so far, it has been observed that the upper limits of stability appear to be dependent solely on $\gamma$, the meniscus size and the external field conditions $E_0$, and current emitted appears to be mostly determined by operational  field conditions only ($\hat{E},\hat{Z},\hat{p}_r$), when $r_0 >> r^*$, and very weakly dependent on temperature changes.

\begin{figure}
    \centering
    \includegraphics[width=\linewidth]{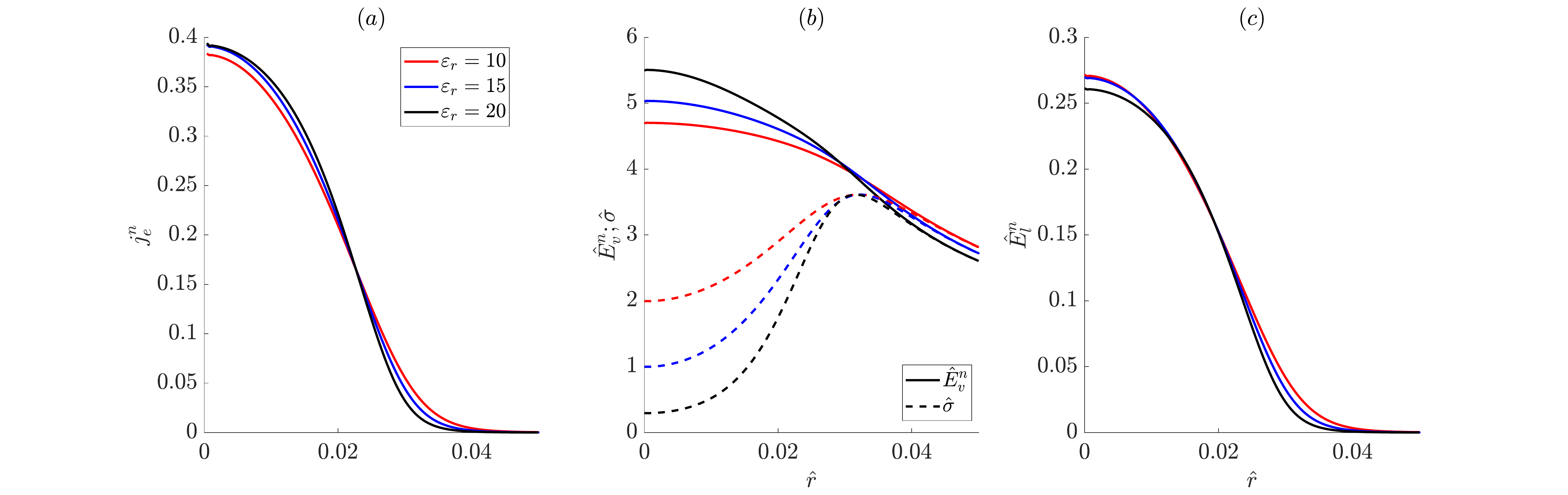}
    \caption{Subfigure a) shows the non-dimensional current density along the emission region in the meniscus interface for different $\varepsilon_r$. Normal electric fields in the vacuum and interfacial charge in b). Normal electric fields in the liquid in c). Non-dimensional numbers dependent on $\varepsilon_r$ were updated as: $We = 10^{-6}$, $Ca =0.017$, $\chi = 1.21\cdot 10^{-3}$, $H = 0.078$, $Gz =0.016$ for $\varepsilon_r = 15$, and   $We = 5.65 \cdot 10^{-7}$, $Ca =0.013$, $\chi = 9.05\cdot 10^{-4}$, $H = 0.044$, $Gz =0.012$ for $\varepsilon_r = 20$.}
    \label{fig:var_er}
\end{figure}

At the $\Delta G = 1$ eV considered in this paper, hydrodynamic stresses play a minor role and the highest variability in the emission conditions and equilibrium configurations will mostly be given by parameters governing the electric problem, namely $\varepsilon_r$. Figure \ref{fig:var_er} shows the current density, normal electric fields and interfacial charge along the emission region for $\varepsilon_r = 10,15,20$, where most of the ionic liquids lie. Similar to what happens with the temperature increase, the effect of a higher charge relaxation time with $\varepsilon_r$, is balanced by higher electric fields in the liquid to yield almost equal currents. Notice how interfacial charge departs from relaxation when the $\varepsilon_r$ increases. Recall the charge relaxtion time $\tau_e = \frac{\varepsilon_0 \varepsilon_r}{\kappa_0}$. From the results shown, a maximum value of $\varepsilon_r$ is predicted beyond which charges cannot travel fast enough to the interface for the scale of the characteristic emission time $\tau_r$, and emission vanishes.

\begin{figure}
    \centering
    \includegraphics[width=\linewidth]{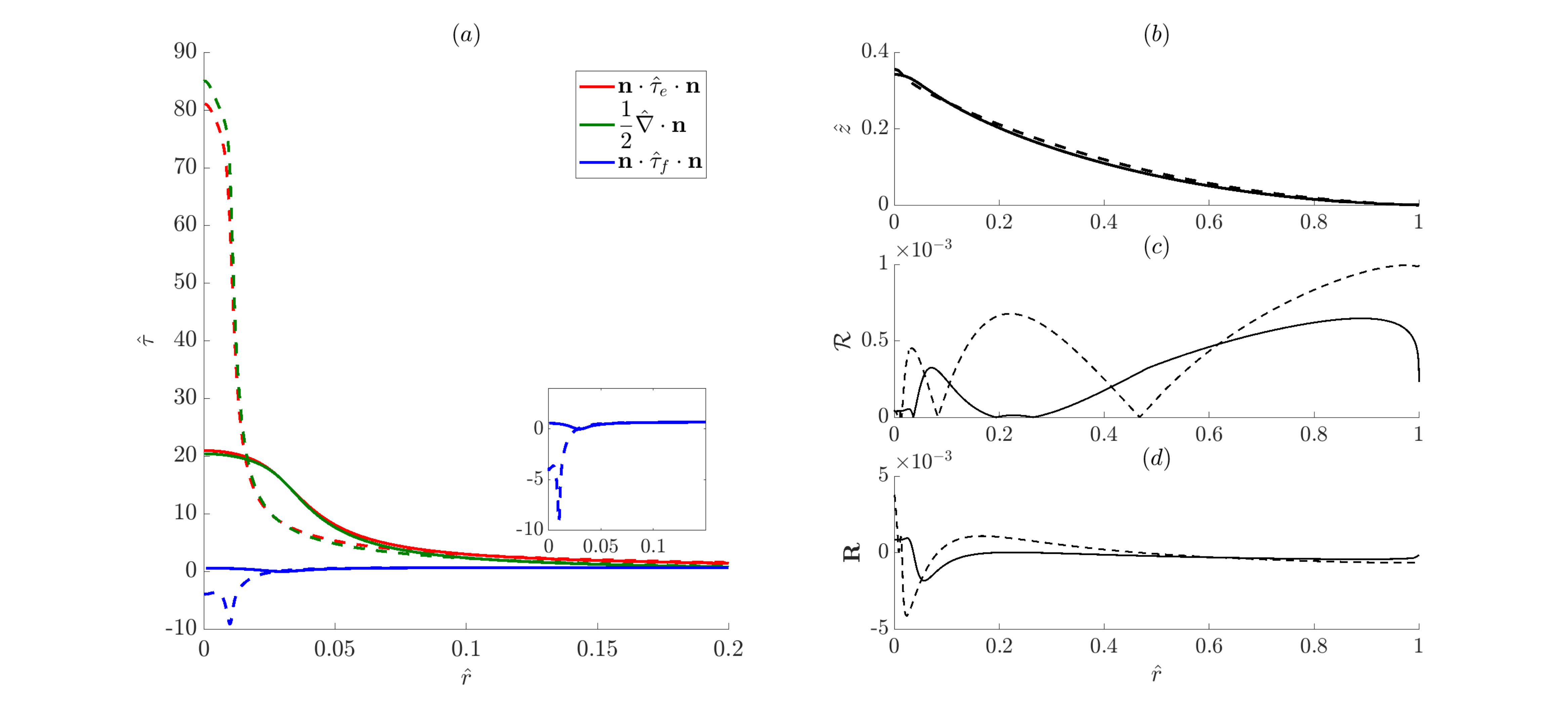}
    \caption{Subfigure a) shows the balance of stresses in the normal direction for the cases of $\Delta G = 1$ eV (solid) and $\Delta G = 1.3$ eV in dashed. Results are shown up to $r=0.2$ to reinforce the values at the emission region. Electric stresses are shown in red, surface tension in green, and hydrodynamic viscous stresses in blue. Equilibrium shapes are shown in  b). Relative and absolute residuals are shown in subfigures c) and d). The same dimensional contact line radius was used of 2 $\mu m$, which corresponds to $\hat{R} = 42.8$ when using $\Delta G = 1$ eV, and $\hat{R} =121.0$ when using $\Delta G = 1.3$ eV. The non-dimensional electric field is $\hat{E}_0 = 0.77$. The non-dimensional hydraulic impedance is $\hat{Z} = 0.105$. Non-dimensional numbers dependent on $\Delta G$ were updated for $\Delta G = 1.3$ eV as: $K_c = 6.37\cdot10^{-4}$, $\psi = 50.18$ $Ca =0.044$, $H = 0.062$, $Gz =0.041$.}
    \label{fig:deltaGreviewed}
\end{figure}

It is also worth mentioning that accurate values of $\Delta G$ are not very well known for ionic liquids. Variations in $\Delta G$ affect the critical field to the square power (eq. \ref{eq:E_critical}) and reduce the value of $r^*$ at a power 4 rate (eq. \ref{eq:r_star}). The sensitivity of the results to increments of $\Delta G$ is substantial and can be seen in figure \ref{fig:deltaGreviewed}, where the balance of normal stresses (subfigure a) and equilibrium shapes (subfigure b) are shown for two meniscus of equal radii and different $\Delta G$ (1 and 1.3 eV in solid and dashed, respectively). Moderate variations of $\Delta G$ originate equilibrium shapes with almost 4 times the magnitude of the normal stresses in the emission region. It is worth mentioning how hydrodynamic stresses start to become relevant in the emission region at higher values of $\Delta G$, yet keeping the total current emitted constant, and invariant to changes in this property.

\section{Conclusions}
\label{sec:conclusions}
A simulation framework based on the equations of electrohydrodynamics has been extended from \cite{Coffman2019ElectrohydrodynamicsField} and applied to explore the static stability of an ionic liquid meniscus experiencing pure ion evaporation. The dependencies of this process on the external field $\hat{E}_0$, meniscus size $\hat{R}$ and hydraulic impedance coefficient $\hat{Z}$ have been analyzed in detail through a comprehensive set of simulation runs. Four regions in the parameter space have been identified, three of which are found to be statically stable. One of them is characterized at low fields with no current emission (\textbf{I}). The rest are characterized by the evaporation of charge (\textbf{II}, \textbf{IV}). 

Region $\textbf{II}$ is characterized by non-dimensional radius higher than $\hat{R} > \hat{R}_{crit}$. Within this region, a set of solutions with cylindrical bumps was identified for combinations of external electric fields larger than $\hat{E}_0 \approx 1.1$ (\textbf{II.b}). These \textbf{II.b} menisci are prone to be dynamically unstable due to pinch off effects not captured by the axially-symmetric formulation in this work. The existence of solutions in \textbf{II} is conditioned to a minimum hydraulic impedance and limited by a maximum current output $I_{max}$ on the order of $I^*$, mostly dependent on the temperature of the ionic liquid.  In addition to the identification of this $I^*$ limit, a possible meniscus bifurcation boundary is found that restricts external fields generating a maximum electric pressure of $2 \frac{2\gamma}{r_0}$, independent of the hydraulic impedance $\hat{Z}$ and external reservoir pressure $\hat{p}_r$. A narrow range of electric fields exists between non-emitting region \textbf{I}) and emitting region \textbf{II}  where hysteretic solutions can be found for the same input impedance and meniscus size. 

A different stable region is identified for meniscus radii below $\hat{R}_{crit}$ (\textbf{IV}), where emission is supported for a continuous range of electric fields that is counter intuitively reduced at high hydraulic impedances.

The reduction of the viscosity coefficient is identified as the sole contributor to the increase of current observed at higher ionic liquid temperatures, as current output is found to depend only on the hydraulic impedance, external field, reservoir pressure and meniscus size. In cases where these parameters are fixed, higher electrical conductivities resulting from heated ionic liquids  play a negligible role due to a better  charge relaxation.

It is necessary to take the energy transport phenomena into account to prevent an underestimation of the ranges of $\hat{R}-\hat{E}_0$ in which pure-ion emitting equilibrium solutions exist. Furthermore, energy transport reveals that Ohmic heating is dissipated mostly via conduction through the emitter structure, regardless of the current emitted. This reinforces the notion that electrosprays in the pure-ion mode run mostly \textit{cold} when the thermal diffusivity of the electrode is substantially larger than that of the ionic liquid. Interestingly, the temperature of the extracted ions is several hundred degrees higher than the liquid bulk \citep{Miller2020MeasurementSources,FernandezDeLaMora2020MeasuringAnalyzer}. 
This disparity is likely due to the molecular stretching and vibrating processes occurring during the emission process, as suggested by molecular dynamics simulations \citep{Coles2012InvestigatingBeams}  and by experimental measurements of the energy loss during the emission process  \citep{Lozano2006EnergySource}. 

This work provides more details of the numerical procedure and provides a substantial extension to the analysis introduced in \cite{Coffman2019ElectrohydrodynamicsField}. However, the model still neglects space charge and does not resolve the Debye layer. The model is also constrained to a simplified planar geometry of the emitter structure and yields steady-state, axially-symmetric solutions and is therefore unable to capture three dimensional bifurcating transitions. A proper eigenmode study should be done to go beyond the static stability analysis performed here and infer global stability boundaries of these menisci. It is expected that the dynamic stability domains will not be very different from the ones computed in this study (at least the ones that lie in region \textbf{II.a}), due to the negligible inertial effects that characterize the ionic liquid flow in these systems. 

Some of these limitations could be removed through the development of a plume model to investigate the effects of space charge on the electric field, which would be required to extend this computational approach to liquid metals. In addition, the resolution of the Debye layer, implementation of more realistic geometries (curved electrodes), and less constrained operational modes (meniscus pinned at any location on the electrode) are left as future work.

Despite the limitations of the model, the findings described in this work reveal the existence of a hard limit in the external field and current throughput above which static pure-ion emission cannot be sustained. These findings appear to confirm experimental observations reported in the literature, where emission stability exists only in a relatively narrow range of electric fields. Such range seems to be incompatible with the cone-jet mode at sufficient hydraulic impedance and $\eta$ values lower than $\sim1$  \citep{FernandezDeLaMora1994TheCones}. The insensitivity of the upper bound of this range to any upstream operational condition, namely hydraulic impedance, bulk temperature of the ionic liquid or input pressure confers some sense of universality in the description of the stability for ionic liquid ion sources. The validity of these results could have a definite impact on the design of engineering devices, for instance by selecting emitter geometries that promote the formation of such a small meniscus working near the upper edge of the stability limit to obtain the highest possible current in the pure ionic mode.
\section{Acknowledgements}
 The authors would like to thank Prof. Manuel Martínez-Sánchez for his insights in analyzing the results of this paper and Obra Social La Caixa and NASA Grant 80NSSC19K0211 for their funding support. PCL acknowledges the support of the Miguel Alemán-Velasco Foundation.
\appendix
\section{Function Space Definitions}
The function spaces used to derive the variational forms of the electrohydrodynamic model are defined here.
Let $\mathcal{L}^p_\alpha \left(\Omega\right)$ be the weighted function space such that:
\begin{equation}
    \mathcal{L}^p_\alpha \left(\mathbf{\Omega}\right) = \{v, \,  \left(\int_{\mathbf{\Omega}} |v|^p \hat{r}^\alpha\right)^{\frac{1}{p}} < \infty \}
\end{equation}
Where $\hat{r}$ is the non-dimensional radial coordinate in the axisymmetric domain $\mathbf{\Omega}$. Let $\mathcal{H}^1\left(\mathbf{\Omega}\right)$ be a Hilbert space of functions such that:
\begin{align}
\begin{split}
    \mathcal{H}^1\left(\mathbf{\Omega}\right) = \{v \,:\; v \in \mathcal{L}^2_1\left(\mathbf{\Omega}\right), \frac{\partial v}{\partial \hat{r}} \in \mathcal{L}^2_1\left(\mathbf{\Omega}\right),    \frac{\partial v}{\partial \hat{z}} \in \mathcal{L}^2_1\left(\mathbf{\Omega}\right)\}
    \end{split}
\end{align}
\begin{align}
\begin{split}
    \mathcal{H}^{\frac{1}{2}}\left(\Gamma\right) = \{ v \,:\, v \in \mathcal{L}^2_1(\Gamma) \;|
    \; \exists \tilde v \in \mathcal{H}^1(\mathbf{\Omega})\, \colon v = tr(\tilde v) \}
    \end{split}
    \end{align}
    
    The latter subspace reads as the space of restrictions to $\Gamma \subseteq \partial\mathbf{\Omega}$ of functions of $\mathcal{H}^1\left(\mathbf{\Omega}\right)$. That is, $v \in \mathcal{H}^{\frac{1}{2}}\left(\Gamma\right)$ means that there exists at least a function $\tilde v \in \mathcal{H}^1\left(\mathbf{\Omega}\right)$ such that $\tilde v=v$ on $\Gamma$. 

\begin{align}
\begin{split}
    \mathcal{V}\left(\mathbf{\Omega},\Gamma_*\right) = \{v \,:\; v \in \mathcal{H}^1 \left(\mathbf{\Omega}\right), v = 0 \quad \text{on} \quad \Gamma_* \}
    \end{split}
\end{align}
\begin{equation}
    \mathcal{S}\left(\mathbf{\mathbf{\Omega}},\Gamma_{*}\right) = \{v \,:\; v \in \mathcal{H}^1 \left(\mathbf{\Omega}\right), v = g \quad \text{on} \quad \Gamma_* \}
\end{equation}
\begin{equation}
    V^1\left(\mathbf{\Omega}\right) =\mathcal{H}^1 \left(\mathbf{\Omega}\right) \cap \mathcal{L}^2_{-1}\left(\mathbf{\Omega}\right)
\end{equation}

\begin{align}
\begin{split}
    \mathcal{\chi}\left(\mathbf{\Omega},\Gamma_*\right) = \{\vec{\mathbf{v}} = \left(v_r,v_z\right)\;: \vec{\mathbf{v}} \in V^1\left(\mathbf{\Omega}\right) \times \mathcal{H}^1\left(\mathbf{\Omega}\right), \; \vec{\mathbf{v}} = 0 \quad \text{on} \quad \Gamma_* \}
\end{split}
\end{align}
Where $\Gamma_* \subseteq \partial \mathbf{\Omega}$ is the part of $\partial \mathbf{\Omega}$ where Dirichlet boundary conditions equal to function $g$ are imposed. 
\section{Variational Forms}
\label{sec:Annex_Variational_Forms}
The variational formulation of the electric problem at iteration $k$ consists of finding $\left(\hat{\phi}^k, \hat{\sigma}^k\right)$ in $\mathcal{S}\left(\mathbf{\Omega}_l\cup\mathbf{\Omega}_v, \Gamma_*\right) \times \mathcal{H}^{\frac{1}{2}}\left(\Gamma_M\right)$ such that:
\begin{align}
\begin{split}
    F\left(\hat{\phi}^k,\hat{\sigma}^k; v,\bar{\lambda}\right) = \int_{\mathbf{\Omega}_l} \varepsilon_r \hat{r} \hat{\nabla} \hat{\phi} \cdot \hat{\nabla} v \; d\mathbf{\Omega}_l + \int_{\mathbf{\mathbf{\Omega}}_v}  \hat{r} \hat{\nabla} \hat{\phi} \cdot \hat{\nabla} v \; d\mathbf{\Omega}_v \\ - \int_{\mathbf{\Omega}_l} \hat{r}\hat{\rho}_m^{k-1}\; v \; d\mathbf{\Omega}_l
    - \int_{\Gamma_M}  \hat{r} \hat{\sigma}^k \; v \; d \Gamma_M - \int_{\Gamma_M} \hat{r} \hat{\sigma}^k \; \bar{\lambda} \; d\Gamma_M \\+ \int_{\Gamma_M}        \hat{r} \frac{\hat{K}^{k-1}\left(-\hat{\nabla}\hat{\phi}^{v^k} \cdot \mathbf{n}\right) + \varepsilon_r \hat{j}^{k-1}_{conv}}{\hat{K}^{k-1} + \frac{\hat{T}^{k-1}}{\chi}\exp{\left(\frac{\psi}{\hat{T}^{k-1}}\left(1 - \hat{R}^{\frac{-1}{4}}\sqrt{-\hat{\nabla}\hat{\phi}^{v^k} \cdot \mathbf{n}}\right)\right)}} \; \bar{\lambda} \; d\Gamma_M 
    = 0 \\ \quad \forall \left(v,\hat{\lambda}\right) \in \mathcal{V}\left(\mathbf{\Omega}_l\cup\mathbf{\Omega}_v, \Gamma_*\right) \times \mathcal{H}^{\frac{1}{2}}\left(\Gamma_M\right)\\
\end{split}
\label{eq:weak_form_electric_def}
\end{align}
Where according to eq. \ref{eq:rho_m}:
\begin{equation}
    \hat{\rho}_m^{k-1} = \varepsilon_r\frac{\hat{\nabla} \hat{K}^{k-1}\cdot \hat{\nabla} \hat{\phi}^k}{\hat{K}^{k-1}}
\end{equation}
Where $\Gamma_* = \Gamma_I \cup \Gamma_D \cup \Gamma_R$ and $g$ are set according to the boundary conditions in \ref{eq:Electric_BC}. 

System \ref{eq:weak_form_electric_def} is highly non-linear and can be solved using standard Newton iterations. More details of the Jacobian form of system \ref{eq:weak_form_electric_def} can be read in \cite{Gallud2019AThesis}.

The variational formulation of the fluid problem at iteration $k$ consists of finding $\left(\hat{\mathbf{u}}^k,  \hat{p}^k,\mathbf{n}\cdot\hat{\tau}^k_f \cdot \mathbf{n}\right)$ in $\mathcal{\chi}\left(\mathbf{\Omega}_l,\Gamma^l_D\right) \times \mathcal{H}^1\left(\mathbf{\Omega}_l\right)\times\mathcal{H}^{\frac{1}{2}}\left(\Gamma_M\right)$ such that:
\begin{equation}
    \begin{aligned}
    a \left(\hat{\mathbf{u}}^k,\mathbf{w}\right) +  d \left(\hat{\mathbf{u}}^{k-1},\hat{\mathbf{u}}^k,\mathbf{w}\right) + b\left(\hat{\mathbf{w}},\hat{p}^k\right) + c\left(\mathbf{w},\mathbf{n}\cdot\hat{\tau}^k_f \cdot \mathbf{n}\right)\\ =  -l\left(\mathbf{t}\cdot \left(\hat{\mathbf{\tau}}^{v^k}_e-\hat{\mathbf{\tau}}^{l^k}_e\right) \cdot \mathbf{n},\mathbf{w}\cdot\mathbf{t}\right) - 2\int_{\mathbf{\Omega}_l} \hat{r} \hat{\rho}_m^{k-1}\; \hat{\nabla} \hat{\phi}^k \cdot \hat{\mathbf{w}} \; d\mathbf{\Omega}_l\\
    b\left(\hat{\mathbf{u}}^k,q\right) = 0 \\
    c\left(\mathbf{u}^k,\lambda\right) = l\left(j^{e^k}_n,\lambda\right)
    \\ \forall \mathbf{w}, q, \lambda \in \mathcal{\chi}\left(\mathbf{\Omega}_l,\Gamma^l_D\right)\times \mathcal{H}^1 \left(\mathbf{\Omega}_l\right) \times \mathcal{H}^{\frac{1}{2}}\left(\Gamma_M\right).
    \end{aligned}
 \label{eq:weak_form_fluid_def}
\end{equation}
Where:

\begin{equation*}
\begin{aligned}
a \left(\hat{\mathbf{u}},\mathbf{w}\right) =  \int_{ \mathbf{\Omega}_l} \hat{r} \frac{\varepsilon_r Ca \hat{\mu}^{k-1}}{\hat{R}^{\frac{1}{2}}} \left(\hat{\nabla} \hat{\mathbf{u}} + \hat{\nabla} \hat{\mathbf{u}}^T  \right) : \left(\hat{\nabla} {\mathbf{w}} + \hat{\nabla} {\mathbf{w}}^T  \right) \; d  \mathbf{\Omega}_l 
 + \int_{\mathbf{\Omega}_l} 2 \frac{\varepsilon_r Ca \hat{\mu}^{k-1}}{\hat{R}^{\frac{1}{2}}}\frac{\hat{u}_r w_r}{\hat{r}}\; d \mathbf{\Omega}_l
 \end{aligned}
 \label{eq:weak_fluid}
\end{equation*}
\begin{equation*}
   d \left(\hat{\mathbf{u}},\hat{\mathbf{u}},\mathbf{w}\right) = \int_{\mathbf{\Omega}_l} \hat{r} \varepsilon^2_r We \left[\left(\hat{\mathbf{u}}\cdot \hat{\nabla} \right) \hat{\mathbf{u}}\right] \cdot \hat{\mathbf{w}} \; d\mathbf{\Omega}_l
\end{equation*}
\begin{equation*}
   b \left(\hat{\mathbf{u}},q\right) = -\int_{\mathbf{\Omega}_l} \hat{\nabla} \cdot \left(\hat{r}\hat{\mathbf{u}}\right) \; q \; d\mathbf{\Omega}_l
\end{equation*}
\begin{equation*}
   c \left(\hat{\mathbf{u}},\lambda\right) = -\int_{\Gamma_M} \hat{r} \hat{\mathbf{u}} \cdot \mathbf{n} \; \lambda \; d\Gamma_M
\end{equation*}
\begin{equation*}
   l \left(h,\lambda\right) = -\int_{\Gamma_M} \hat{r} h \; \lambda \; d\Gamma_M
\end{equation*}
The variational formulation of the energy problem at iteration $k$ consists of finding $\left(\hat{T}^k\right)$ in $\mathcal{S}\left(\mathbf{\Omega}_l,\Gamma_I \cup \Gamma^l_D\right)$ such that:
\begin{equation}
\begin{aligned}
 \int_{\mathbf{\Omega}_l} \frac{\hat{r}}{\varepsilon^2 H \sqrt{\hat{R}}}\hat{\nabla} \hat{T}^k\cdot\hat{\nabla}v \; d\mathbf{\Omega}_l + \int_{\mathbf{\Omega}_l}  \hat{r} \Lambda \;\hat{T}^k\;\hat{\nabla}\hat{\phi}^{k}\cdot\hat{\nabla}\hat{\phi}^{k}\; v \; d\mathbf{\Omega}_l + \int_{\mathbf{\Omega}_l} \hat{r} \frac{Gz}{ \varepsilon_r H \sqrt{\hat{R}}}\hat{\mathbf{u}}^{k}\cdot\hat{\nabla}\hat{T}^k\; v \; d\mathbf{\Omega}_l \\= \int_{\mathbf{\Omega}_l} \hat{r}\left(1 - \Lambda\right)\hat{\nabla}\hat{\phi}^{k}\cdot\hat{\nabla}\hat{\phi}^{k}  \;v\; d\mathbf{\Omega}_l + \int_{\mathbf{\Omega}_l} \hat{r}\frac{Ca K_C \varepsilon_r}{\hat{R}^2} \hat{\mu}  \hat{e}^{k^2}_{ij}  \;v\; d\mathbf{\Omega}_l \qquad
 \forall v \in \mathcal{V}\left(\mathbf{\Omega}_l, \Gamma_M \cup \Gamma^l_D \right)
 \end{aligned}
\label{eq:weak_energy}
\end{equation}

The equation eq. \ref{eq:weak_energy} is non-linear in $T$, since the model for $\hat{\mu} = \frac{1}{1+\Lambda\left(\hat{T}-1\right)}$. 

\section{Interpretation of the calculation of $\rho_m$ and $\sigma$}
\label{sec:Annex_Interpretation_rhom}

Consider the full electric problem in the bulk liquid posed in this paper (eqs. \ref{eq:max_faraday},\ref{eq:poisson_liq},\ref{eq:charge_conserv_bulk}) for the unknowns $\mathbf{E}, \rho_{sc}$, where the Debye layer is included as a part of the domain where the solution is sought. Ideally, the solution to this problem involves the calculation of the whole space charge distribution $\rho_{sc}$ in the bulk liquid domain and Debye layer. The Taylor-Melcher leaky dielectric model \citep{Saville1997ELECTROHYDRODYNAMICS:TheModel} approximates the steady state solution to this problem by considering that the fluid is quasi-neutral ($\rho_{sc} = 0$) in the majority of the liquid domain, except for the larger variation of $\rho_{sc}$ existing in the Debye layer. Since the Debye layer is generally very narrow in comparison to the lengh-scales of the problem in question, the leaky dielectric model uses the integrated value of $\rho_{sc}$ across the Debye layer as a surface charge $\sigma$ to avoid the resolution of the full charge distribution. In this framework, the Poisson equation yields:

\begin{equation}
    \sigma = \int_{\delta} \rho_{sc} d\delta = \varepsilon_0 E^v_n - \varepsilon_0 \varepsilon_r {E}^l_n
    \end{equation}

In the problem presented in this paper, the bulk fluid cannot be considered quasi-neutral due to gradients in conductivity, and the total charge distribution will extend beyond that present in the Debye layer. To understand this situation, the space charge distribution $\rho_{sc}$ can be considered as the sum of two distributions $\rho_{sc} = \rho_m + \rho_{f}$. The space charge $\rho_{m}$ is only a byproduct of the conductivity gradients in the bulk ($\rho_{m} = 0$ in the Debye layer). The space charge $\rho_f$ is only the free charge originated in the Debye layer that is also subject to evaporation ($\rho_{f} = 0$ in the bulk liquid). One can solve eqs.
\ref{eq:poisson_liq},\ref{eq:max_faraday},\ref{eq:charge_conserv_bulk} separately for the fields originated from the two charge distributions ($\mathbf{E} = \mathbf{E}_m + \mathbf{E}_{f}$). Since the equations are linear, these fields can be added safely. The integrated Poisson equation for $\rho_m$ and $\rho_f$ at the interface yields:
\begin{equation}
    \sigma_m = \varepsilon_0 E^v_{n_m} - \varepsilon_0 \varepsilon_r {E}^l_{n_m}
    \label{eq:sigmam}
\end{equation}
\begin{equation}
    \sigma_f = \varepsilon_0 E^v_{n_f} - \varepsilon_0 \varepsilon_r {E}^l_{n_f}
    \label{sigmaf}
\end{equation}

This separation is consistent with the full problem if providing adequate boundary conditions for the split electric field in the surface charge approximation. If $\sigma_m = 0$, then due to charge conservation at the interface eq. \ref{eq:charge_conserv_meniscus} yields $\kappa E^l_{n_m} = 0$. Inserting this in eq. \ref{eq:sigmam} yields $E^v_{n_m} = 0$ as a boundary condition for the electric field associated to $\rho_m$.

In this paper, the total electric field $\mathbf{E}$ is computed for convenience, as shown in system \ref{eq:weak_form_electric_def}.

 \section{Lumped parameter equation for the pure-ion current emitted by an ionic liquid meniscus}
 \label{sec:Annex_Lumped_Model}
 
 A simplified model is presented here to develop an expression for the current emitted by the meniscus as a function of the electric field in the vacuum side near the emission region $E^v_n$ and also a function of an approximate value of the temperature around the tip.  This approximation is valid for menisci with relatively large non-dimensional contact line radius $\hat{R} > 60$, where the upper limits of stability are apparently determined by a maximum current output, and the electric stress is almost completely balanced by the surface tension stress \cite{Coffman2019ElectrohydrodynamicsField}.
 
 For these reasons, any viscous effect, hydraulic pressure drop along the feeding channel, convective charge transport and temperature gradients are neglected.
 
 The electric fields and current density are non-dimensionalized in equation \ref{eq:current_Envac} by $E^*$ and $j^*$ respectively. This yields:

 \begin{equation}
  \hat{j}^e_n = \frac{\hat{K}\hat{E}^n_v}{1+\frac{F}{\hat{K}}}
  \label{eq:nond_current_Envac}
\end{equation}
Where $F= F\left(\hat{E},\hat{T}\right) = \frac{\tau_e}{\tau_r}\exp{\frac{\psi}{\hat{T}}\left(1-\sqrt{\hat{E}}\right)}$, and $\hat{K} = \hat{K}\left(\hat{T}\right) = 1 +\Lambda\left(\hat{T}-1\right)$.

The non-dimensional equation \ref{eq:charge_conserv_meniscus}, $\hat{j}^e_n = \varepsilon_r \hat{K}\hat{E}^l_n$, is used to get an expression for $\hat{E}^l_n$ as a function of $\hat{E}^v_n$. 

The emission region is modeled as a spherical cap. The non-dimensional equation \ref{eq:equilibrium_stresses_normal} yields:

\begin{equation}
    \hat{E}_n^{v^2} - \frac{F^2\hat{E}_n^{v^2}}{ \varepsilon_r \hat{K}^2\left(1+\frac{F}{\hat{K}}\right)^2} = \frac{1}{\hat{r}_c}
    \label{eq:nond_norm1}
\end{equation}

Where $\hat{r}_c = \frac{r_c}{r^*}$ is the non-dimensional radius of curvature of the spherical cap emission region.

The total current emitted $\hat{I} = \frac{I}{I^*} = \hat{r}^2_c\hat{j}^e_n$, can be used to substitute the radius of curvature in equation \ref{eq:nond_norm1} as a function of $\bar{I}$.

\begin{equation}
     \hat{E}_n^{v^2} - \frac{F^2\hat{E}_n^{v^2}}{ \varepsilon_r \hat{K}^2\left(1+\frac{F}{\hat{K}}\right)^2} = \sqrt{\frac{\hat{K}\hat{E}^v_n}{\hat{I}\left(1+\frac{F}{\hat{K}}\right)}}
    \label{eq:nond_norm2}
\end{equation}
Finally, $\hat{I}$ can be isolated from \ref{eq:nond_norm2}:

\begin{equation}
   \hat{I} = \frac{\hat{K}\hat{E}^v_n}{\left(1+\frac{F}{\hat{K}}\right) \left(\hat{E}_n^{v^2} - \frac{F^2\hat{E}_n^{v^2}}{ \varepsilon_r \hat{K}^2\left(1+\frac{F}{\hat{K}}\right)^2}\right)^2}
    \label{eq:nond_current_def_lumped}
\end{equation}
 
\begin{figure}
    \centering
    \includegraphics[width=0.6\textwidth]{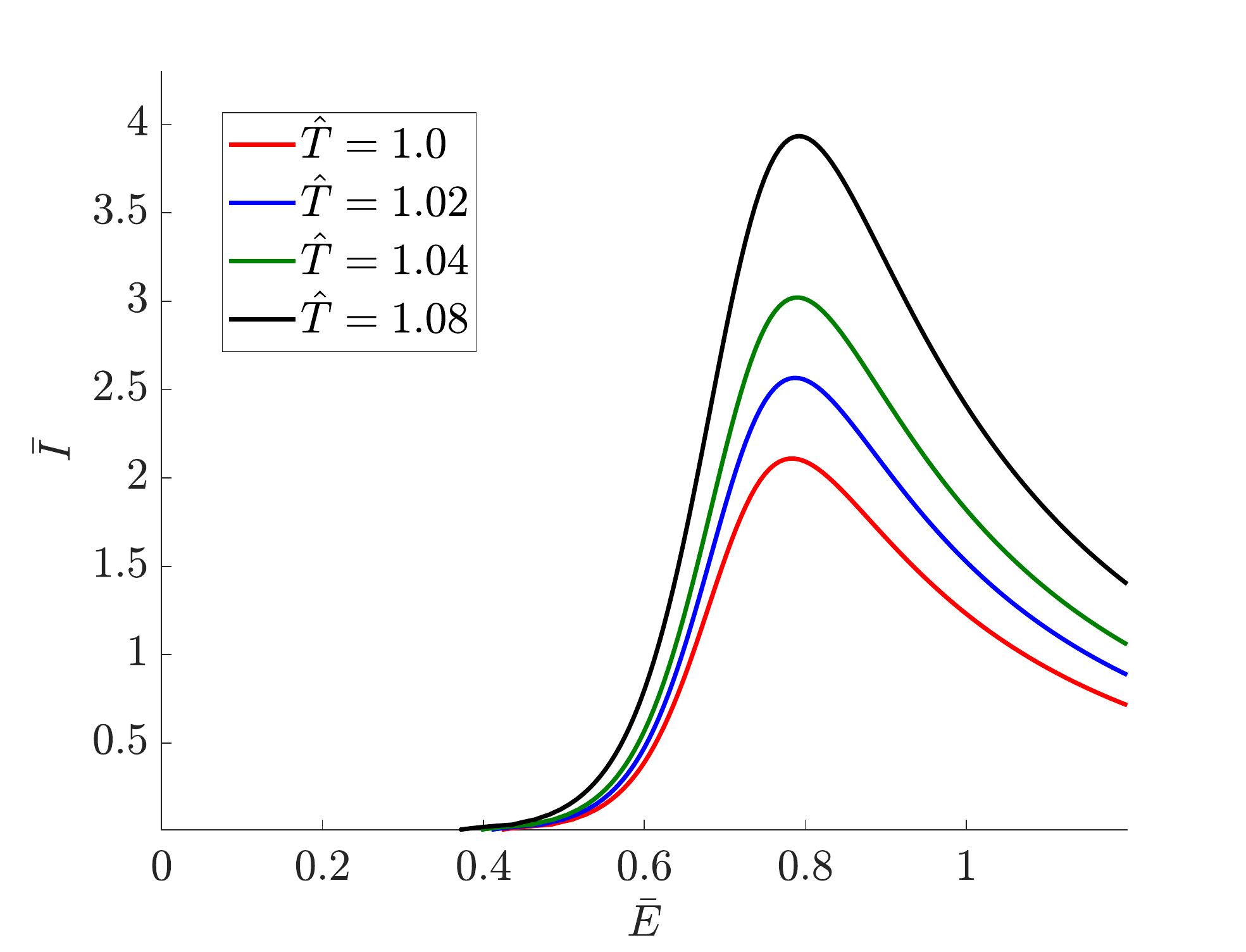}
    \caption{Current emitted for the zero-th dimensional model presented in appendix \ref{sec:Annex_Lumped_Model}. Equation (\ref{eq:nond_current_def_lumped}) presents a maximum at $\bar{E} = \frac{E^n_v}{E^*} \approx 0.78$, at a point close to the field of maximum current observed in figure \ref{fig:IV-EXP} b).} 
    \label{fig:current_lumped}
\end{figure}

 Figure \ref{fig:current_lumped} shows the non-dimensional current emitted using the lumped equation in \ref{eq:nond_current_def_lumped} as a function of the non-dimensional external electric field. It can be observed that this current limit is on the order of the maximum currents observed in figure \ref{fig:IV-CR} for both hydraulic impedance coefficients.

\section{Mesh Convergence Details}
\label{sec:Mesh_Convergence}
In this annex section, we provide details of the mesh used, and numerical data regarding the convergence to the equilibrium shape. The non-dimensional physical parameters for this analysis are the same as the ones used in the results of the paper, and the non-dimensional operational parameters are $\hat{E} = 0.7$, $\hat{R} = 176.8$ and $\hat{Z} = 0.0833$. The non-dimensional parameters used are very close to the limit cases of the results presented in this paper (very high $\hat{Z}$ and $\hat{R}$).

Two different initial solutions are provided to the solver that are very far away from the equilibrium solution. 
The first initial solution is a "flattened" Taylor cone of semiangle $60^\circ$, with constant non-dimensional surface tension stress $\frac{1}{2}\hat{\nabla} \cdot \mathbf{n} = 70$ in the numerical emission region $\left(\hat{r} \in \left[0,\frac{2.5}{\hat{R}}\right]\right)$.

The second initial solution is the equilibrium shape corresponding to $\hat{E}=1.1$.

The procedure is repeated for three different meshes with increasing element size: a coarse mesh, a medium mesh and a fine mesh.

In the coarse mesh, the interface is discretized in 500 points. The points are distributed geometrically, containing 90 points in the aforementioned emission region distance. For the medium mesh, the interface is discretized in 900 points and 150 in the emission region distance. For the fine mesh, the points are 1750 and 250, respectively. No solution converged for a coarser mesh. The numerical parameters used are $\epsilon = 0.01$ for the convergence limit (eq. \ref{eq:residue_tol}) and $\beta = 0.01$.

With regard to the finite element category, second order Lagrange triangular elements were used for the potential $\hat{\phi}$, the velocity $\hat{\mathbf{u}}$ and the temperature $\hat{T}$. First order Lagrange triangular elements were used for the interface charge $\hat{\sigma}$ and the pressure $\hat{p}$. A transfinite mesh was used in the vicinity of the emission region to ensure accuracy of the normal stresses. Out of the numerical emission region, a mesh frontal algorithm was used. 

For the fine mesh, a total of 208792 elements was used for the vacuum domain, and 180679 for the liquid, respectively. For the medium mesh, 105966 and 107181, respectively. For the coarse mesh, 59191 and 72164. The numbers are averaged, since remeshing is done to prevent the quality of mesh from decaying due to large deformations. 

\begin{figure}
    \centering
    \includegraphics[width=\linewidth]{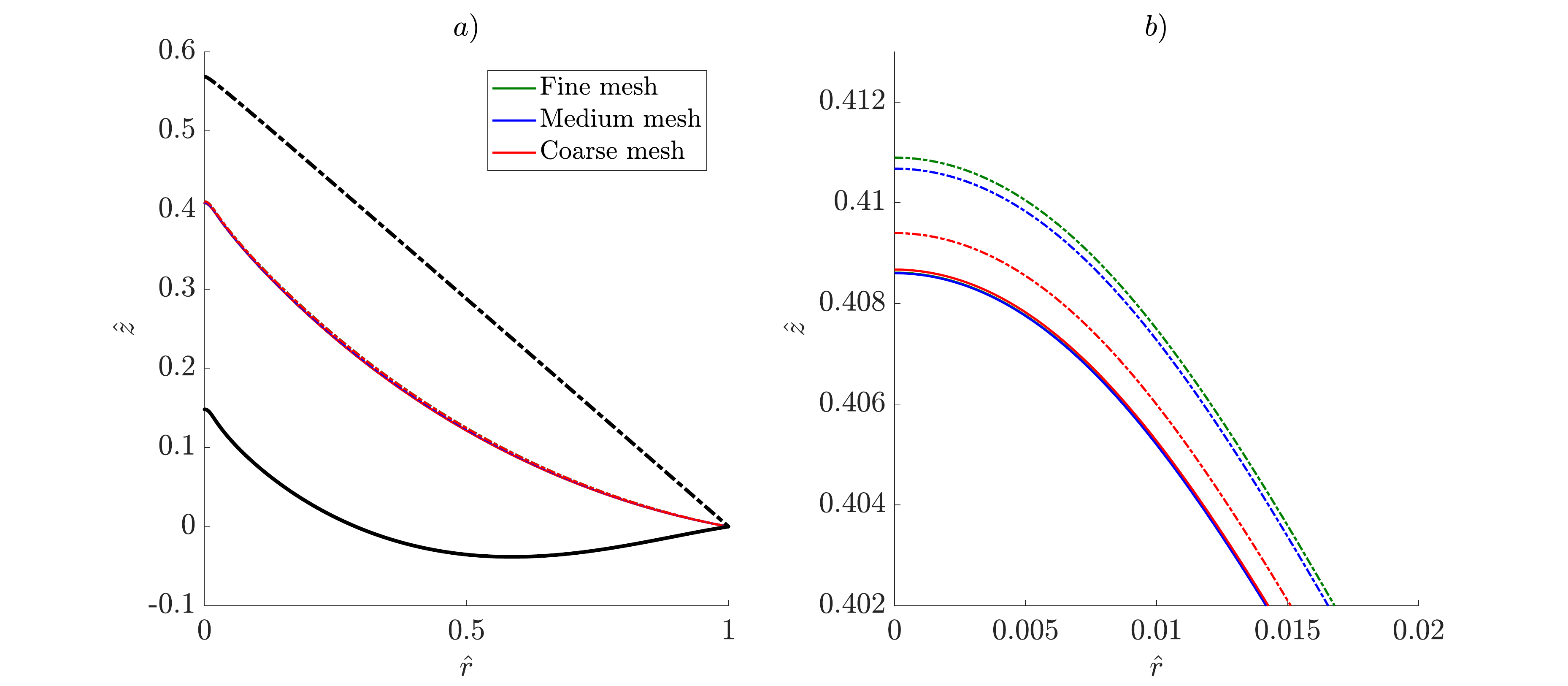}
    \caption{Subfigure $a)$ shows equilibrium shapes (colored) and initial solutions (black) used in the convergence analysis for the three different meshes used. Dashed plots reference the initial solution in the conical shape. Solid plots reference the high field initial solution. Subfigure $b)$ shows a zoom of the equilibrium shapes near the emission region.} 
    \label{fig:eqShapesConvergence}
\end{figure}

Figure \ref{fig:eqShapesConvergence} shows both the initial solutions of the two cases considered in black, and the convergence solutions in colored. It can be observed how despite the initial solutions being very far from each other, they converge to the same solution for the three meshes considered, thus reinforcing the idea that only a statically stable solution exists for given external conditions. Subfigure $b)$ shows that the difference of the solutions as a function of which initial shape was provided is less than $0.4\%$. This variability is within the residue tolerance limit of $\epsilon = 0.01$.

\begin{figure}
    \centering
    \includegraphics[width=\linewidth]{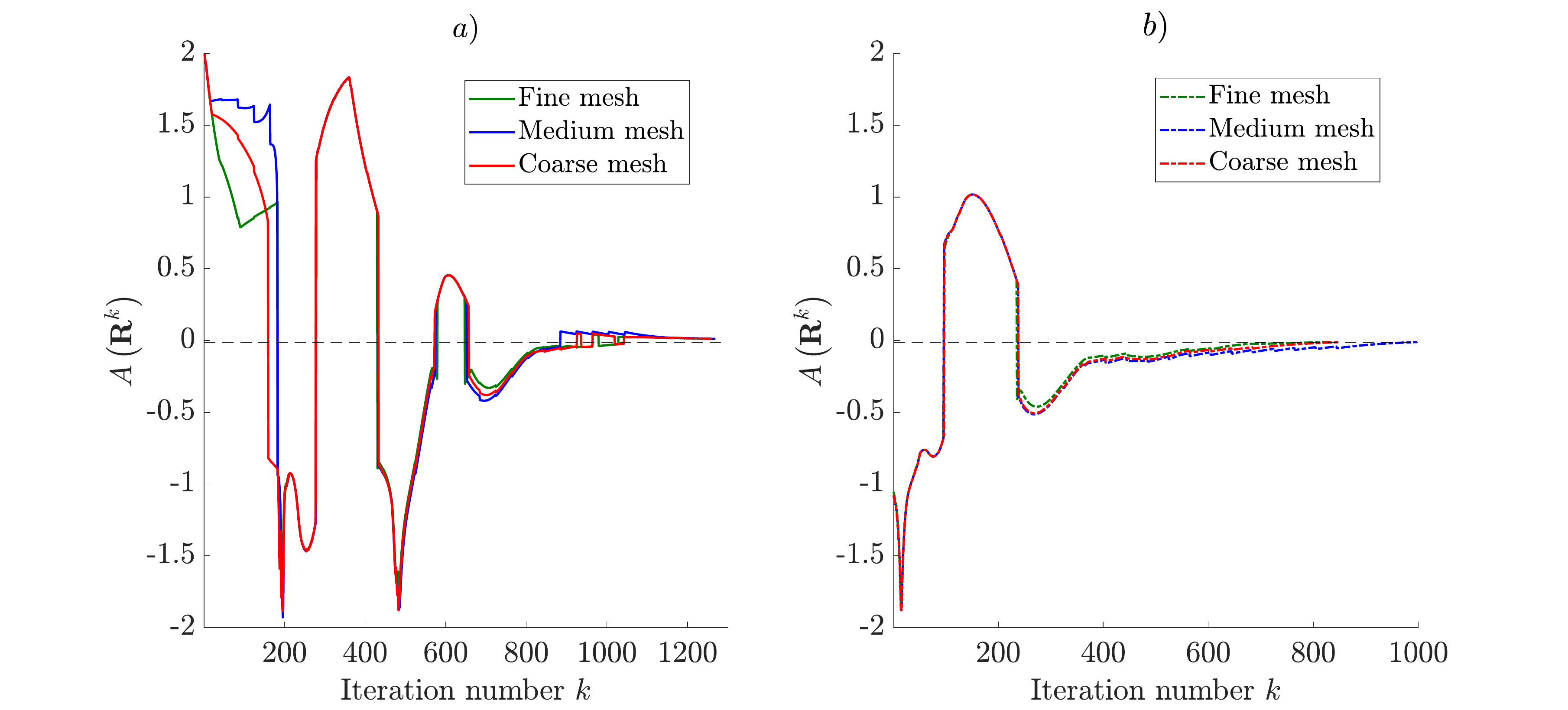}
    \caption{Subfigure $a)$ shows the residue function as a function of the iteration process for the three meshes starting from the high field solution at $\hat{E}$. Subfigure $b)$ shows the results starting with the flattened Taylor cone solution. The dashed lines in both subplots mark the convergence boundaries of $\|\mathbf{R}^k\| < \varepsilon$} 
    \label{fig:iterationConvergence}
\end{figure}

Figure \ref{fig:iterationConvergence} shows the equilibrium residual as a function of the number of iterations $k$. Notice the chaotic behaviour in the first $500$ iterations probably caused because the initial solutions are very far from equilibrium. The convergence trajectory is very similar for the three meshes considered. The finer mesh converges  earlier, but at the expense of more computational time.

\section*{Declaration of Interests}
The authors report no conflict of interest.

\bibliographystyle{jfm}
\bibliography{references}

\end{document}